\documentclass[aps,prd,showpacs,twocolumn,superscriptaddress,nofootinbib,preprintnumbers]{revtex4-1} 

\makeatletter
\def\p@subsection{}
\makeatother

\usepackage{color,graphicx,amsmath,amssymb,lipsum}
\usepackage{grffile}
\usepackage[colorlinks=true,citecolor=blue,linkcolor=blue,urlcolor=blue, backref=false,pdfborder={0 0 0}]{hyperref}

\usepackage[normalem]{ulem}
\usepackage[utf8]{inputenc} 
\usepackage{mathtools}

\usepackage{float}

\usepackage{ulem}
\usepackage{diagbox}

\usepackage[dvipsnames]{xcolor}
\usepackage{mathrsfs}

\newcommand{\be}{\begin{equation}}
\newcommand{\ee}{\end{equation}}
\newcommand{\beqa}{\begin{eqnarray}}
\newcommand{\eeqa}{\end{eqnarray}}

\renewcommand\k{{\bf k}}
\newcommand\q{{\bf q}}

\renewcommand\a{\alpha}

\def\d{\partial}
\newcommand{\bseq}{\begin{subequations}}
\newcommand{\eseq}{\end{subequations}}

\renewcommand{\ln}{\mathop{\rm ln}\nolimits}

\newcommand{\khat}{\hat{k}}
\newcommand{\qhat}{\hat{q}}

\def\gsim{\raise0.3ex\hbox{$\;>$\kern-0.75em\raise-1.1ex\hbox{$\sim\;$}}}
\def\lsim{\raise0.3ex\hbox{$\;<$\kern-0.75em\raise-1.1ex\hbox{$\sim\;$}}}

\def\beqn#1{\begin{equation}\label{#1}}
\def\eeqn{\end{equation}}

\def\beqa#1{\begin{eqnarray}\label{#1}}
\def\eeqa{\end{eqnarray}}

\def\kmax{{k_\text{max}}}
\def\hMpc{h{\text{Mpc}}^{-1}}
\def\Mpch{h^{-1}{\text{Mpc}}}

\def\Z2{$\mathcal{Z_2}$}

\newcommand {\ignore}[1]{}

\begin{document}

\preprint{MIT-CTP/5721}

\title{Fundamental physics with the Lyman-alpha forest: 
constraints on the growth of structure
and neutrino masses  
from 
SDSS 
with effective field theory
}

\author{Mikhail M. Ivanov}
\email{ivanov99@mit.edu}
\affiliation{Center for Theoretical Physics, Massachusetts Institute of Technology, 
Cambridge, MA 02139, USA}

\author{Michael W. Toomey}
\email{mtoomey@mit.edu}
\affiliation{Center for Theoretical Physics, Massachusetts Institute of Technology, 
Cambridge, MA 02139, USA}

\author{Naim~G{\" o}ksel~Kara{\c c}ayl{\i}}
\email{karacayli.1@osu.edu}
\affiliation{Center for Cosmology and AstroParticle Physics, The Ohio State University, 191 West Woodruff Avenue, Columbus, OH 43210, USA}
\affiliation{Department of Astronomy, The Ohio State University, 4055 McPherson Laboratory, 140 W 18th Avenue, Columbus, OH 43210, USA}
\affiliation{Department of Physics, The Ohio State University, 191 West Woodruff Avenue, Columbus, OH 43210, USA}

\begin{abstract} 
We present an effective field theory (EFT) approach to extract fundamental 
cosmological parameters from the Lyman-alpha forest flux fluctuations as an alternative to the standard simulation-based 
techniques. As a first application, we re-analyze the publicly available 
one-dimensional Lyman-alpha flux power spectrum data from the Sloan Digital Sky Survey. 
Our analysis relies on informative priors on EFT parameters 
which we extract from a combination of public hydrodynamic simulation and emulator data.
Assuming the concordance cosmological model, our one-parameter analysis yields 
a $2\%$ measurement of the late time 
mass fluctuation amplitude
$\sigma_8 = 0.841\pm 0.017$, or equivalently, 
the structure growth parameter $S_8 = 0.852\pm 0.017$, 
consistent with the standard  
cosmology. This result is obtained 
assuming that non-linear EFT parameters are cosmology-independent functions of the 
linear bias parameter.
When this assumption is loosened, 
the limit degrades by a factor of 3, suggesting that informative
priors are necessary for competitive constraints.
Combining our EFT likelihood
with Planck + baryon acoustic oscillation data, we find a new constraint 
on the total neutrino mass, $\sum m_\nu<$ 0.08 eV (at 95\% CL).
Our study defines priorities for the development of EFT methods 
and  
sets the benchmark for cosmological analyses of the Lyman-alpha forest
data from the Dark Energy Spectroscopic Instrument.
\end{abstract}

\maketitle

\textit{Introduction.}
The distribution of galaxies and matter on
cosmic scales, known as large-scale 
structure, is a key probe of cosmological physics.
In particular, the cosmic web has been instrumental in discoveries and 
subsequent studies of dark matter and dark energy, thereby 
helping establish  
the standard $\Lambda$ Cold Dark Matter ($\Lambda$CDM) model of cosmology. 
Within $\Lambda$CDM, structure formation has a hierarchical pattern: 
small structures form in the early Universe at high redshifts while 
larger structures form at later times.
Some of the earliest and smallest clustering structures  
available for observations are neutral hydrogen 
clouds that fill the intergalactic medium at redshifts $z\gtrsim 2$.
These clouds absorb light from 
distant quasars, creating characteristic combs in their 
spectra, known as the Lyman-alpha  
(Ly$\alpha$)
forest~\cite{1965ApJ...142.1633G,1997ApJ...488..532C,Croft:1997jf,Croft:2000hs,Hui:1997dp,Gnedin:1997td,McDonald:1999dt,SDSS:2004kjl,Viel:2005ha}.
Fluctuations in the transmitted flux from quasars 
trace the underlying dark matter density, 
providing a probe of
fundamental physics at scales 
of a few Megaparsec and redshifts 
$2\lesssim z\lesssim 5$.
Thanks to this advantageous property, 
the Ly$\alpha$ forest data have provided an excellent probe of 
$\Lambda$CDM, 
dark matter, neutrino masses, primordial features, 
and many other extensions of  
the concordance model~\cite{SDSS:2004kjl,Seljak:2006bg,Viel:2005ha,Viel:2005qj,Viel:2013fqw,Slosar:2011mq,Slosar:2013fi,Boyarsky:2008xj,Palanque-Delabrouille:2015pga,Palanque-Delabrouille:2014jca,Baur:2015jsy,Baur:2017stq,Irsic:2017yje,Palanque-Delabrouille:2019iyz,Boyarsky:2018tvu,Goldstein:2023gnw}.

Previous observational and theoretical efforts 
in Ly$\alpha$ forest physics have culminated 
in percent-level precision  
measurements of the baryon acoustic oscillations, flux power spectrum
(FPS) and three-dimensional correlations of the 
Ly$\alpha$ forest flux 
with the Sloan Digital Sky Survey (SDSS)~\cite{Slosar:2013fi,BOSS:2013rpr,Chabanier:2018rga,eBOSS:2020tmo}, and FPS measurements ~\cite{Irsic:2017sop, waltherNewPrecisionP1d2018, dayPowerSpectrumFlux2019, Karacayli:2021jeg} with high-resolution spectrographs
such as X-Shooter~\cite{vernetXshooterNewWide2011, lopezXQ100LegacySurvey2016}, HIRES~\cite{vogtHIRESHighresolutionEchelle1994, omearaSecondDataRelease2017}, and UVES~\cite{dekkerDesignConstructionPerformance2000, murphyUVESSpectralQuasar2019}.
The number of quasar spectra with the Ly$\alpha$ forest 
will soon increase multifold with 
the ongoing Dark Energy
Spectroscopic Instrument (DESI)~\cite{Karacayli:2023afs,DESI:2023xwh}.

With great statistical power comes great 
responsibility for modeling accuracy.
This is particularly important given 
that the effective slope 
and amplitude of the linear 
matter power spectrum inferred 
from the SDSS 1D FPS
are found to be in some
tension 
with the $\Lambda$CDM model~\cite{Palanque-Delabrouille:2015pga,Chabanier:2018rga,Goldstein:2023gnw,Rogers:2023upm,Fernandez:2023grg}.
To reconcile this potential discrepancy new physics interpretations of this tension have been
suggested, e.g. in~\cite{Hooper:2022byl,Goldstein:2023gnw,He:2023oke,Rogers:2023upm}.
Given the significance of 
these results, it is imperative to independently test the
modeling assumptions behind Ly$\alpha$ forest power spectrum analyses.

The standard approach to the Ly$\alpha$
physics has relied on emulators trained on hydrodynamic simulations~\cite{Borde:2014xsa,Palanque-Delabrouille:2015pga,Rossi:2014wsa,Bolton:2016bfs,Bird:2018efe,Pedersen:2019ieb,Pedersen:2020kaw,Fernandez:2023grg}.
These simulations are based on detailed models 
of underlying astrophysics, e.g. star formation, 
stellar feedback, reionization. 
Powerful as they are,
these simulations are 
currently 
limited by resolution~\cite{Chabanier:2024knr}. 
Emulators also 
face additional uncertainties in the numerical interpolation over a grid of simulations~\cite{Cabayol-Garcia:2023ygj,Fernandez:2023grg}.

In this \textit{Letter}, we explore an alternative approach to the 
Ly$\alpha$ forest analysis based 
on ideas borrowed from particle physics.
The perturbative or effective field theory (EFT) approaches to large-scale structure 
develop the description of cosmic web dynamics 
that is agnostic about the underlying astrophysical processes~\cite{Baumann:2010tm,Carrasco:2012cv,Desjacques:2016bnm,Ivanov:2022mrd} (see~\cite{Seljak:2012tp,Cieplak:2015kra,Garny:2018byk,Garny:2020rom,Givans:2020sez,Chen:2021rnb,Givans:2022qgb,Fuss:2022zyt,Ivanov:2023yla} for perturbative calculations 
of the Ly$\alpha$ forest).
In our context, 
the only two assumptions that EFT makes are that the 
flux absorption fluctuations trace the underlying 
dark matter density and non-linear corrections are 
perturbative~\cite{Ivanov:2023yla}. 
EFT does not make additional assumptions about gas physics. In particular, 
it does not impose any temperature-density relation. 
The relationship between flux fluctuations and the dark matter density field should be 
described by an analytic function, which can be Taylor expanded
on large scales. Then the terms in this Taylor series are parameterized 
using the fundamental symmetries of the problem, such as 
the rotations around the line-of-sight direction and the equivalence 
principle. 
Perturbative EFT approaches have been especially 
successful in galaxy clustering analyses, where they have become a standard 
tool for parameter estimation~\cite{BOSS:2016wmc,eBOSS:2020yzd,Ivanov:2019pdj,DAmico:2019fhj,Chen:2021wdi}. 
This \textit{Letter} presents 
an application of the
similar approach to the Ly$\alpha$ data.

An important advantage of EFT is its flexibility: a typical 
calculation of relevant non-linear corrections to the Ly$\alpha$
power spectrum takes about 1 second~\cite{Chudaykin:2020aoj},
which allows one to efficiently 
explore cosmological models without the need to re-run hydrodynamical 
simulations for new parameters added. 
In contrast to emulators, the EFT theory model can be consistently computed for every point in parameter space, thereby removing any uncertainty associated with the 
grid interpolation.
This is especially valuable for extended cosmological models proposed as a resolution 
to the above tension between $\Lambda$CDM and the SDSS Ly$\alpha$ data.
A rigorous interpretation of the Ly$\alpha$ data in the context of such models 
requires one to self-consistently recompute the FPS,
which can be easily done with EFT.

This \textit{Letter} lays the groundwork for systematic full-shape 
analyses of Ly$\alpha$ data with EFT.
Here, we re-analyze the publicly available SDSS 
data on the 1D FPS
using the EFT for Ly$\alpha$ data developed  
in~\cite{Ivanov:2023yla}. 
Our work pursues several goals. 
First, we develop an efficient toolkit to probe fundamental cosmology with 
the Ly$\alpha$ forest and test it against the real data.
Second, we investigate if the SDSS Ly$\alpha$ data are consistent
with $\Lambda$CDM within the EFT approach.
Third, we identify key directions for the future 
development of the Ly$\alpha$ 
EFT full-shape analysis.

\textit{Data.} Our analysis is based on quasar spectra from 
the fourteenth data release (DR14) of SDSS~\cite{Chabanier:2018rga}, 
which consists of the Baryon Oscillation Spectroscopic Survey (BOSS) and extended-BOSS (eBOSS)  
samples. The 1D FPS 
measurements from the SDSS quasars 
in the redshift range $2.2<z<4.6$ are publicly available.
In our analysis, 
we will focus on a more narrow redshift range 
$3.4\leq z\leq 4.2$, encompassing 
6 equally spaced redshift bins.
We do not use the lower redshift bins because 
their description 
requires an accurate calibration of the 
bias parameters that is not possible with publicly available simulations. 
We discard the highest two redshift bins in this work 
in order to avoid 
potential sensitivity to inhomogeneous reionization~\cite{cainIGMreionization2024}.
To build the power spectrum likelihood, we use the public 
covariance matrices that include additional diagonal corrections
to account for systematic effects due to
finite resolution, active galactic nuclei feedback, damped Ly$\alpha$ systems etc.~\cite{Chabanier:2018rga}.

\textit{Theory model.} 
Let us start our theory background with a brief discussion 
of the galaxy clustering in redshift space
at the lowest (linear) order in dark matter perturbations~\cite{Bernardeau:2001qr,Desjacques:2016bnm,Desjacques:2018pfv}.
Causality, the equivalence principle, and rotational symmetry 
imply that on the largest scales, the galaxy overdensity
field $\delta_g=n_g/\bar n_g-1$ ($n_g$ being the galaxy number density, and 
overbar denotes spatial averaging), should be linearly 
proportional to the dark matter overdensity $\delta^{(1)}$ computed 
in linear theory, 
\be 
\label{eq:linb}
\delta_g=b_1 \delta^{(1)}\,.
\ee 
$b_1$ above is the linear bias parameter,
which is treated as a nuisance parameter 
in EFT. 
Since Eq.~\eqref{eq:linb} is written in the galaxy rest frame, to describe the galaxy distribution seen by the observer $\delta^{(s)}_g$, one has to 
transform $\delta_g$ in Eq.~\eqref{eq:linb} to the observer's frame, which introduces redshift-space distortions~\cite{Kaiser:1987qv,Scoccimarro:2004tg}. 
At the linear level, the map between the two frames yields 
\be 
\label{eq:linrsd}
\delta^{(s)}_g=b_1 \delta^{(1)}-\frac{\partial_\parallel v^{(1)}_\parallel}{aH}\,,
\ee 
where $v^{(1)}_i$ is the peculiar velocity of galaxies, $H$ is the  
Hubble parameter, $a$ is the metric scale factor, and the subscript $\parallel$ denotes
the line-of-sight projection.
Note that Eq.~\eqref{eq:linrsd} does not introduce any new free parameters.
The coefficient in front of the line-of-sight velocity 
gradient is protected by the 
equivalence principle, 
because
the galaxies should ``fall"
in the gravitational potential the same way 
as dark matter.
This fact has been the 
rationale behind 
using redshift-space distortions as a cosmological probe~\cite{BOSS:2016wmc}.

In the context of the Ly$\alpha$ forest, the direct observables are
absorption lines that depend on the local physics along the observer's 
line-of-sight. The probability of absorption depends 
on peculiar velocity gradients along the line of sight,  
making it necessary to include such terms in the expansion already
in the rest frame of hydrogen clouds~\cite{McDonald:1999dt}.
In EFT, this implies that 
the Ly$\alpha$ forest bias expansion has only the azimuthal symmetry w.r.t. rotations around the line-of-sight, 
as opposed to the galaxy distribution, which  enjoys the full spherical symmetry,
imposing the linear bias relation Eq.~\eqref{eq:linb}.
In particular, in linear theory 
the Ly$\alpha$
flux 
fluctuations may depend now
on an operator $\d_\parallel^2\Phi\propto \partial_\parallel v^{(1)}_\parallel $ ($\Phi$ stands for the Newton's potential), i.e. the linear bias relation takes the form~\cite{Desjacques:2018pfv,Ivanov:2023yla}
\be 
\label{eq:linrsdLA}
\delta^{(s)}_g=b_1 \delta^{(1)}+b_\eta \frac{\partial_\parallel v^{(1)}_\parallel}{aH}\,,
\ee
where $b_\eta$ is a new bias parameter
that captures the correlation of the Ly$\alpha$ fluctuations 
with projections of the dark matter line-of-sight  velocity gradient. Eq.~\eqref{eq:linrsdLA} is still subject to the redshift-space mapping, 
which will produce 
an unobservable shift of $b_\eta$ by unity. 

EFT systematically describes non-linear effects
in the Ly$\alpha$ forest by higher-order operators
constructed from the relevant degrees of freedom, i.e. the 
density, velocity, and tidal fields. The structure of these operators
is fixed by symmetries only, just like in Eq.~\eqref{eq:linrsdLA}. 
As such, EFT automatically implements constraints 
dictated by the equivalence principle 
for non-linear operators stemming from the
redshift-space mapping~\cite{Ivanov:2023yla}.

The summary of the EFT modeling is given
in Supplemental Material (which
includes references to~\cite{Simonovic:2017mhp,Villasenor:2022aiy,1989MNRAS.238..293L,Ivanov:2024hgq,Irsic:2018hhg,Chudaykin:2022nru,fauchergiguereMeasurementOpacity2008, turnerLyaForestMeanFluxFromDesiY12024}).
We use the one-loop EFT model for the 3D power
spectrum, which depends on 11 EFT parameters,
which we collectively call $b_{\mathcal{O}}$.
The 1D power spectrum involves 2
additional nuisance parameters $C_{0,2}$ that 
renormalize the UV sensitivity
of 1D projection integrals~\cite{1989MNRAS.238..293L}. 
Additionally, they capture
physical small-scale
stochasticity   whose effect is significant for 1D
correlations~\cite{Viel:2005ha,Irsic:2018hhg}.
Finally, the Ly$\alpha$ data from SDSS is modulated by 
several systematic effects, including the Si~III oscillations. These cannot be described by EFT, and we 
follow the standard approach for their theoretical modeling~\cite{Chabanier:2018rga}, 
also see Supplemental Material.

\begin{figure*}
\includegraphics[width=0.39\textwidth]{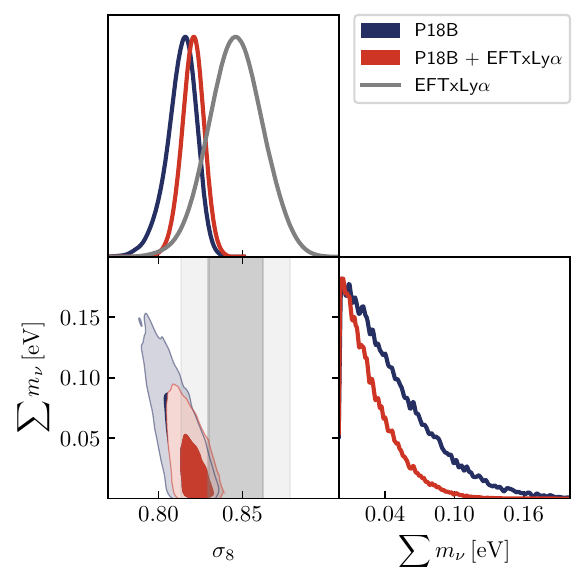}
\includegraphics[width=0.59\textwidth]{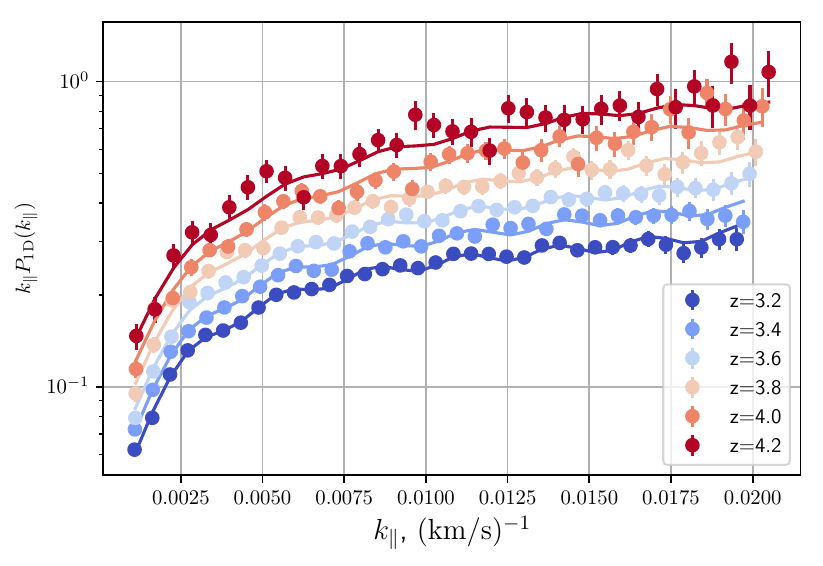}
   \caption{\textit{Left panel:} constraints on the mass fluctuation amplitude
   $\sigma_8$ from the EFT-based full-shape Ly$\alpha$ likelihood (EFTxLy$\alpha$) 
   presented in this work. 
   We also display 
   \textit{Planck} 18 CMB + BOSS DR12 BAO (P18B) results for $\sigma_8$ and the sum of neutrino 
   masses $\sum m_\nu$, and the results of the joint analysis of the P18B + EFTxLy$\alpha$
   data. \textit{Right panel:} 1-dimensional Ly$\alpha$ flux power spectrum data 
   from SDSS DR14 in the range $3.1<z<4.3$ (central values of redshifts are quoted), 
   along with EFT best-fit curves. Errorbars shown correspond to 
   diagonal elements of covariance matrices with systematic weights included.
    } \label{fig:money}
\end{figure*}

\textit{Analysis details.} 
Our EFT theory involves many nuisance parameters 
whose redshift-dependence 
is not determined by EFT. 
To estimate the impact of the EFT coefficients
on parameter estimation, we run a series of Fisher 
matrix analyses, described in Supplemental Material. 
We have found that 
we can achieve good constraints on $\sigma_8$
and maintain a significant level of flexibility 
if we keep the linear bias parameters $\{b_1,b_{\eta}\}$
free in the fit, while the non-linear EFT 
parameters are assumed to be
deterministic functions of $b_1$.
This approach is motivated by the expectation 
that the cosmological signal is dominated by linear fluctuations, 
and hence the linear 
bias parameters are most important to fit the data.
Our analysis confirms this assertion \textit{a posteriori}.
This is the strategy that we adopt in what follows. 

The non-linear bias parameters can be extracted from large sets
of high-fidelity hydrodynamic simulations of 3D Ly$\alpha$
forest data. 
3D information is crucial in 
order to break degeneracies between the EFT parameters
that are present for 1D FPS.
We extract the EFT parameters from the public power spectra measurements
from the state-of-the-art public Sherwood hydrodynamic simulation~\cite{Bolton:2016bfs,Givans:2022qgb}. 
We use relationships between non-linear 
bias parameters and $b_1$, $b_{\mathcal{O}}(b_1)$, which are more robust than the bias 
parameter values themselves. 
Setting priors on bias parameters in terms of functions of $b_1$
also makes our analysis more flexible to 
possible inaccuracies in astrophysics models used in simulations, by allowing for values
that are not fully determined by the simulations
themselves. 
This approach is motivated by galaxy clustering analysis, 
where semi-analytic functions $b_{\mathcal{O}}(b_1)$
are often used in cosmological analyses, see e.g.~\cite{Beutler:2016arn,Cabass:2022epm,Cabass:2022ymb,Ivanov:2024hgq}.
In line with these arguments, we have found that the EFT parameters from the Sherwood data 
exhibit significant
correlations with $b_1$, 
which can be well captured with simple interpolating functions, presented
in Supplemental Material.
We refer to functions $b_{\mathcal{O}}(b_1)$ as ``Sherwood prior.'' They
are rather strong as the bias parameters
are fully determined by $b_1$. However, they still allow for parameters 
beyond the Sherwood ones, implying some amount of flexibility.

Our Sherwood priors 
are subject
to statistical noise due to sample variance
and degeneracies between EFT parameters at the power spectrum level.
One can 
propagate this scatter 
to parameter constraints, but this is too conservative
since it is
mostly an artifact
of sample variance
rather than the Ly$\alpha$ physics. 
An alternative best-case scenario is that
the fundamental $b_{\mathcal{O}}(b_1)$ relations
are fully deterministic and cosmology-independent,
which is motivated by the 
phenomenology of
$b_{\mathcal{O}}(b_1)$ relations of dark matter halos
and galaxies. We assume this 
best-case scenario in our baseline analysis 
to benchmark the constraining power of EFT. 
This approach calls for the mitigation
of the effects of the scatter in the simulation-based priors.

Additionally,
the public power spectra from the Sherwood suite  
do not match those of SDSS for overlapping redshifts.
The bulk of this discrepancy can be 
compensated by re-normalizing the mean flux
in post-processing, but some discrepancies remain.
We compensate for both  uncertainties in our priors
at the level of the 1D correlations 
by re-calibrating the $C_n$ counterterms from emulator data.
To that end, we have produced synthetic 1D spectra 
that match the real SDSS data and
cover the relevant redshift range using the \texttt{LaCE} emulator~\cite{Pedersen:2020kaw}.
After the additional calibration, we find that the EFT model with 
3D non-linear bias parameters fixed to priors from Sherwood and 
1D counterterms determined by functions $C_n(b_1)$ ($n=0,2$)
extracted from a combination of Sherwood and \texttt{LaCE} data
provides a good description of the SDSS data, as
estimated based on the $\chi^2$ statistic. 
We adopt this modeling choice as our baseline.

We provide tests of this baseline analysis in Supplemental Material. There we show
that our main results remain stable after
an appropriate renormalization of the Sherwood's 
mean flux in post-processing, and the inclusion
of $k^2$ -type EFT counterterms.
In addition, we 
carry out a conservative
analysis with
marginalization over the scatter in the Sherwood
priors. This  
highlights the flexibility of EFT as compared to emulators: EFT allows
for marginalization over 
astrophysical uncertainties in an  agnostic fashion.

Several options are available for the description of 
the redshift evolution of remaining free EFT parameters $b_1,b_\eta$. 
The most conservative choice is to  
fit the EFT parameters independently in each bin, see e.g.~\cite{Ivanov:2019pdj}. 
An alternative is to assume a smooth evolution
of the free EFT parameters as a function of redshift as was done e.g. in~\cite{Garny:2020rom}. 
We found that the former option works better (in terms of the $\chi^2$ statistic), 
and it also provides more flexibility.
This baseline choice results in 12 free parameters 
for 6 redshift bins in the fit.

As far as the cosmological parameters are concerned, 
our Fisher matrix analysis suggests that 
adding other parameters to the fit will lead to strong degeneracies that cannot be 
broken at the level of the 1D FPS data, with our baseline settings.
Thus, our main analysis will have free $\sigma_8$ and 
other cosmological parameters fixed to a fiducial cosmology 
consistent with the $\textit{Planck}$ baseline $\Lambda$CDM model.
These are $\{\omega_b,\omega_{cdm},n_s,h,\sum m_\nu\}=\{0.02215,0.12,0.961,0.678,0\}$.

\textit{Results.} 
To start off, we validate our pipeline
on mock data generated with the \texttt{LaCE} emulator. 
Applying our pipeline to the 
\texttt{LaCE} mocks in the range $3.2\leq z\leq 4.2$,
we have found that 
after re-calibration of $C_n$,
it recovers the input value of $\sigma_8$
with $0.3\%$ accuracy, which corresponds to $\approx 15\%$
of the statistical error. This suggests that the 
theory systematic error 
is negligible.

Applying the same pipeline to the actual SDSS data 
we find the marginalized constraint 
$\sigma_8 = 0.841\pm 0.017$ in the baseline analysis.
The best-fit $\chi^2=170$ for $197$ degrees of freedom indicates an excellent fit to data.
The best-fit models 
and data for relevant 
redshift bins are shown in the right panel of Fig.~\ref{fig:money}.
Using our fiducial value $\Omega_m = 0.3082$, our $\sigma_8$
constraint can be converted into a limit on $S_8 = (\Omega_m/0.3)^{1/2}\sigma_8$.
We find $S_8 = 0.852\pm 0.017$, which agrees well with the 
Planck CMB result $0.834 \pm 0.016$~\cite{Aghanim:2018eyx}.
Our complementary conservative analysis 
nominally finds 
$\sigma_8=0.69~(0.75)\pm 0.05$
for the mean (best-fit) 
values and $1\sigma$ errors, 
which is broadly consistent 
with \textit{Planck} and LSS probes~\cite{eBOSS:2020yzd}, including the emulator-based
analysis of eBOSS Ly$\a$
from~\cite{Fernandez:2023grg}. 
While this measurement 
is largely consistent with our baseline, we 
note that it should not 
be over-interpreted 
as it is
subject to significant 
prior effects.

Combining our EFTxLy$\alpha$ measurement
with the \textit{Planck} 2018 base likelihood,
and adding the BAO data from BOSS DR12 as in Ref.~\cite{Ivanov:2019hqk},
we obtain the limit on the sum of neutrino masses \mbox{$\sum m_\nu<0.080$ eV} 
(at $95\%$CL), see the left panel of Fig.~\ref{fig:money} for the 1D and 2D marginalized 
contours. (Other cosmological parameters of $\nu \Lambda$CDM model 
were also consistently varied 
in all likelihoods
but not shown
on the plot for compactness.) 
This is one of the strongest cosmological 
constraints 
on the neutrino mass to date, c.f.~\cite{Aghanim:2018eyx,Ivanov:2019hqk,eBOSS:2020yzd,DESI:2024mwx}. In the
conservative analysis the constraint degrades to \mbox{$\sum m_\nu<0.15$ eV}.

Finally, we include additional cosmological parameters
in our analysis. Varying both $\sigma_8$ and $n_s$ we find a strong 
degeneracy between these parameters, in agreement
with our Fisher matrix analysis.
The data are not able to produce competitive constraints on either of these parameters individually with 
our baseline analysis settings. 
It will be interesting to see
if informative priors on $b_1$ and $b_\eta$, or more constraining 
data sets, e.g. 1D FPS from DESI, 
could break this degeneracy.

\textit{Discussion and conclusions.} 
We have presented the first EFT-based full-shape 
cosmological parameter 
analysis of the Ly$\alpha$ forest 1D clustering data 
from SDSS. From SDSS 1D Ly$\alpha$ data alone, 
we find a $2\%$ constraint on the mass
clustering amplitude $\sigma_8$. Combining our likelihood 
with data on CMB anisotropies from \textit{Planck} 2018  
and BAO, we derive one of the strongest constraints on the
total neutrino mass to date, 
$\sum m_\nu < 0.08$~eV. These results show the 
potential power of the 1D FPS
when analyzed with EFT.

Our results indicate that within 
the EFT framework, the SDSS Ly$\alpha$ data are 
in good agreement with the baseline Planck 
$\Lambda$CDM cosmological model. This weakens the case for new physics interpretations
of the apparent tension between $\Lambda$CDM and Ly$\alpha$ data in the literature.
This suggests that new ingredients beyond those
used in the Ly$\alpha$ emulator of~\cite{Chabanier:2018rga} may be needed to reconcile 
the SDSS 1D Ly$\alpha$ forest data 
with the $\Lambda$CDM~\cite{Fernandez:2023grg}.

We emphasize that our analysis should be taken in 
conjunction with our assumptions about the EFT nuisance parameters. We have used priors on  EFT parameters from Sherwood simulations,
and priors on 1D counterterms from the 1D power spectrum mocks produced 
with the \texttt{LaCE} emulator. These priors play a significant role in our analysis.
The marginalization over EFT parameters within more
conservative priors significantly degrades constraining 
power, see Supplemental Material.  
Further studies of the EFT parameters with large sets of high-fidelity hydrodynamical simulations with different thermal histories
are needed to fully validate our priors.
Similar precision measurements of 
EFT parameters have been previously carried out for 
halos and galaxies~\cite{Lazeyras:2017hxw,Abidi:2018eyd,Barreira:2020ekm,
Barreira:2021ukk,Lazeyras:2021dar,Ivanov:2024hgq,Cabass:2024wob}. 
In particular, 
recent studies have used informative priors on galaxy EFT 
parameters to improve cosmological constraints~\cite{Cabass:2022epm,Ivanov:2024hgq,Cabass:2024wob,Ivanov:2024xgb}. 
Performing a similar study is 
the main priority of the Ly$\alpha$ EFT program.

Furthermore, EFT can consistently model different Ly$\alpha$ measurements such as 3D Ly$\alpha$ autocorrelation and Ly$\alpha$-quasar cross-correlation functions~\cite{duMasdesBourboux:2020pck, desiKp6BaoLya2024}, 3D power spectrum~\cite{abdulkarimMeasumentPxLya2023, debelsunceP3DLya2024}, and cross-correlations with CMB~\cite{douxFirstDetectionCosmic2016, karacayliCmbxLyaP1d2024}. 
Combining these statistics with 1D FPS can also
improve our current results. 
For example, the degeneracy between EFT bias parameters and $\sigma_8$ can be broken by CMB lensing map and Ly$\alpha$ cross-correlations, and the inclusion of 3D information can alleviate the 
$b_1-b_\eta$ degeneracy.

We note that 
our EFTxLy$\alpha$ approach can be readily 
applied to extended cosmological models,
including the scenario 
motivated by the 
Hubble and $S_8$ tensions, see e.g.~\cite{Abdalla:2022yfr,Goldstein:2023gnw,He:2023oke}. These, and other 
research directions described above, are left for future work.


\textit{Acknowledgments.}
We thank Andrei Cuceau, Vid Irsic, Andreu Font-Ribera,
and 
Sokratis Trifinopoulos
for useful discussions,  and Jonas Chaves-Montero, Jahmour Givans, and Andreu Font-Ribera for providing
the Sherwood transmission files. We thank 
the anonymous Referees for valuable comments that 
improved our work. 
MCMC chains produced as part of this work 
are generated with 
\texttt{Montepython}~\cite{Audren:2012wb,Brinckmann:2018cvx}.
This material is based upon work supported by the U.S. Department of Energy, Office of Science, Office of High Energy Physics of U.S. Department of Energy under grant Contract Number  DE-SC0012567. MWT  acknowledges financial support from the Simons Foundation (Grant Number 929255).


\bibliography{short.bib}

\begin{thebibliography}{104}%
\makeatletter
\providecommand \@ifxundefined [1]{%
 \@ifx{#1\undefined}
}%
\providecommand \@ifnum [1]{%
 \ifnum #1\expandafter \@firstoftwo
 \else \expandafter \@secondoftwo
 \fi
}%
\providecommand \@ifx [1]{%
 \ifx #1\expandafter \@firstoftwo
 \else \expandafter \@secondoftwo
 \fi
}%
\providecommand \natexlab [1]{#1}%
\providecommand \enquote  [1]{``#1''}%
\providecommand \bibnamefont  [1]{#1}%
\providecommand \bibfnamefont [1]{#1}%
\providecommand \citenamefont [1]{#1}%
\providecommand \href@noop [0]{\@secondoftwo}%
\providecommand \href [0]{\begingroup \@sanitize@url \@href}%
\providecommand \@href[1]{\@@startlink{#1}\@@href}%
\providecommand \@@href[1]{\endgroup#1\@@endlink}%
\providecommand \@sanitize@url [0]{\catcode `\\12\catcode `\$12\catcode
  `\&12\catcode `\#12\catcode `\^12\catcode `\_12\catcode `\%12\relax}%
\providecommand \@@startlink[1]{}%
\providecommand \@@endlink[0]{}%
\providecommand \url  [0]{\begingroup\@sanitize@url \@url }%
\providecommand \@url [1]{\endgroup\@href {#1}{\urlprefix }}%
\providecommand \urlprefix  [0]{URL }%
\providecommand \Eprint [0]{\href }%
\providecommand \doibase [0]{http://dx.doi.org/}%
\providecommand \selectlanguage [0]{\@gobble}%
\providecommand \bibinfo  [0]{\@secondoftwo}%
\providecommand \bibfield  [0]{\@secondoftwo}%
\providecommand \translation [1]{[#1]}%
\providecommand \BibitemOpen [0]{}%
\providecommand \bibitemStop [0]{}%
\providecommand \bibitemNoStop [0]{.\EOS\space}%
\providecommand \EOS [0]{\spacefactor3000\relax}%
\providecommand \BibitemShut  [1]{\csname bibitem#1\endcsname}%
\let\auto@bib@innerbib\@empty
\bibitem [{\citenamefont {{Gunn}}\ and\ \citenamefont
  {{Peterson}}(1965)}]{1965ApJ...142.1633G}%
  \BibitemOpen
  \bibfield  {author} {\bibinfo {author} {\bibfnamefont {J.~E.}\ \bibnamefont
  {{Gunn}}}\ and\ \bibinfo {author} {\bibfnamefont {B.~A.}\ \bibnamefont
  {{Peterson}}},\ }\href {\doibase 10.1086/148444} {\bibfield  {journal}
  {\bibinfo  {journal} {APJ}\ }\textbf {\bibinfo {volume} {142}},\ \bibinfo
  {pages} {1633} (\bibinfo {year} {1965})}\BibitemShut {NoStop}%
\bibitem [{\citenamefont {{Croft}}\ \emph {et~al.}(1997)\citenamefont
  {{Croft}}, \citenamefont {{Weinberg}}, \citenamefont {{Katz}},\ and\
  \citenamefont {{Hernquist}}}]{1997ApJ...488..532C}%
  \BibitemOpen
  \bibfield  {author} {\bibinfo {author} {\bibfnamefont {R.~A.~C.}\
  \bibnamefont {{Croft}}}, \bibinfo {author} {\bibfnamefont {D.~H.}\
  \bibnamefont {{Weinberg}}}, \bibinfo {author} {\bibfnamefont
  {N.}~\bibnamefont {{Katz}}}, \ and\ \bibinfo {author} {\bibfnamefont
  {L.}~\bibnamefont {{Hernquist}}},\ }\href {\doibase 10.1086/304723}
  {\bibfield  {journal} {\bibinfo  {journal} {APJ}\ }\textbf {\bibinfo {volume}
  {488}},\ \bibinfo {pages} {532} (\bibinfo {year} {1997})},\ \Eprint
  {http://arxiv.org/abs/astro-ph/9611053} {arXiv:astro-ph/9611053 [astro-ph]}
  \BibitemShut {NoStop}%
\bibitem [{\citenamefont {Croft}\ \emph {et~al.}(1998)\citenamefont {Croft},
  \citenamefont {Weinberg}, \citenamefont {Katz},\ and\ \citenamefont
  {Hernquist}}]{Croft:1997jf}%
  \BibitemOpen
  \bibfield  {author} {\bibinfo {author} {\bibfnamefont {R.~A.~C.}\
  \bibnamefont {Croft}}, \bibinfo {author} {\bibfnamefont {D.~H.}\ \bibnamefont
  {Weinberg}}, \bibinfo {author} {\bibfnamefont {N.}~\bibnamefont {Katz}}, \
  and\ \bibinfo {author} {\bibfnamefont {L.}~\bibnamefont {Hernquist}},\ }\href
  {\doibase 10.1086/305289} {\bibfield  {journal} {\bibinfo  {journal}
  {Astrophys. J.}\ }\textbf {\bibinfo {volume} {495}},\ \bibinfo {pages} {44}
  (\bibinfo {year} {1998})},\ \Eprint {http://arxiv.org/abs/astro-ph/9708018}
  {arXiv:astro-ph/9708018} \BibitemShut {NoStop}%
\bibitem [{\citenamefont {Croft}\ \emph {et~al.}(2002)\citenamefont {Croft},
  \citenamefont {Weinberg}, \citenamefont {Bolte}, \citenamefont {Burles},
  \citenamefont {Hernquist}, \citenamefont {Katz}, \citenamefont {Kirkman},\
  and\ \citenamefont {Tytler}}]{Croft:2000hs}%
  \BibitemOpen
  \bibfield  {author} {\bibinfo {author} {\bibfnamefont {R.~A.~C.}\
  \bibnamefont {Croft}}, \bibinfo {author} {\bibfnamefont {D.~H.}\ \bibnamefont
  {Weinberg}}, \bibinfo {author} {\bibfnamefont {M.}~\bibnamefont {Bolte}},
  \bibinfo {author} {\bibfnamefont {S.}~\bibnamefont {Burles}}, \bibinfo
  {author} {\bibfnamefont {L.}~\bibnamefont {Hernquist}}, \bibinfo {author}
  {\bibfnamefont {N.}~\bibnamefont {Katz}}, \bibinfo {author} {\bibfnamefont
  {D.}~\bibnamefont {Kirkman}}, \ and\ \bibinfo {author} {\bibfnamefont
  {D.}~\bibnamefont {Tytler}},\ }\href {\doibase 10.1086/344099} {\bibfield
  {journal} {\bibinfo  {journal} {Astrophys. J.}\ }\textbf {\bibinfo {volume}
  {581}},\ \bibinfo {pages} {20} (\bibinfo {year} {2002})},\ \Eprint
  {http://arxiv.org/abs/astro-ph/0012324} {arXiv:astro-ph/0012324} \BibitemShut
  {NoStop}%
\bibitem [{\citenamefont {Hui}\ and\ \citenamefont
  {Gnedin}(1997)}]{Hui:1997dp}%
  \BibitemOpen
  \bibfield  {author} {\bibinfo {author} {\bibfnamefont {L.}~\bibnamefont
  {Hui}}\ and\ \bibinfo {author} {\bibfnamefont {N.~Y.}\ \bibnamefont
  {Gnedin}},\ }\href {\doibase 10.1093/mnras/292.1.27} {\bibfield  {journal}
  {\bibinfo  {journal} {Mon. Not. Roy. Astron. Soc.}\ }\textbf {\bibinfo
  {volume} {292}},\ \bibinfo {pages} {27} (\bibinfo {year} {1997})},\ \Eprint
  {http://arxiv.org/abs/astro-ph/9612232} {arXiv:astro-ph/9612232} \BibitemShut
  {NoStop}%
\bibitem [{\citenamefont {Gnedin}\ and\ \citenamefont
  {Hui}(1998)}]{Gnedin:1997td}%
  \BibitemOpen
  \bibfield  {author} {\bibinfo {author} {\bibfnamefont {N.~Y.}\ \bibnamefont
  {Gnedin}}\ and\ \bibinfo {author} {\bibfnamefont {L.}~\bibnamefont {Hui}},\
  }\href {\doibase 10.1046/j.1365-8711.1998.01249.x} {\bibfield  {journal}
  {\bibinfo  {journal} {Mon. Not. Roy. Astron. Soc.}\ }\textbf {\bibinfo
  {volume} {296}},\ \bibinfo {pages} {44} (\bibinfo {year} {1998})},\ \Eprint
  {http://arxiv.org/abs/astro-ph/9706219} {arXiv:astro-ph/9706219} \BibitemShut
  {NoStop}%
\bibitem [{\citenamefont {McDonald}\ \emph {et~al.}(2000)\citenamefont
  {McDonald}, \citenamefont {Miralda-Escude}, \citenamefont {Rauch},
  \citenamefont {Sargent}, \citenamefont {Barlow}, \citenamefont {Cen},\ and\
  \citenamefont {Ostriker}}]{McDonald:1999dt}%
  \BibitemOpen
  \bibfield  {author} {\bibinfo {author} {\bibfnamefont {P.}~\bibnamefont
  {McDonald}}, \bibinfo {author} {\bibfnamefont {J.}~\bibnamefont
  {Miralda-Escude}}, \bibinfo {author} {\bibfnamefont {M.}~\bibnamefont
  {Rauch}}, \bibinfo {author} {\bibfnamefont {W.~L.~W.}\ \bibnamefont
  {Sargent}}, \bibinfo {author} {\bibfnamefont {T.~A.}\ \bibnamefont {Barlow}},
  \bibinfo {author} {\bibfnamefont {R.}~\bibnamefont {Cen}}, \ and\ \bibinfo
  {author} {\bibfnamefont {J.~P.}\ \bibnamefont {Ostriker}},\ }\href {\doibase
  10.1086/317079} {\bibfield  {journal} {\bibinfo  {journal} {Astrophys. J.}\
  }\textbf {\bibinfo {volume} {543}},\ \bibinfo {pages} {1} (\bibinfo {year}
  {2000})},\ \Eprint {http://arxiv.org/abs/astro-ph/9911196}
  {arXiv:astro-ph/9911196} \BibitemShut {NoStop}%
\bibitem [{\citenamefont {McDonald}\ \emph {et~al.}(2006)\citenamefont
  {McDonald} \emph {et~al.}}]{SDSS:2004kjl}%
  \BibitemOpen
  \bibfield  {author} {\bibinfo {author} {\bibfnamefont {P.}~\bibnamefont
  {McDonald}} \emph {et~al.} (\bibinfo {collaboration} {SDSS}),\ }\href
  {\doibase 10.1086/444361} {\bibfield  {journal} {\bibinfo  {journal}
  {Astrophys. J. Suppl.}\ }\textbf {\bibinfo {volume} {163}},\ \bibinfo {pages}
  {80} (\bibinfo {year} {2006})},\ \Eprint
  {http://arxiv.org/abs/astro-ph/0405013} {arXiv:astro-ph/0405013} \BibitemShut
  {NoStop}%
\bibitem [{\citenamefont {Viel}\ and\ \citenamefont
  {Haehnelt}(2006)}]{Viel:2005ha}%
  \BibitemOpen
  \bibfield  {author} {\bibinfo {author} {\bibfnamefont {M.}~\bibnamefont
  {Viel}}\ and\ \bibinfo {author} {\bibfnamefont {M.~G.}\ \bibnamefont
  {Haehnelt}},\ }\href {\doibase 10.1111/j.1365-2966.2005.09703.x} {\bibfield
  {journal} {\bibinfo  {journal} {Mon. Not. Roy. Astron. Soc.}\ }\textbf
  {\bibinfo {volume} {365}},\ \bibinfo {pages} {231} (\bibinfo {year}
  {2006})},\ \Eprint {http://arxiv.org/abs/astro-ph/0508177}
  {arXiv:astro-ph/0508177} \BibitemShut {NoStop}%
\bibitem [{\citenamefont {Seljak}\ \emph {et~al.}(2006)\citenamefont {Seljak},
  \citenamefont {Slosar},\ and\ \citenamefont {McDonald}}]{Seljak:2006bg}%
  \BibitemOpen
  \bibfield  {author} {\bibinfo {author} {\bibfnamefont {U.}~\bibnamefont
  {Seljak}}, \bibinfo {author} {\bibfnamefont {A.}~\bibnamefont {Slosar}}, \
  and\ \bibinfo {author} {\bibfnamefont {P.}~\bibnamefont {McDonald}},\ }\href
  {\doibase 10.1088/1475-7516/2006/10/014} {\bibfield  {journal} {\bibinfo
  {journal} {JCAP}\ }\textbf {\bibinfo {volume} {10}},\ \bibinfo {pages} {014}
  (\bibinfo {year} {2006})},\ \Eprint {http://arxiv.org/abs/astro-ph/0604335}
  {arXiv:astro-ph/0604335} \BibitemShut {NoStop}%
\bibitem [{\citenamefont {Viel}\ \emph {et~al.}(2005)\citenamefont {Viel},
  \citenamefont {Lesgourgues}, \citenamefont {Haehnelt}, \citenamefont
  {Matarrese},\ and\ \citenamefont {Riotto}}]{Viel:2005qj}%
  \BibitemOpen
  \bibfield  {author} {\bibinfo {author} {\bibfnamefont {M.}~\bibnamefont
  {Viel}}, \bibinfo {author} {\bibfnamefont {J.}~\bibnamefont {Lesgourgues}},
  \bibinfo {author} {\bibfnamefont {M.~G.}\ \bibnamefont {Haehnelt}}, \bibinfo
  {author} {\bibfnamefont {S.}~\bibnamefont {Matarrese}}, \ and\ \bibinfo
  {author} {\bibfnamefont {A.}~\bibnamefont {Riotto}},\ }\href {\doibase
  10.1103/PhysRevD.71.063534} {\bibfield  {journal} {\bibinfo  {journal} {Phys.
  Rev. D}\ }\textbf {\bibinfo {volume} {71}},\ \bibinfo {pages} {063534}
  (\bibinfo {year} {2005})},\ \Eprint {http://arxiv.org/abs/astro-ph/0501562}
  {arXiv:astro-ph/0501562} \BibitemShut {NoStop}%
\bibitem [{\citenamefont {Viel}\ \emph {et~al.}(2013)\citenamefont {Viel},
  \citenamefont {Becker}, \citenamefont {Bolton},\ and\ \citenamefont
  {Haehnelt}}]{Viel:2013fqw}%
  \BibitemOpen
  \bibfield  {author} {\bibinfo {author} {\bibfnamefont {M.}~\bibnamefont
  {Viel}}, \bibinfo {author} {\bibfnamefont {G.~D.}\ \bibnamefont {Becker}},
  \bibinfo {author} {\bibfnamefont {J.~S.}\ \bibnamefont {Bolton}}, \ and\
  \bibinfo {author} {\bibfnamefont {M.~G.}\ \bibnamefont {Haehnelt}},\ }\href
  {\doibase 10.1103/PhysRevD.88.043502} {\bibfield  {journal} {\bibinfo
  {journal} {Phys. Rev. D}\ }\textbf {\bibinfo {volume} {88}},\ \bibinfo
  {pages} {043502} (\bibinfo {year} {2013})},\ \Eprint
  {http://arxiv.org/abs/1306.2314} {arXiv:1306.2314 [astro-ph.CO]} \BibitemShut
  {NoStop}%
\bibitem [{\citenamefont {Slosar}\ \emph {et~al.}(2011)\citenamefont {Slosar}
  \emph {et~al.}}]{Slosar:2011mq}%
  \BibitemOpen
  \bibfield  {author} {\bibinfo {author} {\bibfnamefont {A.}~\bibnamefont
  {Slosar}} \emph {et~al.},\ }\href {\doibase 10.1088/1475-7516/2011/09/001}
  {\bibfield  {journal} {\bibinfo  {journal} {JCAP}\ }\textbf {\bibinfo
  {volume} {09}},\ \bibinfo {pages} {001} (\bibinfo {year} {2011})},\ \Eprint
  {http://arxiv.org/abs/1104.5244} {arXiv:1104.5244 [astro-ph.CO]} \BibitemShut
  {NoStop}%
\bibitem [{\citenamefont {Slosar}\ \emph {et~al.}(2013)\citenamefont {Slosar}
  \emph {et~al.}}]{Slosar:2013fi}%
  \BibitemOpen
  \bibfield  {author} {\bibinfo {author} {\bibfnamefont {A.}~\bibnamefont
  {Slosar}} \emph {et~al.},\ }\href {\doibase 10.1088/1475-7516/2013/04/026}
  {\bibfield  {journal} {\bibinfo  {journal} {JCAP}\ }\textbf {\bibinfo
  {volume} {04}},\ \bibinfo {pages} {026} (\bibinfo {year} {2013})},\ \Eprint
  {http://arxiv.org/abs/1301.3459} {arXiv:1301.3459 [astro-ph.CO]} \BibitemShut
  {NoStop}%
\bibitem [{\citenamefont {Boyarsky}\ \emph {et~al.}(2009)\citenamefont
  {Boyarsky}, \citenamefont {Lesgourgues}, \citenamefont {Ruchayskiy},\ and\
  \citenamefont {Viel}}]{Boyarsky:2008xj}%
  \BibitemOpen
  \bibfield  {author} {\bibinfo {author} {\bibfnamefont {A.}~\bibnamefont
  {Boyarsky}}, \bibinfo {author} {\bibfnamefont {J.}~\bibnamefont
  {Lesgourgues}}, \bibinfo {author} {\bibfnamefont {O.}~\bibnamefont
  {Ruchayskiy}}, \ and\ \bibinfo {author} {\bibfnamefont {M.}~\bibnamefont
  {Viel}},\ }\href {\doibase 10.1088/1475-7516/2009/05/012} {\bibfield
  {journal} {\bibinfo  {journal} {JCAP}\ }\textbf {\bibinfo {volume} {05}},\
  \bibinfo {pages} {012} (\bibinfo {year} {2009})},\ \Eprint
  {http://arxiv.org/abs/0812.0010} {arXiv:0812.0010 [astro-ph]} \BibitemShut
  {NoStop}%
\bibitem [{\citenamefont {Palanque-Delabrouille}\ \emph
  {et~al.}(2015{\natexlab{a}})\citenamefont {Palanque-Delabrouille} \emph
  {et~al.}}]{Palanque-Delabrouille:2015pga}%
  \BibitemOpen
  \bibfield  {author} {\bibinfo {author} {\bibfnamefont {N.}~\bibnamefont
  {Palanque-Delabrouille}} \emph {et~al.},\ }\href {\doibase
  10.1088/1475-7516/2015/11/011} {\bibfield  {journal} {\bibinfo  {journal}
  {JCAP}\ }\textbf {\bibinfo {volume} {1511}},\ \bibinfo {pages} {011}
  (\bibinfo {year} {2015}{\natexlab{a}})},\ \Eprint
  {http://arxiv.org/abs/1506.05976} {arXiv:1506.05976 [astro-ph.CO]}
  \BibitemShut {NoStop}%
\bibitem [{\citenamefont {Palanque-Delabrouille}\ \emph
  {et~al.}(2015{\natexlab{b}})\citenamefont {Palanque-Delabrouille} \emph
  {et~al.}}]{Palanque-Delabrouille:2014jca}%
  \BibitemOpen
  \bibfield  {author} {\bibinfo {author} {\bibfnamefont {N.}~\bibnamefont
  {Palanque-Delabrouille}} \emph {et~al.},\ }\href {\doibase
  10.1088/1475-7516/2015/02/045} {\bibfield  {journal} {\bibinfo  {journal}
  {JCAP}\ }\textbf {\bibinfo {volume} {02}},\ \bibinfo {pages} {045} (\bibinfo
  {year} {2015}{\natexlab{b}})},\ \Eprint {http://arxiv.org/abs/1410.7244}
  {arXiv:1410.7244 [astro-ph.CO]} \BibitemShut {NoStop}%
\bibitem [{\citenamefont {Baur}\ \emph {et~al.}(2016)\citenamefont {Baur},
  \citenamefont {Palanque-Delabrouille}, \citenamefont {Y\`eche}, \citenamefont
  {Magneville},\ and\ \citenamefont {Viel}}]{Baur:2015jsy}%
  \BibitemOpen
  \bibfield  {author} {\bibinfo {author} {\bibfnamefont {J.}~\bibnamefont
  {Baur}}, \bibinfo {author} {\bibfnamefont {N.}~\bibnamefont
  {Palanque-Delabrouille}}, \bibinfo {author} {\bibfnamefont {C.}~\bibnamefont
  {Y\`eche}}, \bibinfo {author} {\bibfnamefont {C.}~\bibnamefont {Magneville}},
  \ and\ \bibinfo {author} {\bibfnamefont {M.}~\bibnamefont {Viel}},\ }\href
  {\doibase 10.1088/1475-7516/2016/08/012} {\bibfield  {journal} {\bibinfo
  {journal} {JCAP}\ }\textbf {\bibinfo {volume} {08}},\ \bibinfo {pages} {012}
  (\bibinfo {year} {2016})},\ \Eprint {http://arxiv.org/abs/1512.01981}
  {arXiv:1512.01981 [astro-ph.CO]} \BibitemShut {NoStop}%
\bibitem [{\citenamefont {Baur}\ \emph {et~al.}(2017)\citenamefont {Baur},
  \citenamefont {Palanque-Delabrouille}, \citenamefont {Yeche}, \citenamefont
  {Boyarsky}, \citenamefont {Ruchayskiy}, \citenamefont {Armengaud},\ and\
  \citenamefont {Lesgourgues}}]{Baur:2017stq}%
  \BibitemOpen
  \bibfield  {author} {\bibinfo {author} {\bibfnamefont {J.}~\bibnamefont
  {Baur}}, \bibinfo {author} {\bibfnamefont {N.}~\bibnamefont
  {Palanque-Delabrouille}}, \bibinfo {author} {\bibfnamefont {C.}~\bibnamefont
  {Yeche}}, \bibinfo {author} {\bibfnamefont {A.}~\bibnamefont {Boyarsky}},
  \bibinfo {author} {\bibfnamefont {O.}~\bibnamefont {Ruchayskiy}}, \bibinfo
  {author} {\bibfnamefont {E.}~\bibnamefont {Armengaud}}, \ and\ \bibinfo
  {author} {\bibfnamefont {J.}~\bibnamefont {Lesgourgues}},\ }\href {\doibase
  10.1088/1475-7516/2017/12/013} {\bibfield  {journal} {\bibinfo  {journal}
  {JCAP}\ }\textbf {\bibinfo {volume} {12}},\ \bibinfo {pages} {013} (\bibinfo
  {year} {2017})},\ \Eprint {http://arxiv.org/abs/1706.03118} {arXiv:1706.03118
  [astro-ph.CO]} \BibitemShut {NoStop}%
\bibitem [{\citenamefont {Ir\v{s}i\v{c}}\ \emph
  {et~al.}(2017{\natexlab{a}})\citenamefont {Ir\v{s}i\v{c}}, \citenamefont
  {Viel}, \citenamefont {Haehnelt}, \citenamefont {Bolton},\ and\ \citenamefont
  {Becker}}]{Irsic:2017yje}%
  \BibitemOpen
  \bibfield  {author} {\bibinfo {author} {\bibfnamefont {V.}~\bibnamefont
  {Ir\v{s}i\v{c}}}, \bibinfo {author} {\bibfnamefont {M.}~\bibnamefont {Viel}},
  \bibinfo {author} {\bibfnamefont {M.~G.}\ \bibnamefont {Haehnelt}}, \bibinfo
  {author} {\bibfnamefont {J.~S.}\ \bibnamefont {Bolton}}, \ and\ \bibinfo
  {author} {\bibfnamefont {G.~D.}\ \bibnamefont {Becker}},\ }\href {\doibase
  10.1103/PhysRevLett.119.031302} {\bibfield  {journal} {\bibinfo  {journal}
  {Phys. Rev. Lett.}\ }\textbf {\bibinfo {volume} {119}},\ \bibinfo {pages}
  {031302} (\bibinfo {year} {2017}{\natexlab{a}})},\ \Eprint
  {http://arxiv.org/abs/1703.04683} {arXiv:1703.04683 [astro-ph.CO]}
  \BibitemShut {NoStop}%
\bibitem [{\citenamefont {Palanque-Delabrouille}\ \emph
  {et~al.}(2020)\citenamefont {Palanque-Delabrouille}, \citenamefont {Y\`eche},
  \citenamefont {Sch\"oneberg}, \citenamefont {Lesgourgues}, \citenamefont
  {Walther}, \citenamefont {Chabanier},\ and\ \citenamefont
  {Armengaud}}]{Palanque-Delabrouille:2019iyz}%
  \BibitemOpen
  \bibfield  {author} {\bibinfo {author} {\bibfnamefont {N.}~\bibnamefont
  {Palanque-Delabrouille}}, \bibinfo {author} {\bibfnamefont {C.}~\bibnamefont
  {Y\`eche}}, \bibinfo {author} {\bibfnamefont {N.}~\bibnamefont
  {Sch\"oneberg}}, \bibinfo {author} {\bibfnamefont {J.}~\bibnamefont
  {Lesgourgues}}, \bibinfo {author} {\bibfnamefont {M.}~\bibnamefont
  {Walther}}, \bibinfo {author} {\bibfnamefont {S.}~\bibnamefont {Chabanier}},
  \ and\ \bibinfo {author} {\bibfnamefont {E.}~\bibnamefont {Armengaud}},\
  }\href {\doibase 10.1088/1475-7516/2020/04/038} {\bibfield  {journal}
  {\bibinfo  {journal} {JCAP}\ }\textbf {\bibinfo {volume} {04}},\ \bibinfo
  {pages} {038} (\bibinfo {year} {2020})},\ \Eprint
  {http://arxiv.org/abs/1911.09073} {arXiv:1911.09073 [astro-ph.CO]}
  \BibitemShut {NoStop}%
\bibitem [{\citenamefont {Boyarsky}\ \emph {et~al.}(2019)\citenamefont
  {Boyarsky}, \citenamefont {Drewes}, \citenamefont {Lasserre}, \citenamefont
  {Mertens},\ and\ \citenamefont {Ruchayskiy}}]{Boyarsky:2018tvu}%
  \BibitemOpen
  \bibfield  {author} {\bibinfo {author} {\bibfnamefont {A.}~\bibnamefont
  {Boyarsky}}, \bibinfo {author} {\bibfnamefont {M.}~\bibnamefont {Drewes}},
  \bibinfo {author} {\bibfnamefont {T.}~\bibnamefont {Lasserre}}, \bibinfo
  {author} {\bibfnamefont {S.}~\bibnamefont {Mertens}}, \ and\ \bibinfo
  {author} {\bibfnamefont {O.}~\bibnamefont {Ruchayskiy}},\ }\href {\doibase
  10.1016/j.ppnp.2018.07.004} {\bibfield  {journal} {\bibinfo  {journal} {Prog.
  Part. Nucl. Phys.}\ }\textbf {\bibinfo {volume} {104}},\ \bibinfo {pages} {1}
  (\bibinfo {year} {2019})},\ \Eprint {http://arxiv.org/abs/1807.07938}
  {arXiv:1807.07938 [hep-ph]} \BibitemShut {NoStop}%
\bibitem [{\citenamefont {Goldstein}\ \emph {et~al.}(2023)\citenamefont
  {Goldstein}, \citenamefont {Hill}, \citenamefont {Ir\v{s}i\v{c}},\ and\
  \citenamefont {Sherwin}}]{Goldstein:2023gnw}%
  \BibitemOpen
  \bibfield  {author} {\bibinfo {author} {\bibfnamefont {S.}~\bibnamefont
  {Goldstein}}, \bibinfo {author} {\bibfnamefont {J.~C.}\ \bibnamefont {Hill}},
  \bibinfo {author} {\bibfnamefont {V.}~\bibnamefont {Ir\v{s}i\v{c}}}, \ and\
  \bibinfo {author} {\bibfnamefont {B.~D.}\ \bibnamefont {Sherwin}},\
  }\href@noop {} {\  (\bibinfo {year} {2023})},\ \Eprint
  {http://arxiv.org/abs/2303.00746} {arXiv:2303.00746 [astro-ph.CO]}
  \BibitemShut {NoStop}%
\bibitem [{\citenamefont {Palanque-Delabrouille}\ \emph
  {et~al.}(2013)\citenamefont {Palanque-Delabrouille} \emph
  {et~al.}}]{BOSS:2013rpr}%
  \BibitemOpen
  \bibfield  {author} {\bibinfo {author} {\bibfnamefont {N.}~\bibnamefont
  {Palanque-Delabrouille}} \emph {et~al.} (\bibinfo {collaboration} {BOSS}),\
  }\href {\doibase 10.1051/0004-6361/201322130} {\bibfield  {journal} {\bibinfo
   {journal} {Astron. Astrophys.}\ }\textbf {\bibinfo {volume} {559}},\
  \bibinfo {pages} {A85} (\bibinfo {year} {2013})},\ \Eprint
  {http://arxiv.org/abs/1306.5896} {arXiv:1306.5896 [astro-ph.CO]} \BibitemShut
  {NoStop}%
\bibitem [{\citenamefont {Chabanier}\ \emph {et~al.}(2019)\citenamefont
  {Chabanier} \emph {et~al.}}]{Chabanier:2018rga}%
  \BibitemOpen
  \bibfield  {author} {\bibinfo {author} {\bibfnamefont {S.}~\bibnamefont
  {Chabanier}} \emph {et~al.},\ }\href {\doibase 10.1088/1475-7516/2019/07/017}
  {\bibfield  {journal} {\bibinfo  {journal} {JCAP}\ }\textbf {\bibinfo
  {volume} {07}},\ \bibinfo {pages} {017} (\bibinfo {year} {2019})},\ \Eprint
  {http://arxiv.org/abs/1812.03554} {arXiv:1812.03554 [astro-ph.CO]}
  \BibitemShut {NoStop}%
\bibitem [{\citenamefont {du~Mas~des Bourboux}\ \emph
  {et~al.}(2020{\natexlab{a}})\citenamefont {du~Mas~des Bourboux} \emph
  {et~al.}}]{eBOSS:2020tmo}%
  \BibitemOpen
  \bibfield  {author} {\bibinfo {author} {\bibfnamefont {H.}~\bibnamefont
  {du~Mas~des Bourboux}} \emph {et~al.} (\bibinfo {collaboration} {eBOSS}),\
  }\href {\doibase 10.3847/1538-4357/abb085} {\bibfield  {journal} {\bibinfo
  {journal} {Astrophys. J.}\ }\textbf {\bibinfo {volume} {901}},\ \bibinfo
  {pages} {153} (\bibinfo {year} {2020}{\natexlab{a}})},\ \Eprint
  {http://arxiv.org/abs/2007.08995} {arXiv:2007.08995 [astro-ph.CO]}
  \BibitemShut {NoStop}%
\bibitem [{\citenamefont {Ir\v{s}i\v{c}}\ \emph
  {et~al.}(2017{\natexlab{b}})\citenamefont {Ir\v{s}i\v{c}} \emph
  {et~al.}}]{Irsic:2017sop}%
  \BibitemOpen
  \bibfield  {author} {\bibinfo {author} {\bibfnamefont {V.}~\bibnamefont
  {Ir\v{s}i\v{c}}} \emph {et~al.},\ }\href {\doibase 10.1093/mnras/stw3372}
  {\bibfield  {journal} {\bibinfo  {journal} {Mon. Not. Roy. Astron. Soc.}\
  }\textbf {\bibinfo {volume} {466}},\ \bibinfo {pages} {4332} (\bibinfo {year}
  {2017}{\natexlab{b}})},\ \Eprint {http://arxiv.org/abs/1702.01761}
  {arXiv:1702.01761 [astro-ph.CO]} \BibitemShut {NoStop}%
\bibitem [{\citenamefont {{Walther}}\ \emph {et~al.}(2018)\citenamefont
  {{Walther}}, \citenamefont {{Hennawi}}, \citenamefont {{Hiss}}, \citenamefont
  {{O{\~n}orbe}}, \citenamefont {{Lee}}, \citenamefont {{Rorai}},\ and\
  \citenamefont {{O'Meara}}}]{waltherNewPrecisionP1d2018}%
  \BibitemOpen
  \bibfield  {author} {\bibinfo {author} {\bibfnamefont {M.}~\bibnamefont
  {{Walther}}}, \bibinfo {author} {\bibfnamefont {J.~F.}\ \bibnamefont
  {{Hennawi}}}, \bibinfo {author} {\bibfnamefont {H.}~\bibnamefont {{Hiss}}},
  \bibinfo {author} {\bibfnamefont {J.}~\bibnamefont {{O{\~n}orbe}}}, \bibinfo
  {author} {\bibfnamefont {K.-G.}\ \bibnamefont {{Lee}}}, \bibinfo {author}
  {\bibfnamefont {A.}~\bibnamefont {{Rorai}}}, \ and\ \bibinfo {author}
  {\bibfnamefont {J.}~\bibnamefont {{O'Meara}}},\ }\href {\doibase
  10.3847/1538-4357/aa9c81} {\bibfield  {journal} {\bibinfo  {journal} {\apj}\
  }\textbf {\bibinfo {volume} {852}},\ \bibinfo {eid} {22} (\bibinfo {year}
  {2018})},\ \Eprint {http://arxiv.org/abs/1709.07354} {arXiv:1709.07354
  [astro-ph.CO]} \BibitemShut {NoStop}%
\bibitem [{\citenamefont {Day}\ \emph {et~al.}(2019)\citenamefont {Day},
  \citenamefont {Tytler},\ and\ \citenamefont
  {Kambalur}}]{dayPowerSpectrumFlux2019}%
  \BibitemOpen
  \bibfield  {author} {\bibinfo {author} {\bibfnamefont {A.}~\bibnamefont
  {Day}}, \bibinfo {author} {\bibfnamefont {D.}~\bibnamefont {Tytler}}, \ and\
  \bibinfo {author} {\bibfnamefont {B.}~\bibnamefont {Kambalur}},\ }\href
  {\doibase 10.1093/mnras/stz2214} {\bibfield  {journal} {\bibinfo  {journal}
  {Monthly Notices of the Royal Astronomical Society}\ }\textbf {\bibinfo
  {volume} {489}},\ \bibinfo {pages} {2536} (\bibinfo {year}
  {2019})}\BibitemShut {NoStop}%
\bibitem [{\citenamefont {Kara\c{c}ayl\i{}}\ \emph {et~al.}(2022)\citenamefont
  {Kara\c{c}ayl\i{}} \emph {et~al.}}]{Karacayli:2021jeg}%
  \BibitemOpen
  \bibfield  {author} {\bibinfo {author} {\bibfnamefont {N.~G.}\ \bibnamefont
  {Kara\c{c}ayl\i{}}} \emph {et~al.},\ }\href {\doibase 10.1093/mnras/stab3201}
  {\bibfield  {journal} {\bibinfo  {journal} {Mon. Not. Roy. Astron. Soc.}\
  }\textbf {\bibinfo {volume} {509}},\ \bibinfo {pages} {2842} (\bibinfo {year}
  {2022})},\ \Eprint {http://arxiv.org/abs/2108.10870} {arXiv:2108.10870
  [astro-ph.CO]} \BibitemShut {NoStop}%
\bibitem [{\citenamefont {Vernet}\ \emph {et~al.}(2011)\citenamefont {Vernet},
  \citenamefont {Dekker}, \citenamefont {D'Odorico}, \citenamefont {Kaper},
  \citenamefont {Kjaergaard}, \citenamefont {Hammer}, \citenamefont {Randich},
  \citenamefont {Zerbi}, \citenamefont {Groot}, \citenamefont {Hjorth},
  \citenamefont {Guinouard}, \citenamefont {Navarro}, \citenamefont {Adolfse},
  \citenamefont {Albers}, \citenamefont {Amans}, \citenamefont {Andersen},
  \citenamefont {Andersen}, \citenamefont {Binetruy}, \citenamefont {Bristow},
  \citenamefont {Castillo}, \citenamefont {Chemla}, \citenamefont
  {Christensen}, \citenamefont {Conconi}, \citenamefont {Conzelmann},
  \citenamefont {Dam}, \citenamefont {{de Caprio}}, \citenamefont {{de Ugarte
  Postigo}}, \citenamefont {Delabre}, \citenamefont {{di Marcantonio}},
  \citenamefont {Downing}, \citenamefont {Elswijk}, \citenamefont {Finger},
  \citenamefont {Fischer}, \citenamefont {Flores}, \citenamefont {Fran{\c
  c}ois}, \citenamefont {Goldoni}, \citenamefont {Guglielmi}, \citenamefont
  {Haigron}, \citenamefont {Hanenburg}, \citenamefont {Hendriks}, \citenamefont
  {Horrobin}, \citenamefont {Horville}, \citenamefont {Jessen}, \citenamefont
  {Kerber}, \citenamefont {Kern}, \citenamefont {Kiekebusch}, \citenamefont
  {Kleszcz}, \citenamefont {Klougart}, \citenamefont {Kragt}, \citenamefont
  {Larsen}, \citenamefont {Lizon}, \citenamefont {Lucuix}, \citenamefont
  {Mainieri}, \citenamefont {Manuputy}, \citenamefont {Martayan}, \citenamefont
  {Mason}, \citenamefont {Mazzoleni}, \citenamefont {Michaelsen}, \citenamefont
  {Modigliani}, \citenamefont {Moehler}, \citenamefont {M{\o}ller},
  \citenamefont {Norup~S{\o}rensen}, \citenamefont {N{\o}rregaard},
  \citenamefont {P{\'e}roux}, \citenamefont {Patat}, \citenamefont {Pena},
  \citenamefont {Pragt}, \citenamefont {Reinero}, \citenamefont {Rigal},
  \citenamefont {Riva}, \citenamefont {Roelfsema}, \citenamefont {Royer},
  \citenamefont {Sacco}, \citenamefont {Santin}, \citenamefont {Schoenmaker},
  \citenamefont {Spano}, \citenamefont {Sweers}, \citenamefont {Ter~Horst},
  \citenamefont {Tintori}, \citenamefont {Tromp}, \citenamefont {{van Dael}},
  \citenamefont {{van der Vliet}}, \citenamefont {Venema}, \citenamefont
  {Vidali}, \citenamefont {Vinther}, \citenamefont {Vola}, \citenamefont
  {Winters}, \citenamefont {Wistisen}, \citenamefont {Wulterkens},\ and\
  \citenamefont {Zacchei}}]{vernetXshooterNewWide2011}%
  \BibitemOpen
  \bibfield  {author} {\bibinfo {author} {\bibfnamefont {J.}~\bibnamefont
  {Vernet}}, \bibinfo {author} {\bibfnamefont {H.}~\bibnamefont {Dekker}},
  \bibinfo {author} {\bibfnamefont {S.}~\bibnamefont {D'Odorico}}, \bibinfo
  {author} {\bibfnamefont {L.}~\bibnamefont {Kaper}}, \bibinfo {author}
  {\bibfnamefont {P.}~\bibnamefont {Kjaergaard}}, \bibinfo {author}
  {\bibfnamefont {F.}~\bibnamefont {Hammer}}, \bibinfo {author} {\bibfnamefont
  {S.}~\bibnamefont {Randich}}, \bibinfo {author} {\bibfnamefont
  {F.}~\bibnamefont {Zerbi}}, \bibinfo {author} {\bibfnamefont {P.~J.}\
  \bibnamefont {Groot}}, \bibinfo {author} {\bibfnamefont {J.}~\bibnamefont
  {Hjorth}}, \bibinfo {author} {\bibfnamefont {I.}~\bibnamefont {Guinouard}},
  \bibinfo {author} {\bibfnamefont {R.}~\bibnamefont {Navarro}}, \bibinfo
  {author} {\bibfnamefont {T.}~\bibnamefont {Adolfse}}, \bibinfo {author}
  {\bibfnamefont {P.~W.}\ \bibnamefont {Albers}}, \bibinfo {author}
  {\bibfnamefont {J.-P.}\ \bibnamefont {Amans}}, \bibinfo {author}
  {\bibfnamefont {J.~J.}\ \bibnamefont {Andersen}}, \bibinfo {author}
  {\bibfnamefont {M.~I.}\ \bibnamefont {Andersen}}, \bibinfo {author}
  {\bibfnamefont {P.}~\bibnamefont {Binetruy}}, \bibinfo {author}
  {\bibfnamefont {P.}~\bibnamefont {Bristow}}, \bibinfo {author} {\bibfnamefont
  {R.}~\bibnamefont {Castillo}}, \bibinfo {author} {\bibfnamefont
  {F.}~\bibnamefont {Chemla}}, \bibinfo {author} {\bibfnamefont
  {L.}~\bibnamefont {Christensen}}, \bibinfo {author} {\bibfnamefont
  {P.}~\bibnamefont {Conconi}}, \bibinfo {author} {\bibfnamefont
  {R.}~\bibnamefont {Conzelmann}}, \bibinfo {author} {\bibfnamefont
  {J.}~\bibnamefont {Dam}}, \bibinfo {author} {\bibfnamefont {V.}~\bibnamefont
  {{de Caprio}}}, \bibinfo {author} {\bibfnamefont {A.}~\bibnamefont {{de
  Ugarte Postigo}}}, \bibinfo {author} {\bibfnamefont {B.}~\bibnamefont
  {Delabre}}, \bibinfo {author} {\bibfnamefont {P.}~\bibnamefont {{di
  Marcantonio}}}, \bibinfo {author} {\bibfnamefont {M.}~\bibnamefont
  {Downing}}, \bibinfo {author} {\bibfnamefont {E.}~\bibnamefont {Elswijk}},
  \bibinfo {author} {\bibfnamefont {G.}~\bibnamefont {Finger}}, \bibinfo
  {author} {\bibfnamefont {G.}~\bibnamefont {Fischer}}, \bibinfo {author}
  {\bibfnamefont {H.}~\bibnamefont {Flores}}, \bibinfo {author} {\bibfnamefont
  {P.}~\bibnamefont {Fran{\c c}ois}}, \bibinfo {author} {\bibfnamefont
  {P.}~\bibnamefont {Goldoni}}, \bibinfo {author} {\bibfnamefont
  {L.}~\bibnamefont {Guglielmi}}, \bibinfo {author} {\bibfnamefont
  {R.}~\bibnamefont {Haigron}}, \bibinfo {author} {\bibfnamefont
  {H.}~\bibnamefont {Hanenburg}}, \bibinfo {author} {\bibfnamefont
  {I.}~\bibnamefont {Hendriks}}, \bibinfo {author} {\bibfnamefont
  {M.}~\bibnamefont {Horrobin}}, \bibinfo {author} {\bibfnamefont
  {D.}~\bibnamefont {Horville}}, \bibinfo {author} {\bibfnamefont {N.~C.}\
  \bibnamefont {Jessen}}, \bibinfo {author} {\bibfnamefont {F.}~\bibnamefont
  {Kerber}}, \bibinfo {author} {\bibfnamefont {L.}~\bibnamefont {Kern}},
  \bibinfo {author} {\bibfnamefont {M.}~\bibnamefont {Kiekebusch}}, \bibinfo
  {author} {\bibfnamefont {P.}~\bibnamefont {Kleszcz}}, \bibinfo {author}
  {\bibfnamefont {J.}~\bibnamefont {Klougart}}, \bibinfo {author}
  {\bibfnamefont {J.}~\bibnamefont {Kragt}}, \bibinfo {author} {\bibfnamefont
  {H.~H.}\ \bibnamefont {Larsen}}, \bibinfo {author} {\bibfnamefont {J.-L.}\
  \bibnamefont {Lizon}}, \bibinfo {author} {\bibfnamefont {C.}~\bibnamefont
  {Lucuix}}, \bibinfo {author} {\bibfnamefont {V.}~\bibnamefont {Mainieri}},
  \bibinfo {author} {\bibfnamefont {R.}~\bibnamefont {Manuputy}}, \bibinfo
  {author} {\bibfnamefont {C.}~\bibnamefont {Martayan}}, \bibinfo {author}
  {\bibfnamefont {E.}~\bibnamefont {Mason}}, \bibinfo {author} {\bibfnamefont
  {R.}~\bibnamefont {Mazzoleni}}, \bibinfo {author} {\bibfnamefont
  {N.}~\bibnamefont {Michaelsen}}, \bibinfo {author} {\bibfnamefont
  {A.}~\bibnamefont {Modigliani}}, \bibinfo {author} {\bibfnamefont
  {S.}~\bibnamefont {Moehler}}, \bibinfo {author} {\bibfnamefont
  {P.}~\bibnamefont {M{\o}ller}}, \bibinfo {author} {\bibfnamefont
  {A.}~\bibnamefont {Norup~S{\o}rensen}}, \bibinfo {author} {\bibfnamefont
  {P.}~\bibnamefont {N{\o}rregaard}}, \bibinfo {author} {\bibfnamefont
  {C.}~\bibnamefont {P{\'e}roux}}, \bibinfo {author} {\bibfnamefont
  {F.}~\bibnamefont {Patat}}, \bibinfo {author} {\bibfnamefont
  {E.}~\bibnamefont {Pena}}, \bibinfo {author} {\bibfnamefont {J.}~\bibnamefont
  {Pragt}}, \bibinfo {author} {\bibfnamefont {C.}~\bibnamefont {Reinero}},
  \bibinfo {author} {\bibfnamefont {F.}~\bibnamefont {Rigal}}, \bibinfo
  {author} {\bibfnamefont {M.}~\bibnamefont {Riva}}, \bibinfo {author}
  {\bibfnamefont {R.}~\bibnamefont {Roelfsema}}, \bibinfo {author}
  {\bibfnamefont {F.}~\bibnamefont {Royer}}, \bibinfo {author} {\bibfnamefont
  {G.}~\bibnamefont {Sacco}}, \bibinfo {author} {\bibfnamefont
  {P.}~\bibnamefont {Santin}}, \bibinfo {author} {\bibfnamefont
  {T.}~\bibnamefont {Schoenmaker}}, \bibinfo {author} {\bibfnamefont
  {P.}~\bibnamefont {Spano}}, \bibinfo {author} {\bibfnamefont
  {E.}~\bibnamefont {Sweers}}, \bibinfo {author} {\bibfnamefont
  {R.}~\bibnamefont {Ter~Horst}}, \bibinfo {author} {\bibfnamefont
  {M.}~\bibnamefont {Tintori}}, \bibinfo {author} {\bibfnamefont
  {N.}~\bibnamefont {Tromp}}, \bibinfo {author} {\bibfnamefont
  {P.}~\bibnamefont {{van Dael}}}, \bibinfo {author} {\bibfnamefont
  {H.}~\bibnamefont {{van der Vliet}}}, \bibinfo {author} {\bibfnamefont
  {L.}~\bibnamefont {Venema}}, \bibinfo {author} {\bibfnamefont
  {M.}~\bibnamefont {Vidali}}, \bibinfo {author} {\bibfnamefont
  {J.}~\bibnamefont {Vinther}}, \bibinfo {author} {\bibfnamefont
  {P.}~\bibnamefont {Vola}}, \bibinfo {author} {\bibfnamefont {R.}~\bibnamefont
  {Winters}}, \bibinfo {author} {\bibfnamefont {D.}~\bibnamefont {Wistisen}},
  \bibinfo {author} {\bibfnamefont {G.}~\bibnamefont {Wulterkens}}, \ and\
  \bibinfo {author} {\bibfnamefont {A.}~\bibnamefont {Zacchei}},\ }\href
  {\doibase 10.1051/0004-6361/201117752} {\bibfield  {journal} {\bibinfo
  {journal} {Astronomy and Astrophysics}\ }\textbf {\bibinfo {volume} {536}},\
  \bibinfo {pages} {A105} (\bibinfo {year} {2011})}\BibitemShut {NoStop}%
\bibitem [{\citenamefont {L{\'o}pez}\ \emph {et~al.}(2016)\citenamefont
  {L{\'o}pez}, \citenamefont {D'Odorico}, \citenamefont {Ellison},
  \citenamefont {Becker}, \citenamefont {Christensen}, \citenamefont {Cupani},
  \citenamefont {Denney}, \citenamefont {P{\^a}ris}, \citenamefont {Worseck},
  \citenamefont {Berg}, \citenamefont {Cristiani}, \citenamefont
  {{Dessauges-Zavadsky}}, \citenamefont {Haehnelt}, \citenamefont {Hamann},
  \citenamefont {Hennawi}, \citenamefont {Ir{\v s}i{\v c}}, \citenamefont
  {Kim}, \citenamefont {L{\'o}pez}, \citenamefont {Lund~Saust}, \citenamefont
  {M{\'e}nard}, \citenamefont {Perrotta}, \citenamefont {Prochaska},
  \citenamefont {{S{\'a}nchez-Ram{\'i}rez}}, \citenamefont {Vestergaard},
  \citenamefont {Viel},\ and\ \citenamefont
  {Wisotzki}}]{lopezXQ100LegacySurvey2016}%
  \BibitemOpen
  \bibfield  {author} {\bibinfo {author} {\bibfnamefont {S.}~\bibnamefont
  {L{\'o}pez}}, \bibinfo {author} {\bibfnamefont {V.}~\bibnamefont
  {D'Odorico}}, \bibinfo {author} {\bibfnamefont {S.~L.}\ \bibnamefont
  {Ellison}}, \bibinfo {author} {\bibfnamefont {G.~D.}\ \bibnamefont {Becker}},
  \bibinfo {author} {\bibfnamefont {L.}~\bibnamefont {Christensen}}, \bibinfo
  {author} {\bibfnamefont {G.}~\bibnamefont {Cupani}}, \bibinfo {author}
  {\bibfnamefont {K.~D.}\ \bibnamefont {Denney}}, \bibinfo {author}
  {\bibfnamefont {I.}~\bibnamefont {P{\^a}ris}}, \bibinfo {author}
  {\bibfnamefont {G.}~\bibnamefont {Worseck}}, \bibinfo {author} {\bibfnamefont
  {T.~A.~M.}\ \bibnamefont {Berg}}, \bibinfo {author} {\bibfnamefont
  {S.}~\bibnamefont {Cristiani}}, \bibinfo {author} {\bibfnamefont
  {M.}~\bibnamefont {{Dessauges-Zavadsky}}}, \bibinfo {author} {\bibfnamefont
  {M.}~\bibnamefont {Haehnelt}}, \bibinfo {author} {\bibfnamefont
  {F.}~\bibnamefont {Hamann}}, \bibinfo {author} {\bibfnamefont
  {J.}~\bibnamefont {Hennawi}}, \bibinfo {author} {\bibfnamefont
  {V.}~\bibnamefont {Ir{\v s}i{\v c}}}, \bibinfo {author} {\bibfnamefont
  {T.-S.}\ \bibnamefont {Kim}}, \bibinfo {author} {\bibfnamefont
  {P.}~\bibnamefont {L{\'o}pez}}, \bibinfo {author} {\bibfnamefont
  {R.}~\bibnamefont {Lund~Saust}}, \bibinfo {author} {\bibfnamefont
  {B.}~\bibnamefont {M{\'e}nard}}, \bibinfo {author} {\bibfnamefont
  {S.}~\bibnamefont {Perrotta}}, \bibinfo {author} {\bibfnamefont {J.~X.}\
  \bibnamefont {Prochaska}}, \bibinfo {author} {\bibfnamefont {R.}~\bibnamefont
  {{S{\'a}nchez-Ram{\'i}rez}}}, \bibinfo {author} {\bibfnamefont
  {M.}~\bibnamefont {Vestergaard}}, \bibinfo {author} {\bibfnamefont
  {M.}~\bibnamefont {Viel}}, \ and\ \bibinfo {author} {\bibfnamefont
  {L.}~\bibnamefont {Wisotzki}},\ }\href {\doibase 10.1051/0004-6361/201628161}
  {\bibfield  {journal} {\bibinfo  {journal} {Astronomy and Astrophysics}\
  }\textbf {\bibinfo {volume} {594}},\ \bibinfo {pages} {A91} (\bibinfo {year}
  {2016})}\BibitemShut {NoStop}%
\bibitem [{\citenamefont {Vogt}\ \emph {et~al.}(1994)\citenamefont {Vogt},
  \citenamefont {Allen}, \citenamefont {Bigelow}, \citenamefont {Bresee},
  \citenamefont {Brown}, \citenamefont {Cantrall}, \citenamefont {Conrad},
  \citenamefont {Couture}, \citenamefont {Delaney}, \citenamefont {Epps},
  \citenamefont {Hilyard}, \citenamefont {Hilyard}, \citenamefont {Horn},
  \citenamefont {Jern}, \citenamefont {Kanto}, \citenamefont {Keane},
  \citenamefont {Kibrick}, \citenamefont {Lewis}, \citenamefont {Osborne},
  \citenamefont {Pardeilhan}, \citenamefont {Pfister}, \citenamefont
  {Ricketts}, \citenamefont {Robinson}, \citenamefont {Stover}, \citenamefont
  {Tucker}, \citenamefont {Ward},\ and\ \citenamefont
  {Wei}}]{vogtHIRESHighresolutionEchelle1994}%
  \BibitemOpen
  \bibfield  {author} {\bibinfo {author} {\bibfnamefont {S.~S.}\ \bibnamefont
  {Vogt}}, \bibinfo {author} {\bibfnamefont {S.~L.}\ \bibnamefont {Allen}},
  \bibinfo {author} {\bibfnamefont {B.~C.}\ \bibnamefont {Bigelow}}, \bibinfo
  {author} {\bibfnamefont {L.}~\bibnamefont {Bresee}}, \bibinfo {author}
  {\bibfnamefont {W.~E.}\ \bibnamefont {Brown}}, \bibinfo {author}
  {\bibfnamefont {T.}~\bibnamefont {Cantrall}}, \bibinfo {author}
  {\bibfnamefont {A.}~\bibnamefont {Conrad}}, \bibinfo {author} {\bibfnamefont
  {M.}~\bibnamefont {Couture}}, \bibinfo {author} {\bibfnamefont
  {C.}~\bibnamefont {Delaney}}, \bibinfo {author} {\bibfnamefont {H.~W.}\
  \bibnamefont {Epps}}, \bibinfo {author} {\bibfnamefont {D.}~\bibnamefont
  {Hilyard}}, \bibinfo {author} {\bibfnamefont {D.~F.}\ \bibnamefont
  {Hilyard}}, \bibinfo {author} {\bibfnamefont {E.}~\bibnamefont {Horn}},
  \bibinfo {author} {\bibfnamefont {N.}~\bibnamefont {Jern}}, \bibinfo {author}
  {\bibfnamefont {D.}~\bibnamefont {Kanto}}, \bibinfo {author} {\bibfnamefont
  {M.~J.}\ \bibnamefont {Keane}}, \bibinfo {author} {\bibfnamefont {R.~I.}\
  \bibnamefont {Kibrick}}, \bibinfo {author} {\bibfnamefont {J.~W.}\
  \bibnamefont {Lewis}}, \bibinfo {author} {\bibfnamefont {J.}~\bibnamefont
  {Osborne}}, \bibinfo {author} {\bibfnamefont {G.~H.}\ \bibnamefont
  {Pardeilhan}}, \bibinfo {author} {\bibfnamefont {T.}~\bibnamefont {Pfister}},
  \bibinfo {author} {\bibfnamefont {T.}~\bibnamefont {Ricketts}}, \bibinfo
  {author} {\bibfnamefont {L.~B.}\ \bibnamefont {Robinson}}, \bibinfo {author}
  {\bibfnamefont {R.~J.}\ \bibnamefont {Stover}}, \bibinfo {author}
  {\bibfnamefont {D.}~\bibnamefont {Tucker}}, \bibinfo {author} {\bibfnamefont
  {J.~M.}\ \bibnamefont {Ward}}, \ and\ \bibinfo {author} {\bibfnamefont
  {M.}~\bibnamefont {Wei}},\ }in\ \href {\doibase 10.1117/12.176725} {\emph
  {\bibinfo {booktitle} {Instrumentation in {{Astronomy VIII}}}}},\ Vol.\
  \bibinfo {volume} {2198},\ \bibinfo {editor} {edited by\ \bibinfo {editor}
  {\bibfnamefont {D.~L.}\ \bibnamefont {Crawford}}\ and\ \bibinfo {editor}
  {\bibfnamefont {E.~R.}\ \bibnamefont {Craine}}}\ (\bibinfo  {publisher}
  {SPIE},\ \bibinfo {year} {1994})\ pp.\ \bibinfo {pages}
  {362--375}\BibitemShut {NoStop}%
\bibitem [{\citenamefont {O'Meara}\ \emph {et~al.}(2017)\citenamefont
  {O'Meara}, \citenamefont {Lehner}, \citenamefont {Howk}, \citenamefont
  {Prochaska}, \citenamefont {Fox}, \citenamefont {Peeples}, \citenamefont
  {Tumlinson},\ and\ \citenamefont {O'Shea}}]{omearaSecondDataRelease2017}%
  \BibitemOpen
  \bibfield  {author} {\bibinfo {author} {\bibfnamefont {J.~M.}\ \bibnamefont
  {O'Meara}}, \bibinfo {author} {\bibfnamefont {N.}~\bibnamefont {Lehner}},
  \bibinfo {author} {\bibfnamefont {J.~C.}\ \bibnamefont {Howk}}, \bibinfo
  {author} {\bibfnamefont {J.~X.}\ \bibnamefont {Prochaska}}, \bibinfo {author}
  {\bibfnamefont {A.~J.}\ \bibnamefont {Fox}}, \bibinfo {author} {\bibfnamefont
  {M.~S.}\ \bibnamefont {Peeples}}, \bibinfo {author} {\bibfnamefont
  {J.}~\bibnamefont {Tumlinson}}, \ and\ \bibinfo {author} {\bibfnamefont
  {B.~W.}\ \bibnamefont {O'Shea}},\ }\href {\doibase 10.3847/1538-3881/aa82b8}
  {\bibfield  {journal} {\bibinfo  {journal} {The Astronomical Journal}\
  }\textbf {\bibinfo {volume} {154}},\ \bibinfo {pages} {114} (\bibinfo {year}
  {2017})}\BibitemShut {NoStop}%
\bibitem [{\citenamefont {Dekker}\ \emph {et~al.}(2000)\citenamefont {Dekker},
  \citenamefont {D'Odorico}, \citenamefont {Kaufer}, \citenamefont {Delabre},\
  and\ \citenamefont {Kotzlowski}}]{dekkerDesignConstructionPerformance2000}%
  \BibitemOpen
  \bibfield  {author} {\bibinfo {author} {\bibfnamefont {H.}~\bibnamefont
  {Dekker}}, \bibinfo {author} {\bibfnamefont {S.}~\bibnamefont {D'Odorico}},
  \bibinfo {author} {\bibfnamefont {A.}~\bibnamefont {Kaufer}}, \bibinfo
  {author} {\bibfnamefont {B.}~\bibnamefont {Delabre}}, \ and\ \bibinfo
  {author} {\bibfnamefont {H.}~\bibnamefont {Kotzlowski}},\ }in\ \href
  {\doibase 10.1117/12.395512} {\emph {\bibinfo {booktitle} {Optical and {{IR
  Telescope Instrumentation}} and {{Detectors}}}}},\ Vol.\ \bibinfo {volume}
  {4008},\ \bibinfo {editor} {edited by\ \bibinfo {editor} {\bibfnamefont
  {M.}~\bibnamefont {Iye}}\ and\ \bibinfo {editor} {\bibfnamefont {A.~F.~M.}\
  \bibnamefont {Moorwood}}}\ (\bibinfo  {publisher} {SPIE},\ \bibinfo {year}
  {2000})\ pp.\ \bibinfo {pages} {534--545}\BibitemShut {NoStop}%
\bibitem [{\citenamefont {Murphy}\ \emph {et~al.}(2019)\citenamefont {Murphy},
  \citenamefont {Kacprzak}, \citenamefont {Savorgnan},\ and\ \citenamefont
  {Carswell}}]{murphyUVESSpectralQuasar2019}%
  \BibitemOpen
  \bibfield  {author} {\bibinfo {author} {\bibfnamefont {M.~T.}\ \bibnamefont
  {Murphy}}, \bibinfo {author} {\bibfnamefont {G.~G.}\ \bibnamefont
  {Kacprzak}}, \bibinfo {author} {\bibfnamefont {G.~A.~D.}\ \bibnamefont
  {Savorgnan}}, \ and\ \bibinfo {author} {\bibfnamefont {R.~F.}\ \bibnamefont
  {Carswell}},\ }\href {\doibase 10.1093/mnras/sty2834} {\bibfield  {journal}
  {\bibinfo  {journal} {Mon. Not. Roy. Astron. Soc.}\ }\textbf {\bibinfo
  {volume} {482}},\ \bibinfo {pages} {3458} (\bibinfo {year}
  {2019})}\BibitemShut {NoStop}%
\bibitem [{\citenamefont {Kara\c{c}ayl\i{}}\ \emph {et~al.}(2024)\citenamefont
  {Kara\c{c}ayl\i{}} \emph {et~al.}}]{Karacayli:2023afs}%
  \BibitemOpen
  \bibfield  {author} {\bibinfo {author} {\bibfnamefont {N.~G.}\ \bibnamefont
  {Kara\c{c}ayl\i{}}} \emph {et~al.},\ }\href {\doibase 10.1093/mnras/stae171}
  {\bibfield  {journal} {\bibinfo  {journal} {Mon. Not. Roy. Astron. Soc.}\
  }\textbf {\bibinfo {volume} {528}},\ \bibinfo {pages} {3941} (\bibinfo {year}
  {2024})},\ \Eprint {http://arxiv.org/abs/2306.06316} {arXiv:2306.06316
  [astro-ph.CO]} \BibitemShut {NoStop}%
\bibitem [{\citenamefont {Ravoux}\ \emph {et~al.}(2023)\citenamefont {Ravoux}
  \emph {et~al.}}]{DESI:2023xwh}%
  \BibitemOpen
  \bibfield  {author} {\bibinfo {author} {\bibfnamefont {C.}~\bibnamefont
  {Ravoux}} \emph {et~al.} (\bibinfo {collaboration} {DESI}),\ }\href {\doibase
  10.1093/mnras/stad3008} {\bibfield  {journal} {\bibinfo  {journal} {Mon. Not.
  Roy. Astron. Soc.}\ }\textbf {\bibinfo {volume} {526}},\ \bibinfo {pages}
  {5118} (\bibinfo {year} {2023})},\ \Eprint {http://arxiv.org/abs/2306.06311}
  {arXiv:2306.06311 [astro-ph.CO]} \BibitemShut {NoStop}%
\bibitem [{\citenamefont {Rogers}\ and\ \citenamefont
  {Poulin}(2023)}]{Rogers:2023upm}%
  \BibitemOpen
  \bibfield  {author} {\bibinfo {author} {\bibfnamefont {K.~K.}\ \bibnamefont
  {Rogers}}\ and\ \bibinfo {author} {\bibfnamefont {V.}~\bibnamefont
  {Poulin}},\ }\href@noop {} {\  (\bibinfo {year} {2023})},\ \Eprint
  {http://arxiv.org/abs/2311.16377} {arXiv:2311.16377 [astro-ph.CO]}
  \BibitemShut {NoStop}%
\bibitem [{\citenamefont {Fernandez}\ \emph {et~al.}(2023)\citenamefont
  {Fernandez}, \citenamefont {Bird},\ and\ \citenamefont
  {Ho}}]{Fernandez:2023grg}%
  \BibitemOpen
  \bibfield  {author} {\bibinfo {author} {\bibfnamefont {M.~A.}\ \bibnamefont
  {Fernandez}}, \bibinfo {author} {\bibfnamefont {S.}~\bibnamefont {Bird}}, \
  and\ \bibinfo {author} {\bibfnamefont {M.-F.}\ \bibnamefont {Ho}},\
  }\href@noop {} {\  (\bibinfo {year} {2023})},\ \Eprint
  {http://arxiv.org/abs/2309.03943} {arXiv:2309.03943 [astro-ph.CO]}
  \BibitemShut {NoStop}%
\bibitem [{\citenamefont {Hooper}\ \emph {et~al.}(2022)\citenamefont {Hooper},
  \citenamefont {Sch\"oneberg}, \citenamefont {Murgia}, \citenamefont
  {Archidiacono}, \citenamefont {Lesgourgues},\ and\ \citenamefont
  {Viel}}]{Hooper:2022byl}%
  \BibitemOpen
  \bibfield  {author} {\bibinfo {author} {\bibfnamefont {D.~C.}\ \bibnamefont
  {Hooper}}, \bibinfo {author} {\bibfnamefont {N.}~\bibnamefont
  {Sch\"oneberg}}, \bibinfo {author} {\bibfnamefont {R.}~\bibnamefont
  {Murgia}}, \bibinfo {author} {\bibfnamefont {M.}~\bibnamefont
  {Archidiacono}}, \bibinfo {author} {\bibfnamefont {J.}~\bibnamefont
  {Lesgourgues}}, \ and\ \bibinfo {author} {\bibfnamefont {M.}~\bibnamefont
  {Viel}},\ }\href {\doibase 10.1088/1475-7516/2022/10/032} {\bibfield
  {journal} {\bibinfo  {journal} {JCAP}\ }\textbf {\bibinfo {volume} {10}},\
  \bibinfo {pages} {032} (\bibinfo {year} {2022})},\ \Eprint
  {http://arxiv.org/abs/2206.08188} {arXiv:2206.08188 [astro-ph.CO]}
  \BibitemShut {NoStop}%
\bibitem [{\citenamefont {He}\ \emph {et~al.}(2023)\citenamefont {He},
  \citenamefont {An}, \citenamefont {Ivanov},\ and\ \citenamefont
  {Gluscevic}}]{He:2023oke}%
  \BibitemOpen
  \bibfield  {author} {\bibinfo {author} {\bibfnamefont {A.}~\bibnamefont
  {He}}, \bibinfo {author} {\bibfnamefont {R.}~\bibnamefont {An}}, \bibinfo
  {author} {\bibfnamefont {M.~M.}\ \bibnamefont {Ivanov}}, \ and\ \bibinfo
  {author} {\bibfnamefont {V.}~\bibnamefont {Gluscevic}},\ }\href@noop {} {\
  (\bibinfo {year} {2023})},\ \Eprint {http://arxiv.org/abs/2309.03956}
  {arXiv:2309.03956 [astro-ph.CO]} \BibitemShut {NoStop}%
\bibitem [{\citenamefont {Borde}\ \emph {et~al.}(2014)\citenamefont {Borde},
  \citenamefont {Palanque-Delabrouille}, \citenamefont {Rossi}, \citenamefont
  {Viel}, \citenamefont {Bolton}, \citenamefont {Y\`eche}, \citenamefont
  {LeGoff},\ and\ \citenamefont {Rich}}]{Borde:2014xsa}%
  \BibitemOpen
  \bibfield  {author} {\bibinfo {author} {\bibfnamefont {A.}~\bibnamefont
  {Borde}}, \bibinfo {author} {\bibfnamefont {N.}~\bibnamefont
  {Palanque-Delabrouille}}, \bibinfo {author} {\bibfnamefont {G.}~\bibnamefont
  {Rossi}}, \bibinfo {author} {\bibfnamefont {M.}~\bibnamefont {Viel}},
  \bibinfo {author} {\bibfnamefont {J.~S.}\ \bibnamefont {Bolton}}, \bibinfo
  {author} {\bibfnamefont {C.}~\bibnamefont {Y\`eche}}, \bibinfo {author}
  {\bibfnamefont {J.-M.}\ \bibnamefont {LeGoff}}, \ and\ \bibinfo {author}
  {\bibfnamefont {J.}~\bibnamefont {Rich}},\ }\href {\doibase
  10.1088/1475-7516/2014/07/005} {\bibfield  {journal} {\bibinfo  {journal}
  {JCAP}\ }\textbf {\bibinfo {volume} {07}},\ \bibinfo {pages} {005} (\bibinfo
  {year} {2014})},\ \Eprint {http://arxiv.org/abs/1401.6472} {arXiv:1401.6472
  [astro-ph.CO]} \BibitemShut {NoStop}%
\bibitem [{\citenamefont {Rossi}\ \emph {et~al.}(2014)\citenamefont {Rossi},
  \citenamefont {Palanque-Delabrouille}, \citenamefont {Borde}, \citenamefont
  {Viel}, \citenamefont {Yeche}, \citenamefont {Bolton}, \citenamefont {Rich},\
  and\ \citenamefont {Le~Goff}}]{Rossi:2014wsa}%
  \BibitemOpen
  \bibfield  {author} {\bibinfo {author} {\bibfnamefont {G.}~\bibnamefont
  {Rossi}}, \bibinfo {author} {\bibfnamefont {N.}~\bibnamefont
  {Palanque-Delabrouille}}, \bibinfo {author} {\bibfnamefont {A.}~\bibnamefont
  {Borde}}, \bibinfo {author} {\bibfnamefont {M.}~\bibnamefont {Viel}},
  \bibinfo {author} {\bibfnamefont {C.}~\bibnamefont {Yeche}}, \bibinfo
  {author} {\bibfnamefont {J.~S.}\ \bibnamefont {Bolton}}, \bibinfo {author}
  {\bibfnamefont {J.}~\bibnamefont {Rich}}, \ and\ \bibinfo {author}
  {\bibfnamefont {J.-M.}\ \bibnamefont {Le~Goff}},\ }\href {\doibase
  10.1051/0004-6361/201423507} {\bibfield  {journal} {\bibinfo  {journal}
  {Astron. Astrophys.}\ }\textbf {\bibinfo {volume} {567}},\ \bibinfo {pages}
  {A79} (\bibinfo {year} {2014})},\ \Eprint {http://arxiv.org/abs/1401.6464}
  {arXiv:1401.6464 [astro-ph.CO]} \BibitemShut {NoStop}%
\bibitem [{\citenamefont {Bolton}\ \emph {et~al.}(2017)\citenamefont {Bolton},
  \citenamefont {Puchwein}, \citenamefont {Sijacki}, \citenamefont {Haehnelt},
  \citenamefont {Kim}, \citenamefont {Meiksin}, \citenamefont {Regan},\ and\
  \citenamefont {Viel}}]{Bolton:2016bfs}%
  \BibitemOpen
  \bibfield  {author} {\bibinfo {author} {\bibfnamefont {J.~S.}\ \bibnamefont
  {Bolton}}, \bibinfo {author} {\bibfnamefont {E.}~\bibnamefont {Puchwein}},
  \bibinfo {author} {\bibfnamefont {D.}~\bibnamefont {Sijacki}}, \bibinfo
  {author} {\bibfnamefont {M.~G.}\ \bibnamefont {Haehnelt}}, \bibinfo {author}
  {\bibfnamefont {T.-S.}\ \bibnamefont {Kim}}, \bibinfo {author} {\bibfnamefont
  {A.}~\bibnamefont {Meiksin}}, \bibinfo {author} {\bibfnamefont {J.~A.}\
  \bibnamefont {Regan}}, \ and\ \bibinfo {author} {\bibfnamefont
  {M.}~\bibnamefont {Viel}},\ }\href {\doibase 10.1093/mnras/stw2397}
  {\bibfield  {journal} {\bibinfo  {journal} {Mon. Not. Roy. Astron. Soc.}\
  }\textbf {\bibinfo {volume} {464}},\ \bibinfo {pages} {897} (\bibinfo {year}
  {2017})},\ \Eprint {http://arxiv.org/abs/1605.03462} {arXiv:1605.03462
  [astro-ph.CO]} \BibitemShut {NoStop}%
\bibitem [{\citenamefont {Bird}\ \emph {et~al.}(2019)\citenamefont {Bird},
  \citenamefont {Rogers}, \citenamefont {Peiris}, \citenamefont {Verde},
  \citenamefont {Font-Ribera},\ and\ \citenamefont {Pontzen}}]{Bird:2018efe}%
  \BibitemOpen
  \bibfield  {author} {\bibinfo {author} {\bibfnamefont {S.}~\bibnamefont
  {Bird}}, \bibinfo {author} {\bibfnamefont {K.~K.}\ \bibnamefont {Rogers}},
  \bibinfo {author} {\bibfnamefont {H.~V.}\ \bibnamefont {Peiris}}, \bibinfo
  {author} {\bibfnamefont {L.}~\bibnamefont {Verde}}, \bibinfo {author}
  {\bibfnamefont {A.}~\bibnamefont {Font-Ribera}}, \ and\ \bibinfo {author}
  {\bibfnamefont {A.}~\bibnamefont {Pontzen}},\ }\href {\doibase
  10.1088/1475-7516/2019/02/050} {\bibfield  {journal} {\bibinfo  {journal}
  {JCAP}\ }\textbf {\bibinfo {volume} {02}},\ \bibinfo {pages} {050} (\bibinfo
  {year} {2019})},\ \Eprint {http://arxiv.org/abs/1812.04654} {arXiv:1812.04654
  [astro-ph.CO]} \BibitemShut {NoStop}%
\bibitem [{\citenamefont {Pedersen}\ \emph {et~al.}(2020)\citenamefont
  {Pedersen}, \citenamefont {Font-Ribera}, \citenamefont {Kitching},
  \citenamefont {McDonald}, \citenamefont {Bird}, \citenamefont {Slosar},
  \citenamefont {Rogers},\ and\ \citenamefont {Pontzen}}]{Pedersen:2019ieb}%
  \BibitemOpen
  \bibfield  {author} {\bibinfo {author} {\bibfnamefont {C.}~\bibnamefont
  {Pedersen}}, \bibinfo {author} {\bibfnamefont {A.}~\bibnamefont
  {Font-Ribera}}, \bibinfo {author} {\bibfnamefont {T.~D.}\ \bibnamefont
  {Kitching}}, \bibinfo {author} {\bibfnamefont {P.}~\bibnamefont {McDonald}},
  \bibinfo {author} {\bibfnamefont {S.}~\bibnamefont {Bird}}, \bibinfo {author}
  {\bibfnamefont {A.}~\bibnamefont {Slosar}}, \bibinfo {author} {\bibfnamefont
  {K.~K.}\ \bibnamefont {Rogers}}, \ and\ \bibinfo {author} {\bibfnamefont
  {A.}~\bibnamefont {Pontzen}},\ }\href {\doibase
  10.1088/1475-7516/2020/04/025} {\bibfield  {journal} {\bibinfo  {journal}
  {JCAP}\ }\textbf {\bibinfo {volume} {04}},\ \bibinfo {pages} {025} (\bibinfo
  {year} {2020})},\ \Eprint {http://arxiv.org/abs/1911.09596} {arXiv:1911.09596
  [astro-ph.CO]} \BibitemShut {NoStop}%
\bibitem [{\citenamefont {Pedersen}\ \emph {et~al.}(2021)\citenamefont
  {Pedersen}, \citenamefont {Font-Ribera}, \citenamefont {Rogers},
  \citenamefont {McDonald}, \citenamefont {Peiris}, \citenamefont {Pontzen},\
  and\ \citenamefont {Slosar}}]{Pedersen:2020kaw}%
  \BibitemOpen
  \bibfield  {author} {\bibinfo {author} {\bibfnamefont {C.}~\bibnamefont
  {Pedersen}}, \bibinfo {author} {\bibfnamefont {A.}~\bibnamefont
  {Font-Ribera}}, \bibinfo {author} {\bibfnamefont {K.~K.}\ \bibnamefont
  {Rogers}}, \bibinfo {author} {\bibfnamefont {P.}~\bibnamefont {McDonald}},
  \bibinfo {author} {\bibfnamefont {H.~V.}\ \bibnamefont {Peiris}}, \bibinfo
  {author} {\bibfnamefont {A.}~\bibnamefont {Pontzen}}, \ and\ \bibinfo
  {author} {\bibfnamefont {A.}~\bibnamefont {Slosar}},\ }\href {\doibase
  10.1088/1475-7516/2021/05/033} {\bibfield  {journal} {\bibinfo  {journal}
  {JCAP}\ }\textbf {\bibinfo {volume} {05}},\ \bibinfo {pages} {033} (\bibinfo
  {year} {2021})},\ \Eprint {http://arxiv.org/abs/2011.15127} {arXiv:2011.15127
  [astro-ph.CO]} \BibitemShut {NoStop}%
\bibitem [{\citenamefont {Chabanier}\ \emph {et~al.}(2024)\citenamefont
  {Chabanier}, \citenamefont {Ravoux}, \citenamefont {Latrille}, \citenamefont
  {Sexton}, \citenamefont {Armengaud}, \citenamefont {Bautista}, \citenamefont
  {Dumerchat},\ and\ \citenamefont {Luki\'c}}]{Chabanier:2024knr}%
  \BibitemOpen
  \bibfield  {author} {\bibinfo {author} {\bibfnamefont {S.}~\bibnamefont
  {Chabanier}}, \bibinfo {author} {\bibfnamefont {C.}~\bibnamefont {Ravoux}},
  \bibinfo {author} {\bibfnamefont {L.}~\bibnamefont {Latrille}}, \bibinfo
  {author} {\bibfnamefont {J.}~\bibnamefont {Sexton}}, \bibinfo {author}
  {\bibfnamefont {E.}~\bibnamefont {Armengaud}}, \bibinfo {author}
  {\bibfnamefont {J.}~\bibnamefont {Bautista}}, \bibinfo {author}
  {\bibfnamefont {T.}~\bibnamefont {Dumerchat}}, \ and\ \bibinfo {author}
  {\bibfnamefont {Z.}~\bibnamefont {Luki\'c}},\ }\href {\doibase
  10.1093/mnras/stae2255} {\bibfield  {journal} {\bibinfo  {journal} {Mon. Not.
  Roy. Astron. Soc.}\ }\textbf {\bibinfo {volume} {534}},\ \bibinfo {pages}
  {2674} (\bibinfo {year} {2024})},\ \Eprint {http://arxiv.org/abs/2407.04473}
  {arXiv:2407.04473 [astro-ph.CO]} \BibitemShut {NoStop}%
\bibitem [{\citenamefont {Cabayol-Garcia}\ \emph {et~al.}(2023)\citenamefont
  {Cabayol-Garcia}, \citenamefont {Chaves-Montero}, \citenamefont
  {Font-Ribera},\ and\ \citenamefont {Pedersen}}]{Cabayol-Garcia:2023ygj}%
  \BibitemOpen
  \bibfield  {author} {\bibinfo {author} {\bibfnamefont {L.}~\bibnamefont
  {Cabayol-Garcia}}, \bibinfo {author} {\bibfnamefont {J.}~\bibnamefont
  {Chaves-Montero}}, \bibinfo {author} {\bibfnamefont {A.}~\bibnamefont
  {Font-Ribera}}, \ and\ \bibinfo {author} {\bibfnamefont {C.}~\bibnamefont
  {Pedersen}},\ }\href {\doibase 10.1093/mnras/stad2512} {\bibfield  {journal}
  {\bibinfo  {journal} {Mon. Not. Roy. Astron. Soc.}\ }\textbf {\bibinfo
  {volume} {525}},\ \bibinfo {pages} {3499} (\bibinfo {year} {2023})},\ \Eprint
  {http://arxiv.org/abs/2305.19064} {arXiv:2305.19064 [astro-ph.CO]}
  \BibitemShut {NoStop}%
\bibitem [{\citenamefont {Baumann}\ \emph {et~al.}(2012)\citenamefont
  {Baumann}, \citenamefont {Nicolis}, \citenamefont {Senatore},\ and\
  \citenamefont {Zaldarriaga}}]{Baumann:2010tm}%
  \BibitemOpen
  \bibfield  {author} {\bibinfo {author} {\bibfnamefont {D.}~\bibnamefont
  {Baumann}}, \bibinfo {author} {\bibfnamefont {A.}~\bibnamefont {Nicolis}},
  \bibinfo {author} {\bibfnamefont {L.}~\bibnamefont {Senatore}}, \ and\
  \bibinfo {author} {\bibfnamefont {M.}~\bibnamefont {Zaldarriaga}},\ }\href
  {\doibase 10.1088/1475-7516/2012/07/051} {\bibfield  {journal} {\bibinfo
  {journal} {JCAP}\ }\textbf {\bibinfo {volume} {1207}},\ \bibinfo {pages}
  {051} (\bibinfo {year} {2012})},\ \Eprint {http://arxiv.org/abs/1004.2488}
  {arXiv:1004.2488 [astro-ph.CO]} \BibitemShut {NoStop}%
\bibitem [{\citenamefont {Carrasco}\ \emph {et~al.}(2012)\citenamefont
  {Carrasco}, \citenamefont {Hertzberg},\ and\ \citenamefont
  {Senatore}}]{Carrasco:2012cv}%
  \BibitemOpen
  \bibfield  {author} {\bibinfo {author} {\bibfnamefont {J.~J.~M.}\
  \bibnamefont {Carrasco}}, \bibinfo {author} {\bibfnamefont {M.~P.}\
  \bibnamefont {Hertzberg}}, \ and\ \bibinfo {author} {\bibfnamefont
  {L.}~\bibnamefont {Senatore}},\ }\href {\doibase 10.1007/JHEP09(2012)082}
  {\bibfield  {journal} {\bibinfo  {journal} {JHEP}\ }\textbf {\bibinfo
  {volume} {09}},\ \bibinfo {pages} {082} (\bibinfo {year} {2012})},\ \Eprint
  {http://arxiv.org/abs/1206.2926} {arXiv:1206.2926 [astro-ph.CO]} \BibitemShut
  {NoStop}%
\bibitem [{\citenamefont {Desjacques}\ \emph
  {et~al.}(2018{\natexlab{a}})\citenamefont {Desjacques}, \citenamefont
  {Jeong},\ and\ \citenamefont {Schmidt}}]{Desjacques:2016bnm}%
  \BibitemOpen
  \bibfield  {author} {\bibinfo {author} {\bibfnamefont {V.}~\bibnamefont
  {Desjacques}}, \bibinfo {author} {\bibfnamefont {D.}~\bibnamefont {Jeong}}, \
  and\ \bibinfo {author} {\bibfnamefont {F.}~\bibnamefont {Schmidt}},\ }\href
  {\doibase 10.1016/j.physrep.2017.12.002} {\bibfield  {journal} {\bibinfo
  {journal} {Phys. Rept.}\ }\textbf {\bibinfo {volume} {733}},\ \bibinfo
  {pages} {1} (\bibinfo {year} {2018}{\natexlab{a}})},\ \Eprint
  {http://arxiv.org/abs/1611.09787} {arXiv:1611.09787 [astro-ph.CO]}
  \BibitemShut {NoStop}%
\bibitem [{\citenamefont {Ivanov}(2022)}]{Ivanov:2022mrd}%
  \BibitemOpen
  \bibfield  {author} {\bibinfo {author} {\bibfnamefont {M.~M.}\ \bibnamefont
  {Ivanov}},\ }\href@noop {} {\  (\bibinfo {year} {2022})},\ \Eprint
  {http://arxiv.org/abs/2212.08488} {arXiv:2212.08488 [astro-ph.CO]}
  \BibitemShut {NoStop}%
\bibitem [{\citenamefont {Seljak}(2012)}]{Seljak:2012tp}%
  \BibitemOpen
  \bibfield  {author} {\bibinfo {author} {\bibfnamefont {U.}~\bibnamefont
  {Seljak}},\ }\href {\doibase 10.1088/1475-7516/2012/03/004} {\bibfield
  {journal} {\bibinfo  {journal} {JCAP}\ }\textbf {\bibinfo {volume} {03}},\
  \bibinfo {pages} {004} (\bibinfo {year} {2012})},\ \Eprint
  {http://arxiv.org/abs/1201.0594} {arXiv:1201.0594 [astro-ph.CO]} \BibitemShut
  {NoStop}%
\bibitem [{\citenamefont {Cieplak}\ and\ \citenamefont
  {Slosar}(2016)}]{Cieplak:2015kra}%
  \BibitemOpen
  \bibfield  {author} {\bibinfo {author} {\bibfnamefont {A.~M.}\ \bibnamefont
  {Cieplak}}\ and\ \bibinfo {author} {\bibfnamefont {A.}~\bibnamefont
  {Slosar}},\ }\href {\doibase 10.1088/1475-7516/2016/03/016} {\bibfield
  {journal} {\bibinfo  {journal} {JCAP}\ }\textbf {\bibinfo {volume} {03}},\
  \bibinfo {pages} {016} (\bibinfo {year} {2016})},\ \Eprint
  {http://arxiv.org/abs/1509.07875} {arXiv:1509.07875 [astro-ph.CO]}
  \BibitemShut {NoStop}%
\bibitem [{\citenamefont {Garny}\ \emph {et~al.}(2018)\citenamefont {Garny},
  \citenamefont {Konstandin}, \citenamefont {Sagunski},\ and\ \citenamefont
  {Tulin}}]{Garny:2018byk}%
  \BibitemOpen
  \bibfield  {author} {\bibinfo {author} {\bibfnamefont {M.}~\bibnamefont
  {Garny}}, \bibinfo {author} {\bibfnamefont {T.}~\bibnamefont {Konstandin}},
  \bibinfo {author} {\bibfnamefont {L.}~\bibnamefont {Sagunski}}, \ and\
  \bibinfo {author} {\bibfnamefont {S.}~\bibnamefont {Tulin}},\ }\href
  {\doibase 10.1088/1475-7516/2018/09/011} {\bibfield  {journal} {\bibinfo
  {journal} {JCAP}\ }\textbf {\bibinfo {volume} {09}},\ \bibinfo {pages} {011}
  (\bibinfo {year} {2018})},\ \Eprint {http://arxiv.org/abs/1805.12203}
  {arXiv:1805.12203 [astro-ph.CO]} \BibitemShut {NoStop}%
\bibitem [{\citenamefont {Garny}\ \emph {et~al.}(2021)\citenamefont {Garny},
  \citenamefont {Konstandin}, \citenamefont {Sagunski},\ and\ \citenamefont
  {Viel}}]{Garny:2020rom}%
  \BibitemOpen
  \bibfield  {author} {\bibinfo {author} {\bibfnamefont {M.}~\bibnamefont
  {Garny}}, \bibinfo {author} {\bibfnamefont {T.}~\bibnamefont {Konstandin}},
  \bibinfo {author} {\bibfnamefont {L.}~\bibnamefont {Sagunski}}, \ and\
  \bibinfo {author} {\bibfnamefont {M.}~\bibnamefont {Viel}},\ }\href {\doibase
  10.1088/1475-7516/2021/03/049} {\bibfield  {journal} {\bibinfo  {journal}
  {JCAP}\ }\textbf {\bibinfo {volume} {03}},\ \bibinfo {pages} {049} (\bibinfo
  {year} {2021})},\ \Eprint {http://arxiv.org/abs/2011.03050} {arXiv:2011.03050
  [astro-ph.CO]} \BibitemShut {NoStop}%
\bibitem [{\citenamefont {Givans}\ and\ \citenamefont
  {Hirata}(2020)}]{Givans:2020sez}%
  \BibitemOpen
  \bibfield  {author} {\bibinfo {author} {\bibfnamefont {J.~J.}\ \bibnamefont
  {Givans}}\ and\ \bibinfo {author} {\bibfnamefont {C.~M.}\ \bibnamefont
  {Hirata}},\ }\href {\doibase 10.1103/PhysRevD.102.023515} {\bibfield
  {journal} {\bibinfo  {journal} {Phys. Rev. D}\ }\textbf {\bibinfo {volume}
  {102}},\ \bibinfo {pages} {023515} (\bibinfo {year} {2020})},\ \Eprint
  {http://arxiv.org/abs/2002.12296} {arXiv:2002.12296 [astro-ph.CO]}
  \BibitemShut {NoStop}%
\bibitem [{\citenamefont {Chen}\ \emph {et~al.}(2021)\citenamefont {Chen},
  \citenamefont {Vlah},\ and\ \citenamefont {White}}]{Chen:2021rnb}%
  \BibitemOpen
  \bibfield  {author} {\bibinfo {author} {\bibfnamefont {S.-F.}\ \bibnamefont
  {Chen}}, \bibinfo {author} {\bibfnamefont {Z.}~\bibnamefont {Vlah}}, \ and\
  \bibinfo {author} {\bibfnamefont {M.}~\bibnamefont {White}},\ }\href
  {\doibase 10.1088/1475-7516/2021/05/053} {\bibfield  {journal} {\bibinfo
  {journal} {JCAP}\ }\textbf {\bibinfo {volume} {05}},\ \bibinfo {pages} {053}
  (\bibinfo {year} {2021})},\ \Eprint {http://arxiv.org/abs/2103.13498}
  {arXiv:2103.13498 [astro-ph.CO]} \BibitemShut {NoStop}%
\bibitem [{\citenamefont {Givans}\ \emph {et~al.}(2022)\citenamefont {Givans},
  \citenamefont {Font-Ribera}, \citenamefont {Slosar}, \citenamefont {Seeyave},
  \citenamefont {Pedersen}, \citenamefont {Rogers}, \citenamefont {Garny},
  \citenamefont {Blas},\ and\ \citenamefont {Ir\v{s}i\v{c}}}]{Givans:2022qgb}%
  \BibitemOpen
  \bibfield  {author} {\bibinfo {author} {\bibfnamefont {J.~J.}\ \bibnamefont
  {Givans}}, \bibinfo {author} {\bibfnamefont {A.}~\bibnamefont {Font-Ribera}},
  \bibinfo {author} {\bibfnamefont {A.}~\bibnamefont {Slosar}}, \bibinfo
  {author} {\bibfnamefont {L.}~\bibnamefont {Seeyave}}, \bibinfo {author}
  {\bibfnamefont {C.}~\bibnamefont {Pedersen}}, \bibinfo {author}
  {\bibfnamefont {K.~K.}\ \bibnamefont {Rogers}}, \bibinfo {author}
  {\bibfnamefont {M.}~\bibnamefont {Garny}}, \bibinfo {author} {\bibfnamefont
  {D.}~\bibnamefont {Blas}}, \ and\ \bibinfo {author} {\bibfnamefont
  {V.}~\bibnamefont {Ir\v{s}i\v{c}}},\ }\href {\doibase
  10.1088/1475-7516/2022/09/070} {\bibfield  {journal} {\bibinfo  {journal}
  {JCAP}\ }\textbf {\bibinfo {volume} {09}},\ \bibinfo {pages} {070} (\bibinfo
  {year} {2022})},\ \Eprint {http://arxiv.org/abs/2205.00962} {arXiv:2205.00962
  [astro-ph.CO]} \BibitemShut {NoStop}%
\bibitem [{\citenamefont {Fu\ss{}}\ and\ \citenamefont
  {Garny}(2022)}]{Fuss:2022zyt}%
  \BibitemOpen
  \bibfield  {author} {\bibinfo {author} {\bibfnamefont {L.}~\bibnamefont
  {Fu\ss{}}}\ and\ \bibinfo {author} {\bibfnamefont {M.}~\bibnamefont
  {Garny}},\ }\href@noop {} {\  (\bibinfo {year} {2022})},\ \Eprint
  {http://arxiv.org/abs/2210.06117} {arXiv:2210.06117 [astro-ph.CO]}
  \BibitemShut {NoStop}%
\bibitem [{\citenamefont {Ivanov}(2024)}]{Ivanov:2023yla}%
  \BibitemOpen
  \bibfield  {author} {\bibinfo {author} {\bibfnamefont {M.~M.}\ \bibnamefont
  {Ivanov}},\ }\href {\doibase 10.1103/PhysRevD.109.023507} {\bibfield
  {journal} {\bibinfo  {journal} {Phys. Rev. D}\ }\textbf {\bibinfo {volume}
  {109}},\ \bibinfo {pages} {023507} (\bibinfo {year} {2024})},\ \Eprint
  {http://arxiv.org/abs/2309.10133} {arXiv:2309.10133 [astro-ph.CO]}
  \BibitemShut {NoStop}%
\bibitem [{\citenamefont {Alam}\ \emph {et~al.}(2017)\citenamefont {Alam} \emph
  {et~al.}}]{BOSS:2016wmc}%
  \BibitemOpen
  \bibfield  {author} {\bibinfo {author} {\bibfnamefont {S.}~\bibnamefont
  {Alam}} \emph {et~al.} (\bibinfo {collaboration} {BOSS}),\ }\href {\doibase
  10.1093/mnras/stx721} {\bibfield  {journal} {\bibinfo  {journal} {Mon. Not.
  Roy. Astron. Soc.}\ }\textbf {\bibinfo {volume} {470}},\ \bibinfo {pages}
  {2617} (\bibinfo {year} {2017})},\ \Eprint {http://arxiv.org/abs/1607.03155}
  {arXiv:1607.03155 [astro-ph.CO]} \BibitemShut {NoStop}%
\bibitem [{\citenamefont {Alam}\ \emph {et~al.}(2021)\citenamefont {Alam} \emph
  {et~al.}}]{eBOSS:2020yzd}%
  \BibitemOpen
  \bibfield  {author} {\bibinfo {author} {\bibfnamefont {S.}~\bibnamefont
  {Alam}} \emph {et~al.} (\bibinfo {collaboration} {eBOSS}),\ }\href {\doibase
  10.1103/PhysRevD.103.083533} {\bibfield  {journal} {\bibinfo  {journal}
  {Phys. Rev. D}\ }\textbf {\bibinfo {volume} {103}},\ \bibinfo {pages}
  {083533} (\bibinfo {year} {2021})},\ \Eprint
  {http://arxiv.org/abs/2007.08991} {arXiv:2007.08991 [astro-ph.CO]}
  \BibitemShut {NoStop}%
\bibitem [{\citenamefont {Ivanov}\ \emph
  {et~al.}(2020{\natexlab{a}})\citenamefont {Ivanov}, \citenamefont
  {Simonovi\'c},\ and\ \citenamefont {Zaldarriaga}}]{Ivanov:2019pdj}%
  \BibitemOpen
  \bibfield  {author} {\bibinfo {author} {\bibfnamefont {M.~M.}\ \bibnamefont
  {Ivanov}}, \bibinfo {author} {\bibfnamefont {M.}~\bibnamefont {Simonovi\'c}},
  \ and\ \bibinfo {author} {\bibfnamefont {M.}~\bibnamefont {Zaldarriaga}},\
  }\href {\doibase 10.1088/1475-7516/2020/05/042} {\bibfield  {journal}
  {\bibinfo  {journal} {JCAP}\ }\textbf {\bibinfo {volume} {05}},\ \bibinfo
  {pages} {042} (\bibinfo {year} {2020}{\natexlab{a}})},\ \Eprint
  {http://arxiv.org/abs/1909.05277} {arXiv:1909.05277 [astro-ph.CO]}
  \BibitemShut {NoStop}%
\bibitem [{\citenamefont {D'Amico}\ \emph {et~al.}(2019)\citenamefont
  {D'Amico}, \citenamefont {Gleyzes}, \citenamefont {Kokron}, \citenamefont
  {Markovic}, \citenamefont {Senatore}, \citenamefont {Zhang}, \citenamefont
  {Beutler},\ and\ \citenamefont {Gil-Marín}}]{DAmico:2019fhj}%
  \BibitemOpen
  \bibfield  {author} {\bibinfo {author} {\bibfnamefont {G.}~\bibnamefont
  {D'Amico}}, \bibinfo {author} {\bibfnamefont {J.}~\bibnamefont {Gleyzes}},
  \bibinfo {author} {\bibfnamefont {N.}~\bibnamefont {Kokron}}, \bibinfo
  {author} {\bibfnamefont {D.}~\bibnamefont {Markovic}}, \bibinfo {author}
  {\bibfnamefont {L.}~\bibnamefont {Senatore}}, \bibinfo {author}
  {\bibfnamefont {P.}~\bibnamefont {Zhang}}, \bibinfo {author} {\bibfnamefont
  {F.}~\bibnamefont {Beutler}}, \ and\ \bibinfo {author} {\bibfnamefont
  {H.}~\bibnamefont {Gil-Marín}},\ }\href@noop {} {\  (\bibinfo {year}
  {2019})},\ \Eprint {http://arxiv.org/abs/1909.05271} {arXiv:1909.05271
  [astro-ph.CO]} \BibitemShut {NoStop}%
\bibitem [{\citenamefont {Chen}\ \emph {et~al.}(2022)\citenamefont {Chen},
  \citenamefont {Vlah},\ and\ \citenamefont {White}}]{Chen:2021wdi}%
  \BibitemOpen
  \bibfield  {author} {\bibinfo {author} {\bibfnamefont {S.-F.}\ \bibnamefont
  {Chen}}, \bibinfo {author} {\bibfnamefont {Z.}~\bibnamefont {Vlah}}, \ and\
  \bibinfo {author} {\bibfnamefont {M.}~\bibnamefont {White}},\ }\href
  {\doibase 10.1088/1475-7516/2022/02/008} {\bibfield  {journal} {\bibinfo
  {journal} {JCAP}\ }\textbf {\bibinfo {volume} {02}},\ \bibinfo {pages} {008}
  (\bibinfo {year} {2022})},\ \Eprint {http://arxiv.org/abs/2110.05530}
  {arXiv:2110.05530 [astro-ph.CO]} \BibitemShut {NoStop}%
\bibitem [{\citenamefont {Chudaykin}\ \emph {et~al.}(2020)\citenamefont
  {Chudaykin}, \citenamefont {Ivanov}, \citenamefont {Philcox},\ and\
  \citenamefont {Simonovi\'c}}]{Chudaykin:2020aoj}%
  \BibitemOpen
  \bibfield  {author} {\bibinfo {author} {\bibfnamefont {A.}~\bibnamefont
  {Chudaykin}}, \bibinfo {author} {\bibfnamefont {M.~M.}\ \bibnamefont
  {Ivanov}}, \bibinfo {author} {\bibfnamefont {O.~H.~E.}\ \bibnamefont
  {Philcox}}, \ and\ \bibinfo {author} {\bibfnamefont {M.}~\bibnamefont
  {Simonovi\'c}},\ }\href {\doibase 10.1103/PhysRevD.102.063533} {\bibfield
  {journal} {\bibinfo  {journal} {Phys. Rev. D}\ }\textbf {\bibinfo {volume}
  {102}},\ \bibinfo {pages} {063533} (\bibinfo {year} {2020})},\ \Eprint
  {http://arxiv.org/abs/2004.10607} {arXiv:2004.10607 [astro-ph.CO]}
  \BibitemShut {NoStop}%
\bibitem [{\citenamefont {{Cain}}\ \emph {et~al.}(2024)\citenamefont {{Cain}},
  \citenamefont {{Scannapieco}}, \citenamefont {{McQuinn}}, \citenamefont
  {{D'Aloisio}},\ and\ \citenamefont {{Trac}}}]{cainIGMreionization2024}%
  \BibitemOpen
  \bibfield  {author} {\bibinfo {author} {\bibfnamefont {C.}~\bibnamefont
  {{Cain}}}, \bibinfo {author} {\bibfnamefont {E.}~\bibnamefont
  {{Scannapieco}}}, \bibinfo {author} {\bibfnamefont {M.}~\bibnamefont
  {{McQuinn}}}, \bibinfo {author} {\bibfnamefont {A.}~\bibnamefont
  {{D'Aloisio}}}, \ and\ \bibinfo {author} {\bibfnamefont {H.}~\bibnamefont
  {{Trac}}},\ }\href {\doibase 10.48550/arXiv.2405.02397} {\bibfield  {journal}
  {\bibinfo  {journal} {arXiv e-prints}\ ,\ \bibinfo {eid} {arXiv:2405.02397}}
  (\bibinfo {year} {2024})},\ \Eprint {http://arxiv.org/abs/2405.02397}
  {arXiv:2405.02397 [astro-ph.CO]} \BibitemShut {NoStop}%
\bibitem [{\citenamefont {Bernardeau}\ \emph {et~al.}(2002)\citenamefont
  {Bernardeau}, \citenamefont {Colombi}, \citenamefont {Gaztanaga},\ and\
  \citenamefont {Scoccimarro}}]{Bernardeau:2001qr}%
  \BibitemOpen
  \bibfield  {author} {\bibinfo {author} {\bibfnamefont {F.}~\bibnamefont
  {Bernardeau}}, \bibinfo {author} {\bibfnamefont {S.}~\bibnamefont {Colombi}},
  \bibinfo {author} {\bibfnamefont {E.}~\bibnamefont {Gaztanaga}}, \ and\
  \bibinfo {author} {\bibfnamefont {R.}~\bibnamefont {Scoccimarro}},\ }\href
  {\doibase 10.1016/S0370-1573(02)00135-7} {\bibfield  {journal} {\bibinfo
  {journal} {Phys. Rept.}\ }\textbf {\bibinfo {volume} {367}},\ \bibinfo
  {pages} {1} (\bibinfo {year} {2002})},\ \Eprint
  {http://arxiv.org/abs/astro-ph/0112551} {arXiv:astro-ph/0112551 [astro-ph]}
  \BibitemShut {NoStop}%
\bibitem [{\citenamefont {Desjacques}\ \emph
  {et~al.}(2018{\natexlab{b}})\citenamefont {Desjacques}, \citenamefont
  {Jeong},\ and\ \citenamefont {Schmidt}}]{Desjacques:2018pfv}%
  \BibitemOpen
  \bibfield  {author} {\bibinfo {author} {\bibfnamefont {V.}~\bibnamefont
  {Desjacques}}, \bibinfo {author} {\bibfnamefont {D.}~\bibnamefont {Jeong}}, \
  and\ \bibinfo {author} {\bibfnamefont {F.}~\bibnamefont {Schmidt}},\ }\href
  {\doibase 10.1088/1475-7516/2018/12/035} {\bibfield  {journal} {\bibinfo
  {journal} {JCAP}\ }\textbf {\bibinfo {volume} {1812}},\ \bibinfo {pages}
  {035} (\bibinfo {year} {2018}{\natexlab{b}})},\ \Eprint
  {http://arxiv.org/abs/1806.04015} {arXiv:1806.04015 [astro-ph.CO]}
  \BibitemShut {NoStop}%
\bibitem [{\citenamefont {Kaiser}(1987)}]{Kaiser:1987qv}%
  \BibitemOpen
  \bibfield  {author} {\bibinfo {author} {\bibfnamefont {N.}~\bibnamefont
  {Kaiser}},\ }\href@noop {} {\bibfield  {journal} {\bibinfo  {journal} {Mon.
  Not. Roy. Astron. Soc.}\ }\textbf {\bibinfo {volume} {227}},\ \bibinfo
  {pages} {1} (\bibinfo {year} {1987})}\BibitemShut {NoStop}%
\bibitem [{\citenamefont {Scoccimarro}(2004)}]{Scoccimarro:2004tg}%
  \BibitemOpen
  \bibfield  {author} {\bibinfo {author} {\bibfnamefont {R.}~\bibnamefont
  {Scoccimarro}},\ }\href {\doibase 10.1103/PhysRevD.70.083007} {\bibfield
  {journal} {\bibinfo  {journal} {Phys. Rev.}\ }\textbf {\bibinfo {volume}
  {D70}},\ \bibinfo {pages} {083007} (\bibinfo {year} {2004})},\ \Eprint
  {http://arxiv.org/abs/astro-ph/0407214} {arXiv:astro-ph/0407214 [astro-ph]}
  \BibitemShut {NoStop}%
\bibitem [{\citenamefont {Simonovic}\ \emph {et~al.}(2018)\citenamefont
  {Simonovic}, \citenamefont {Baldauf}, \citenamefont {Zaldarriaga},
  \citenamefont {Carrasco},\ and\ \citenamefont
  {Kollmeier}}]{Simonovic:2017mhp}%
  \BibitemOpen
  \bibfield  {author} {\bibinfo {author} {\bibfnamefont {M.}~\bibnamefont
  {Simonovic}}, \bibinfo {author} {\bibfnamefont {T.}~\bibnamefont {Baldauf}},
  \bibinfo {author} {\bibfnamefont {M.}~\bibnamefont {Zaldarriaga}}, \bibinfo
  {author} {\bibfnamefont {J.~J.}\ \bibnamefont {Carrasco}}, \ and\ \bibinfo
  {author} {\bibfnamefont {J.~A.}\ \bibnamefont {Kollmeier}},\ }\href {\doibase
  10.1088/1475-7516/2018/04/030} {\bibfield  {journal} {\bibinfo  {journal}
  {JCAP}\ }\textbf {\bibinfo {volume} {1804}},\ \bibinfo {pages} {030}
  (\bibinfo {year} {2018})},\ \Eprint {http://arxiv.org/abs/1708.08130}
  {arXiv:1708.08130 [astro-ph.CO]} \BibitemShut {NoStop}%
\bibitem [{\citenamefont {Villasenor}\ \emph {et~al.}(2023)\citenamefont
  {Villasenor}, \citenamefont {Robertson}, \citenamefont {Madau},\ and\
  \citenamefont {Schneider}}]{Villasenor:2022aiy}%
  \BibitemOpen
  \bibfield  {author} {\bibinfo {author} {\bibfnamefont {B.}~\bibnamefont
  {Villasenor}}, \bibinfo {author} {\bibfnamefont {B.}~\bibnamefont
  {Robertson}}, \bibinfo {author} {\bibfnamefont {P.}~\bibnamefont {Madau}}, \
  and\ \bibinfo {author} {\bibfnamefont {E.}~\bibnamefont {Schneider}},\ }\href
  {\doibase 10.1103/PhysRevD.108.023502} {\bibfield  {journal} {\bibinfo
  {journal} {Phys. Rev. D}\ }\textbf {\bibinfo {volume} {108}},\ \bibinfo
  {pages} {023502} (\bibinfo {year} {2023})},\ \Eprint
  {http://arxiv.org/abs/2209.14220} {arXiv:2209.14220 [astro-ph.CO]}
  \BibitemShut {NoStop}%
\bibitem [{\citenamefont {{Lumsden}}\ \emph {et~al.}(1989)\citenamefont
  {{Lumsden}}, \citenamefont {{Heavens}},\ and\ \citenamefont
  {{Peacock}}}]{1989MNRAS.238..293L}%
  \BibitemOpen
  \bibfield  {author} {\bibinfo {author} {\bibfnamefont {S.~L.}\ \bibnamefont
  {{Lumsden}}}, \bibinfo {author} {\bibfnamefont {A.~F.}\ \bibnamefont
  {{Heavens}}}, \ and\ \bibinfo {author} {\bibfnamefont {J.~A.}\ \bibnamefont
  {{Peacock}}},\ }\href {\doibase 10.1093/mnras/238.2.293} {\bibfield
  {journal} {\bibinfo  {journal} {MNRAS}\ }\textbf {\bibinfo {volume} {238}},\
  \bibinfo {pages} {293} (\bibinfo {year} {1989})}\BibitemShut {NoStop}%
\bibitem [{\citenamefont {Ivanov}\ \emph
  {et~al.}(2024{\natexlab{a}})\citenamefont {Ivanov}, \citenamefont
  {Cuesta-Lazaro}, \citenamefont {Mishra-Sharma}, \citenamefont {Obuljen},\
  and\ \citenamefont {Toomey}}]{Ivanov:2024hgq}%
  \BibitemOpen
  \bibfield  {author} {\bibinfo {author} {\bibfnamefont {M.~M.}\ \bibnamefont
  {Ivanov}}, \bibinfo {author} {\bibfnamefont {C.}~\bibnamefont
  {Cuesta-Lazaro}}, \bibinfo {author} {\bibfnamefont {S.}~\bibnamefont
  {Mishra-Sharma}}, \bibinfo {author} {\bibfnamefont {A.}~\bibnamefont
  {Obuljen}}, \ and\ \bibinfo {author} {\bibfnamefont {M.~W.}\ \bibnamefont
  {Toomey}},\ }\href@noop {} {\  (\bibinfo {year} {2024}{\natexlab{a}})},\
  \Eprint {http://arxiv.org/abs/2402.13310} {arXiv:2402.13310 [astro-ph.CO]}
  \BibitemShut {NoStop}%
\bibitem [{\citenamefont {Ir\v{s}i\v{c}}\ and\ \citenamefont
  {McQuinn}(2018)}]{Irsic:2018hhg}%
  \BibitemOpen
  \bibfield  {author} {\bibinfo {author} {\bibfnamefont {V.}~\bibnamefont
  {Ir\v{s}i\v{c}}}\ and\ \bibinfo {author} {\bibfnamefont {M.}~\bibnamefont
  {McQuinn}},\ }\href {\doibase 10.1088/1475-7516/2018/04/026} {\bibfield
  {journal} {\bibinfo  {journal} {JCAP}\ }\textbf {\bibinfo {volume} {04}},\
  \bibinfo {pages} {026} (\bibinfo {year} {2018})},\ \Eprint
  {http://arxiv.org/abs/1801.02671} {arXiv:1801.02671 [astro-ph.CO]}
  \BibitemShut {NoStop}%
\bibitem [{\citenamefont {Chudaykin}\ and\ \citenamefont
  {Ivanov}(2022)}]{Chudaykin:2022nru}%
  \BibitemOpen
  \bibfield  {author} {\bibinfo {author} {\bibfnamefont {A.}~\bibnamefont
  {Chudaykin}}\ and\ \bibinfo {author} {\bibfnamefont {M.~M.}\ \bibnamefont
  {Ivanov}},\ }\href@noop {} {\  (\bibinfo {year} {2022})},\ \Eprint
  {http://arxiv.org/abs/2210.17044} {arXiv:2210.17044 [astro-ph.CO]}
  \BibitemShut {NoStop}%
\bibitem [{\citenamefont {{Faucher-Gigu{\`e}re}}\ \emph
  {et~al.}(2008)\citenamefont {{Faucher-Gigu{\`e}re}}, \citenamefont
  {{Prochaska}}, \citenamefont {{Lidz}}, \citenamefont {{Hernquist}},\ and\
  \citenamefont {{Zaldarriaga}}}]{fauchergiguereMeasurementOpacity2008}%
  \BibitemOpen
  \bibfield  {author} {\bibinfo {author} {\bibfnamefont {C.-A.}\ \bibnamefont
  {{Faucher-Gigu{\`e}re}}}, \bibinfo {author} {\bibfnamefont {J.~X.}\
  \bibnamefont {{Prochaska}}}, \bibinfo {author} {\bibfnamefont
  {A.}~\bibnamefont {{Lidz}}}, \bibinfo {author} {\bibfnamefont
  {L.}~\bibnamefont {{Hernquist}}}, \ and\ \bibinfo {author} {\bibfnamefont
  {M.}~\bibnamefont {{Zaldarriaga}}},\ }\href {\doibase 10.1086/588648}
  {\bibfield  {journal} {\bibinfo  {journal} {\apj}\ }\textbf {\bibinfo
  {volume} {681}},\ \bibinfo {pages} {831} (\bibinfo {year} {2008})},\ \Eprint
  {http://arxiv.org/abs/0709.2382} {arXiv:0709.2382 [astro-ph]} \BibitemShut
  {NoStop}%
\bibitem [{\citenamefont {{Turner}}\ \emph {et~al.}(2024)\citenamefont
  {{Turner}}, \citenamefont {{Martini}}, \citenamefont {{G{\"o}ksel
  Kara{\c{c}}ayl{\i}}}, \citenamefont {{Aguilar}}, \citenamefont {{Ahlen}},
  \citenamefont {{Brooks}}, \citenamefont {{Claybaugh}}, \citenamefont {{de la
  Macorra}}, \citenamefont {{Dey}}, \citenamefont {{Doel}}, \citenamefont
  {{Fanning}}, \citenamefont {{Forero-Romero}}, \citenamefont {{Gontcho}},
  \citenamefont {{Gonzalez-Morales}}, \citenamefont {{Gutierrez}},
  \citenamefont {{Guy}}, \citenamefont {{Herrera-Alcantar}}, \citenamefont
  {{Honscheid}}, \citenamefont {{Juneau}}, \citenamefont {{Kisner}},
  \citenamefont {{Kremin}}, \citenamefont {{Lambert}}, \citenamefont
  {{Landriau}}, \citenamefont {{Le Guillou}}, \citenamefont {{Meisner}},
  \citenamefont {{Miquel}}, \citenamefont {{Moustakas}}, \citenamefont
  {{Mueller}}, \citenamefont {{Mu{\~n}oz-Guti{\'e}rrez}}, \citenamefont
  {{Myers}}, \citenamefont {{Nie}}, \citenamefont {{Niz}}, \citenamefont
  {{Poppett}}, \citenamefont {{Prada}}, \citenamefont {{Rezaie}}, \citenamefont
  {{Rossi}}, \citenamefont {{Sanchez}}, \citenamefont {{Schlafly}},
  \citenamefont {{Schlegel}}, \citenamefont {{Schubnell}}, \citenamefont
  {{Seo}}, \citenamefont {{Sprayberry}}, \citenamefont {{Tarl{\'e}}},
  \citenamefont {{Weaver}},\ and\ \citenamefont
  {{Zou}}}]{turnerLyaForestMeanFluxFromDesiY12024}%
  \BibitemOpen
  \bibfield  {author} {\bibinfo {author} {\bibfnamefont {W.}~\bibnamefont
  {{Turner}}}, \bibinfo {author} {\bibfnamefont {P.}~\bibnamefont {{Martini}}},
  \bibinfo {author} {\bibfnamefont {N.}~\bibnamefont {{G{\"o}ksel
  Kara{\c{c}}ayl{\i}}}}, \bibinfo {author} {\bibfnamefont {J.}~\bibnamefont
  {{Aguilar}}}, \bibinfo {author} {\bibfnamefont {S.}~\bibnamefont {{Ahlen}}},
  \bibinfo {author} {\bibfnamefont {D.}~\bibnamefont {{Brooks}}}, \bibinfo
  {author} {\bibfnamefont {T.}~\bibnamefont {{Claybaugh}}}, \bibinfo {author}
  {\bibfnamefont {A.}~\bibnamefont {{de la Macorra}}}, \bibinfo {author}
  {\bibfnamefont {A.}~\bibnamefont {{Dey}}}, \bibinfo {author} {\bibfnamefont
  {P.}~\bibnamefont {{Doel}}}, \bibinfo {author} {\bibfnamefont
  {K.}~\bibnamefont {{Fanning}}}, \bibinfo {author} {\bibfnamefont {J.~E.}\
  \bibnamefont {{Forero-Romero}}}, \bibinfo {author} {\bibfnamefont {S.~G.~A.}\
  \bibnamefont {{Gontcho}}}, \bibinfo {author} {\bibfnamefont {A.~X.}\
  \bibnamefont {{Gonzalez-Morales}}}, \bibinfo {author} {\bibfnamefont
  {G.}~\bibnamefont {{Gutierrez}}}, \bibinfo {author} {\bibfnamefont
  {J.}~\bibnamefont {{Guy}}}, \bibinfo {author} {\bibfnamefont {H.~K.}\
  \bibnamefont {{Herrera-Alcantar}}}, \bibinfo {author} {\bibfnamefont
  {K.}~\bibnamefont {{Honscheid}}}, \bibinfo {author} {\bibfnamefont
  {S.}~\bibnamefont {{Juneau}}}, \bibinfo {author} {\bibfnamefont
  {T.}~\bibnamefont {{Kisner}}}, \bibinfo {author} {\bibfnamefont
  {A.}~\bibnamefont {{Kremin}}}, \bibinfo {author} {\bibfnamefont
  {A.}~\bibnamefont {{Lambert}}}, \bibinfo {author} {\bibfnamefont
  {M.}~\bibnamefont {{Landriau}}}, \bibinfo {author} {\bibfnamefont
  {L.}~\bibnamefont {{Le Guillou}}}, \bibinfo {author} {\bibfnamefont
  {A.}~\bibnamefont {{Meisner}}}, \bibinfo {author} {\bibfnamefont
  {R.}~\bibnamefont {{Miquel}}}, \bibinfo {author} {\bibfnamefont
  {J.}~\bibnamefont {{Moustakas}}}, \bibinfo {author} {\bibfnamefont
  {E.}~\bibnamefont {{Mueller}}}, \bibinfo {author} {\bibfnamefont
  {A.}~\bibnamefont {{Mu{\~n}oz-Guti{\'e}rrez}}}, \bibinfo {author}
  {\bibfnamefont {A.~D.}\ \bibnamefont {{Myers}}}, \bibinfo {author}
  {\bibfnamefont {J.}~\bibnamefont {{Nie}}}, \bibinfo {author} {\bibfnamefont
  {G.}~\bibnamefont {{Niz}}}, \bibinfo {author} {\bibfnamefont
  {C.}~\bibnamefont {{Poppett}}}, \bibinfo {author} {\bibfnamefont
  {F.}~\bibnamefont {{Prada}}}, \bibinfo {author} {\bibfnamefont
  {M.}~\bibnamefont {{Rezaie}}}, \bibinfo {author} {\bibfnamefont
  {G.}~\bibnamefont {{Rossi}}}, \bibinfo {author} {\bibfnamefont
  {E.}~\bibnamefont {{Sanchez}}}, \bibinfo {author} {\bibfnamefont {E.~F.}\
  \bibnamefont {{Schlafly}}}, \bibinfo {author} {\bibfnamefont
  {D.}~\bibnamefont {{Schlegel}}}, \bibinfo {author} {\bibfnamefont
  {M.}~\bibnamefont {{Schubnell}}}, \bibinfo {author} {\bibfnamefont
  {H.}~\bibnamefont {{Seo}}}, \bibinfo {author} {\bibfnamefont
  {D.}~\bibnamefont {{Sprayberry}}}, \bibinfo {author} {\bibfnamefont
  {G.}~\bibnamefont {{Tarl{\'e}}}}, \bibinfo {author} {\bibfnamefont {B.~A.}\
  \bibnamefont {{Weaver}}}, \ and\ \bibinfo {author} {\bibfnamefont
  {H.}~\bibnamefont {{Zou}}},\ }\href {\doibase 10.48550/arXiv.2405.06743}
  {\bibfield  {journal} {\bibinfo  {journal} {arXiv e-prints}\ ,\ \bibinfo
  {eid} {arXiv:2405.06743}} (\bibinfo {year} {2024})},\ \Eprint
  {http://arxiv.org/abs/2405.06743} {arXiv:2405.06743 [astro-ph.CO]}
  \BibitemShut {NoStop}%
\bibitem [{\citenamefont {Beutler}\ \emph {et~al.}(2017)\citenamefont {Beutler}
  \emph {et~al.}}]{Beutler:2016arn}%
  \BibitemOpen
  \bibfield  {author} {\bibinfo {author} {\bibfnamefont {F.}~\bibnamefont
  {Beutler}} \emph {et~al.} (\bibinfo {collaboration} {BOSS}),\ }\href
  {\doibase 10.1093/mnras/stw3298} {\bibfield  {journal} {\bibinfo  {journal}
  {Mon. Not. Roy. Astron. Soc.}\ }\textbf {\bibinfo {volume} {466}},\ \bibinfo
  {pages} {2242} (\bibinfo {year} {2017})},\ \Eprint
  {http://arxiv.org/abs/1607.03150} {arXiv:1607.03150 [astro-ph.CO]}
  \BibitemShut {NoStop}%
\bibitem [{\citenamefont {Cabass}\ \emph
  {et~al.}(2022{\natexlab{a}})\citenamefont {Cabass}, \citenamefont {Ivanov},
  \citenamefont {Philcox}, \citenamefont {Simonovic},\ and\ \citenamefont
  {Zaldarriaga}}]{Cabass:2022epm}%
  \BibitemOpen
  \bibfield  {author} {\bibinfo {author} {\bibfnamefont {G.}~\bibnamefont
  {Cabass}}, \bibinfo {author} {\bibfnamefont {M.~M.}\ \bibnamefont {Ivanov}},
  \bibinfo {author} {\bibfnamefont {O.~H.~E.}\ \bibnamefont {Philcox}},
  \bibinfo {author} {\bibfnamefont {M.}~\bibnamefont {Simonovic}}, \ and\
  \bibinfo {author} {\bibfnamefont {M.}~\bibnamefont {Zaldarriaga}},\
  }\href@noop {} {\  (\bibinfo {year} {2022}{\natexlab{a}})},\ \Eprint
  {http://arxiv.org/abs/2211.14899} {arXiv:2211.14899 [astro-ph.CO]}
  \BibitemShut {NoStop}%
\bibitem [{\citenamefont {Cabass}\ \emph
  {et~al.}(2022{\natexlab{b}})\citenamefont {Cabass}, \citenamefont {Ivanov},
  \citenamefont {Philcox}, \citenamefont {Simonovi\'c},\ and\ \citenamefont
  {Zaldarriaga}}]{Cabass:2022ymb}%
  \BibitemOpen
  \bibfield  {author} {\bibinfo {author} {\bibfnamefont {G.}~\bibnamefont
  {Cabass}}, \bibinfo {author} {\bibfnamefont {M.~M.}\ \bibnamefont {Ivanov}},
  \bibinfo {author} {\bibfnamefont {O.~H.~E.}\ \bibnamefont {Philcox}},
  \bibinfo {author} {\bibfnamefont {M.}~\bibnamefont {Simonovi\'c}}, \ and\
  \bibinfo {author} {\bibfnamefont {M.}~\bibnamefont {Zaldarriaga}},\
  }\href@noop {} {\  (\bibinfo {year} {2022}{\natexlab{b}})},\ \Eprint
  {http://arxiv.org/abs/2204.01781} {arXiv:2204.01781 [astro-ph.CO]}
  \BibitemShut {NoStop}%
\bibitem [{\citenamefont {Aghanim}\ \emph {et~al.}(2018)\citenamefont {Aghanim}
  \emph {et~al.}}]{Aghanim:2018eyx}%
  \BibitemOpen
  \bibfield  {author} {\bibinfo {author} {\bibfnamefont {N.}~\bibnamefont
  {Aghanim}} \emph {et~al.} (\bibinfo {collaboration} {Planck}),\ }\href@noop
  {} {\  (\bibinfo {year} {2018})},\ \Eprint {http://arxiv.org/abs/1807.06209}
  {arXiv:1807.06209 [astro-ph.CO]} \BibitemShut {NoStop}%
\bibitem [{\citenamefont {Ivanov}\ \emph
  {et~al.}(2020{\natexlab{b}})\citenamefont {Ivanov}, \citenamefont
  {Simonovi\'c},\ and\ \citenamefont {Zaldarriaga}}]{Ivanov:2019hqk}%
  \BibitemOpen
  \bibfield  {author} {\bibinfo {author} {\bibfnamefont {M.~M.}\ \bibnamefont
  {Ivanov}}, \bibinfo {author} {\bibfnamefont {M.}~\bibnamefont {Simonovi\'c}},
  \ and\ \bibinfo {author} {\bibfnamefont {M.}~\bibnamefont {Zaldarriaga}},\
  }\href {\doibase 10.1103/PhysRevD.101.083504} {\bibfield  {journal} {\bibinfo
   {journal} {Phys. Rev. D}\ }\textbf {\bibinfo {volume} {101}},\ \bibinfo
  {pages} {083504} (\bibinfo {year} {2020}{\natexlab{b}})},\ \Eprint
  {http://arxiv.org/abs/1912.08208} {arXiv:1912.08208 [astro-ph.CO]}
  \BibitemShut {NoStop}%
\bibitem [{\citenamefont {Adame}\ \emph {et~al.}(2024)\citenamefont {Adame}
  \emph {et~al.}}]{DESI:2024mwx}%
  \BibitemOpen
  \bibfield  {author} {\bibinfo {author} {\bibfnamefont {A.~G.}\ \bibnamefont
  {Adame}} \emph {et~al.} (\bibinfo {collaboration} {DESI}),\ }\href@noop {} {\
   (\bibinfo {year} {2024})},\ \Eprint {http://arxiv.org/abs/2404.03002}
  {arXiv:2404.03002 [astro-ph.CO]} \BibitemShut {NoStop}%
\bibitem [{\citenamefont {Lazeyras}\ and\ \citenamefont
  {Schmidt}(2018)}]{Lazeyras:2017hxw}%
  \BibitemOpen
  \bibfield  {author} {\bibinfo {author} {\bibfnamefont {T.}~\bibnamefont
  {Lazeyras}}\ and\ \bibinfo {author} {\bibfnamefont {F.}~\bibnamefont
  {Schmidt}},\ }\href {\doibase 10.1088/1475-7516/2018/09/008} {\bibfield
  {journal} {\bibinfo  {journal} {JCAP}\ }\textbf {\bibinfo {volume} {1809}},\
  \bibinfo {pages} {008} (\bibinfo {year} {2018})},\ \Eprint
  {http://arxiv.org/abs/1712.07531} {arXiv:1712.07531 [astro-ph.CO]}
  \BibitemShut {NoStop}%
\bibitem [{\citenamefont {Abidi}\ and\ \citenamefont
  {Baldauf}(2018)}]{Abidi:2018eyd}%
  \BibitemOpen
  \bibfield  {author} {\bibinfo {author} {\bibfnamefont {M.~M.}\ \bibnamefont
  {Abidi}}\ and\ \bibinfo {author} {\bibfnamefont {T.}~\bibnamefont
  {Baldauf}},\ }\href {\doibase 10.1088/1475-7516/2018/07/029} {\bibfield
  {journal} {\bibinfo  {journal} {JCAP}\ }\textbf {\bibinfo {volume} {1807}},\
  \bibinfo {pages} {029} (\bibinfo {year} {2018})},\ \Eprint
  {http://arxiv.org/abs/1802.07622} {arXiv:1802.07622 [astro-ph.CO]}
  \BibitemShut {NoStop}%
\bibitem [{\citenamefont {Barreira}(2020)}]{Barreira:2020ekm}%
  \BibitemOpen
  \bibfield  {author} {\bibinfo {author} {\bibfnamefont {A.}~\bibnamefont
  {Barreira}},\ }\href {\doibase 10.1088/1475-7516/2020/12/031} {\bibfield
  {journal} {\bibinfo  {journal} {JCAP}\ }\textbf {\bibinfo {volume} {12}},\
  \bibinfo {pages} {031} (\bibinfo {year} {2020})},\ \Eprint
  {http://arxiv.org/abs/2009.06622} {arXiv:2009.06622 [astro-ph.CO]}
  \BibitemShut {NoStop}%
\bibitem [{\citenamefont {Barreira}\ \emph {et~al.}(2021)\citenamefont
  {Barreira}, \citenamefont {Lazeyras},\ and\ \citenamefont
  {Schmidt}}]{Barreira:2021ukk}%
  \BibitemOpen
  \bibfield  {author} {\bibinfo {author} {\bibfnamefont {A.}~\bibnamefont
  {Barreira}}, \bibinfo {author} {\bibfnamefont {T.}~\bibnamefont {Lazeyras}},
  \ and\ \bibinfo {author} {\bibfnamefont {F.}~\bibnamefont {Schmidt}},\
  }\href@noop {} {\  (\bibinfo {year} {2021})},\ \Eprint
  {http://arxiv.org/abs/2105.02876} {arXiv:2105.02876 [astro-ph.CO]}
  \BibitemShut {NoStop}%
\bibitem [{\citenamefont {Lazeyras}\ \emph {et~al.}(2021)\citenamefont
  {Lazeyras}, \citenamefont {Barreira},\ and\ \citenamefont
  {Schmidt}}]{Lazeyras:2021dar}%
  \BibitemOpen
  \bibfield  {author} {\bibinfo {author} {\bibfnamefont {T.}~\bibnamefont
  {Lazeyras}}, \bibinfo {author} {\bibfnamefont {A.}~\bibnamefont {Barreira}},
  \ and\ \bibinfo {author} {\bibfnamefont {F.}~\bibnamefont {Schmidt}},\ }\href
  {\doibase 10.1088/1475-7516/2021/10/063} {\bibfield  {journal} {\bibinfo
  {journal} {JCAP}\ }\textbf {\bibinfo {volume} {10}},\ \bibinfo {pages} {063}
  (\bibinfo {year} {2021})},\ \Eprint {http://arxiv.org/abs/2106.14713}
  {arXiv:2106.14713 [astro-ph.CO]} \BibitemShut {NoStop}%
\bibitem [{\citenamefont {Cabass}\ \emph {et~al.}(2024)\citenamefont {Cabass},
  \citenamefont {Philcox}, \citenamefont {Ivanov}, \citenamefont {Akitsu},
  \citenamefont {Chen}, \citenamefont {Simonovi\'c},\ and\ \citenamefont
  {Zaldarriaga}}]{Cabass:2024wob}%
  \BibitemOpen
  \bibfield  {author} {\bibinfo {author} {\bibfnamefont {G.}~\bibnamefont
  {Cabass}}, \bibinfo {author} {\bibfnamefont {O.~H.~E.}\ \bibnamefont
  {Philcox}}, \bibinfo {author} {\bibfnamefont {M.~M.}\ \bibnamefont {Ivanov}},
  \bibinfo {author} {\bibfnamefont {K.}~\bibnamefont {Akitsu}}, \bibinfo
  {author} {\bibfnamefont {S.-F.}\ \bibnamefont {Chen}}, \bibinfo {author}
  {\bibfnamefont {M.}~\bibnamefont {Simonovi\'c}}, \ and\ \bibinfo {author}
  {\bibfnamefont {M.}~\bibnamefont {Zaldarriaga}},\ }\href@noop {} {\
  (\bibinfo {year} {2024})},\ \Eprint {http://arxiv.org/abs/2404.01894}
  {arXiv:2404.01894 [astro-ph.CO]} \BibitemShut {NoStop}%
\bibitem [{\citenamefont {Ivanov}\ \emph
  {et~al.}(2024{\natexlab{b}})\citenamefont {Ivanov}, \citenamefont {Obuljen},
  \citenamefont {Cuesta-Lazaro},\ and\ \citenamefont
  {Toomey}}]{Ivanov:2024xgb}%
  \BibitemOpen
  \bibfield  {author} {\bibinfo {author} {\bibfnamefont {M.~M.}\ \bibnamefont
  {Ivanov}}, \bibinfo {author} {\bibfnamefont {A.}~\bibnamefont {Obuljen}},
  \bibinfo {author} {\bibfnamefont {C.}~\bibnamefont {Cuesta-Lazaro}}, \ and\
  \bibinfo {author} {\bibfnamefont {M.~W.}\ \bibnamefont {Toomey}},\
  }\href@noop {} {\  (\bibinfo {year} {2024}{\natexlab{b}})},\ \Eprint
  {http://arxiv.org/abs/2409.10609} {arXiv:2409.10609 [astro-ph.CO]}
  \BibitemShut {NoStop}%
\bibitem [{\citenamefont {du~Mas~des Bourboux}\ \emph
  {et~al.}(2020{\natexlab{b}})\citenamefont {du~Mas~des Bourboux} \emph
  {et~al.}}]{duMasdesBourboux:2020pck}%
  \BibitemOpen
  \bibfield  {author} {\bibinfo {author} {\bibfnamefont {H.}~\bibnamefont
  {du~Mas~des Bourboux}} \emph {et~al.},\ }\href {\doibase
  10.3847/1538-4357/abb085} {\bibfield  {journal} {\bibinfo  {journal}
  {Astrophys. J.}\ }\textbf {\bibinfo {volume} {901}},\ \bibinfo {pages} {153}
  (\bibinfo {year} {2020}{\natexlab{b}})},\ \Eprint
  {http://arxiv.org/abs/2007.08995} {arXiv:2007.08995 [astro-ph.CO]}
  \BibitemShut {NoStop}%
\bibitem [{\citenamefont {{DESI Collaboration}}\ \emph
  {et~al.}(2024)\citenamefont {{DESI Collaboration}}, \citenamefont {Adame},
  \citenamefont {Aguilar}, \citenamefont {Ahlen}, \citenamefont {Alam},
  \citenamefont {Alexander}, \citenamefont {Alvarez}, \citenamefont {Alves},
  \citenamefont {Anand}, \citenamefont {Andrade}, \citenamefont {Armengaud},
  \citenamefont {Avila}, \citenamefont {Aviles}, \citenamefont {Awan},
  \citenamefont {Bailey}, \citenamefont {Baltay}, \citenamefont {Bault},
  \citenamefont {Bautista}, \citenamefont {Behera}, \citenamefont {BenZvi},
  \citenamefont {Beutler}, \citenamefont {Bianchi}, \citenamefont {Blake},
  \citenamefont {Blum}, \citenamefont {Brieden}, \citenamefont {Brodzeller},
  \citenamefont {Brooks}, \citenamefont {Buckley-Geer}, \citenamefont {Burtin},
  \citenamefont {Calderon}, \citenamefont {Canning}, \citenamefont {Rosell},
  \citenamefont {Cereskaite}, \citenamefont {Cervantes-Cota}, \citenamefont
  {Chabanier}, \citenamefont {Chaussidon}, \citenamefont {Chaves-Montero},
  \citenamefont {Chen}, \citenamefont {Chen}, \citenamefont {Claybaugh},
  \citenamefont {Cole}, \citenamefont {Cuceu}, \citenamefont {Davis},
  \citenamefont {Dawson}, \citenamefont {de~la Cruz}, \citenamefont {de~la
  Macorra}, \citenamefont {de~Mattia}, \citenamefont {Deiosso}, \citenamefont
  {Dey}, \citenamefont {Dey}, \citenamefont {Ding}, \citenamefont {Ding},
  \citenamefont {Doel}, \citenamefont {Edelstein}, \citenamefont
  {Eftekharzadeh}, \citenamefont {Eisenstein}, \citenamefont {Elliott},
  \citenamefont {Fagrelius}, \citenamefont {Fanning}, \citenamefont {Ferraro},
  \citenamefont {Ereza}, \citenamefont {Findlay}, \citenamefont {Flaugher},
  \citenamefont {Font-Ribera}, \citenamefont {Forero-Sánchez}, \citenamefont
  {Forero-Romero}, \citenamefont {Garcia-Quintero}, \citenamefont {Gaztañaga},
  \citenamefont {Gil-Marín}, \citenamefont {Gontcho}, \citenamefont
  {Gonzalez-Morales}, \citenamefont {Gonzalez-Perez}, \citenamefont {Gordon},
  \citenamefont {Green}, \citenamefont {Gruen}, \citenamefont {Gsponer},
  \citenamefont {Gutierrez}, \citenamefont {Guy}, \citenamefont {Hadzhiyska},
  \citenamefont {Hahn}, \citenamefont {Hanif}, \citenamefont
  {Herrera-Alcantar}, \citenamefont {Honscheid}, \citenamefont {Howlett},
  \citenamefont {Huterer}, \citenamefont {Iršič}, \citenamefont {Ishak},
  \citenamefont {Juneau}, \citenamefont {Karaçayli}, \citenamefont {Kehoe},
  \citenamefont {Kent}, \citenamefont {Kirkby}, \citenamefont {Kremin},
  \citenamefont {Krolewski}, \citenamefont {Lai}, \citenamefont {Lan},
  \citenamefont {Landriau}, \citenamefont {Lang}, \citenamefont {Lasker},
  \citenamefont {Goff}, \citenamefont {Guillou}, \citenamefont {Leauthaud},
  \citenamefont {Levi}, \citenamefont {Li}, \citenamefont {Linder},
  \citenamefont {Lodha}, \citenamefont {Magneville}, \citenamefont {Manera},
  \citenamefont {Margala}, \citenamefont {Martini}, \citenamefont {Maus},
  \citenamefont {McDonald}, \citenamefont {Medina-Varela}, \citenamefont
  {Meisner}, \citenamefont {Mena-Fernández}, \citenamefont {Miquel},
  \citenamefont {Moon}, \citenamefont {Moore}, \citenamefont {Moustakas},
  \citenamefont {Mudur}, \citenamefont {Mueller}, \citenamefont
  {Muñoz-Gutiérrez}, \citenamefont {Myers}, \citenamefont {Nadathur},
  \citenamefont {Napolitano}, \citenamefont {Neveux}, \citenamefont {Newman},
  \citenamefont {Nguyen}, \citenamefont {Nie}, \citenamefont {Niz},
  \citenamefont {Noriega}, \citenamefont {Padmanabhan}, \citenamefont
  {Paillas}, \citenamefont {Palanque-Delabrouille}, \citenamefont {Pan},
  \citenamefont {Penmetsa}, \citenamefont {Percival}, \citenamefont {Pieri},
  \citenamefont {Pinon}, \citenamefont {Poppett}, \citenamefont {Porredon},
  \citenamefont {Prada}, \citenamefont {Pérez-Fernández}, \citenamefont
  {Pérez-Ràfols}, \citenamefont {Rabinowitz}, \citenamefont {Raichoor},
  \citenamefont {Ramírez-Pérez}, \citenamefont {Ramirez-Solano},
  \citenamefont {Rashkovetskyi}, \citenamefont {Ravoux}, \citenamefont
  {Rezaie}, \citenamefont {Rich}, \citenamefont {Rocher}, \citenamefont
  {Rockosi}, \citenamefont {Roe}, \citenamefont {Rosado-Marin}, \citenamefont
  {Ross}, \citenamefont {Rossi}, \citenamefont {Ruggeri}, \citenamefont
  {Ruhlmann-Kleider}, \citenamefont {Samushia}, \citenamefont {Sanchez},
  \citenamefont {Saulder}, \citenamefont {Schlafly}, \citenamefont {Schlegel},
  \citenamefont {Schubnell}, \citenamefont {Seo}, \citenamefont {Sharples},
  \citenamefont {Silber}, \citenamefont {Sinigaglia}, \citenamefont {Slosar},
  \citenamefont {Smith}, \citenamefont {Sprayberry}, \citenamefont {Tan},
  \citenamefont {Tarlé}, \citenamefont {Trusov}, \citenamefont {Vaisakh},
  \citenamefont {Valcin}, \citenamefont {Valdes}, \citenamefont
  {Vargas-Magaña}, \citenamefont {Verde}, \citenamefont {Walther},
  \citenamefont {Wang}, \citenamefont {Wang}, \citenamefont {Weaver},
  \citenamefont {Weaverdyck}, \citenamefont {Wechsler}, \citenamefont
  {Weinberg}, \citenamefont {White}, \citenamefont {Yu}, \citenamefont {Yu},
  \citenamefont {Yuan}, \citenamefont {Yèche}, \citenamefont {Zaborowski},
  \citenamefont {Zarrouk}, \citenamefont {Zhang}, \citenamefont {Zhao},
  \citenamefont {Zhao}, \citenamefont {Zhou},\ and\ \citenamefont
  {Zou}}]{desiKp6BaoLya2024}%
  \BibitemOpen
  \bibfield  {author} {\bibinfo {author} {\bibnamefont {{DESI Collaboration}}},
  \bibinfo {author} {\bibfnamefont {A.~G.}\ \bibnamefont {Adame}}, \bibinfo
  {author} {\bibfnamefont {J.}~\bibnamefont {Aguilar}}, \bibinfo {author}
  {\bibfnamefont {S.}~\bibnamefont {Ahlen}}, \bibinfo {author} {\bibfnamefont
  {S.}~\bibnamefont {Alam}}, \bibinfo {author} {\bibfnamefont {D.~M.}\
  \bibnamefont {Alexander}}, \bibinfo {author} {\bibfnamefont {M.}~\bibnamefont
  {Alvarez}}, \bibinfo {author} {\bibfnamefont {O.}~\bibnamefont {Alves}},
  \bibinfo {author} {\bibfnamefont {A.}~\bibnamefont {Anand}}, \bibinfo
  {author} {\bibfnamefont {U.}~\bibnamefont {Andrade}}, \bibinfo {author}
  {\bibfnamefont {E.}~\bibnamefont {Armengaud}}, \bibinfo {author}
  {\bibfnamefont {S.}~\bibnamefont {Avila}}, \bibinfo {author} {\bibfnamefont
  {A.}~\bibnamefont {Aviles}}, \bibinfo {author} {\bibfnamefont
  {H.}~\bibnamefont {Awan}}, \bibinfo {author} {\bibfnamefont {S.}~\bibnamefont
  {Bailey}}, \bibinfo {author} {\bibfnamefont {C.}~\bibnamefont {Baltay}},
  \bibinfo {author} {\bibfnamefont {A.}~\bibnamefont {Bault}}, \bibinfo
  {author} {\bibfnamefont {J.}~\bibnamefont {Bautista}}, \bibinfo {author}
  {\bibfnamefont {J.}~\bibnamefont {Behera}}, \bibinfo {author} {\bibfnamefont
  {S.}~\bibnamefont {BenZvi}}, \bibinfo {author} {\bibfnamefont
  {F.}~\bibnamefont {Beutler}}, \bibinfo {author} {\bibfnamefont
  {D.}~\bibnamefont {Bianchi}}, \bibinfo {author} {\bibfnamefont
  {C.}~\bibnamefont {Blake}}, \bibinfo {author} {\bibfnamefont
  {R.}~\bibnamefont {Blum}}, \bibinfo {author} {\bibfnamefont {S.}~\bibnamefont
  {Brieden}}, \bibinfo {author} {\bibfnamefont {A.}~\bibnamefont {Brodzeller}},
  \bibinfo {author} {\bibfnamefont {D.}~\bibnamefont {Brooks}}, \bibinfo
  {author} {\bibfnamefont {E.}~\bibnamefont {Buckley-Geer}}, \bibinfo {author}
  {\bibfnamefont {E.}~\bibnamefont {Burtin}}, \bibinfo {author} {\bibfnamefont
  {R.}~\bibnamefont {Calderon}}, \bibinfo {author} {\bibfnamefont
  {R.}~\bibnamefont {Canning}}, \bibinfo {author} {\bibfnamefont {A.~C.}\
  \bibnamefont {Rosell}}, \bibinfo {author} {\bibfnamefont {R.}~\bibnamefont
  {Cereskaite}}, \bibinfo {author} {\bibfnamefont {J.~L.}\ \bibnamefont
  {Cervantes-Cota}}, \bibinfo {author} {\bibfnamefont {S.}~\bibnamefont
  {Chabanier}}, \bibinfo {author} {\bibfnamefont {E.}~\bibnamefont
  {Chaussidon}}, \bibinfo {author} {\bibfnamefont {J.}~\bibnamefont
  {Chaves-Montero}}, \bibinfo {author} {\bibfnamefont {S.}~\bibnamefont
  {Chen}}, \bibinfo {author} {\bibfnamefont {X.}~\bibnamefont {Chen}}, \bibinfo
  {author} {\bibfnamefont {T.}~\bibnamefont {Claybaugh}}, \bibinfo {author}
  {\bibfnamefont {S.}~\bibnamefont {Cole}}, \bibinfo {author} {\bibfnamefont
  {A.}~\bibnamefont {Cuceu}}, \bibinfo {author} {\bibfnamefont {T.~M.}\
  \bibnamefont {Davis}}, \bibinfo {author} {\bibfnamefont {K.}~\bibnamefont
  {Dawson}}, \bibinfo {author} {\bibfnamefont {R.}~\bibnamefont {de~la Cruz}},
  \bibinfo {author} {\bibfnamefont {A.}~\bibnamefont {de~la Macorra}}, \bibinfo
  {author} {\bibfnamefont {A.}~\bibnamefont {de~Mattia}}, \bibinfo {author}
  {\bibfnamefont {N.}~\bibnamefont {Deiosso}}, \bibinfo {author} {\bibfnamefont
  {A.}~\bibnamefont {Dey}}, \bibinfo {author} {\bibfnamefont {B.}~\bibnamefont
  {Dey}}, \bibinfo {author} {\bibfnamefont {J.}~\bibnamefont {Ding}}, \bibinfo
  {author} {\bibfnamefont {Z.}~\bibnamefont {Ding}}, \bibinfo {author}
  {\bibfnamefont {P.}~\bibnamefont {Doel}}, \bibinfo {author} {\bibfnamefont
  {J.}~\bibnamefont {Edelstein}}, \bibinfo {author} {\bibfnamefont
  {S.}~\bibnamefont {Eftekharzadeh}}, \bibinfo {author} {\bibfnamefont {D.~J.}\
  \bibnamefont {Eisenstein}}, \bibinfo {author} {\bibfnamefont
  {A.}~\bibnamefont {Elliott}}, \bibinfo {author} {\bibfnamefont
  {P.}~\bibnamefont {Fagrelius}}, \bibinfo {author} {\bibfnamefont
  {K.}~\bibnamefont {Fanning}}, \bibinfo {author} {\bibfnamefont
  {S.}~\bibnamefont {Ferraro}}, \bibinfo {author} {\bibfnamefont
  {J.}~\bibnamefont {Ereza}}, \bibinfo {author} {\bibfnamefont
  {N.}~\bibnamefont {Findlay}}, \bibinfo {author} {\bibfnamefont
  {B.}~\bibnamefont {Flaugher}}, \bibinfo {author} {\bibfnamefont
  {A.}~\bibnamefont {Font-Ribera}}, \bibinfo {author} {\bibfnamefont
  {D.}~\bibnamefont {Forero-Sánchez}}, \bibinfo {author} {\bibfnamefont
  {J.~E.}\ \bibnamefont {Forero-Romero}}, \bibinfo {author} {\bibfnamefont
  {C.}~\bibnamefont {Garcia-Quintero}}, \bibinfo {author} {\bibfnamefont
  {E.}~\bibnamefont {Gaztañaga}}, \bibinfo {author} {\bibfnamefont
  {H.}~\bibnamefont {Gil-Marín}}, \bibinfo {author} {\bibfnamefont {S.~G.~A.}\
  \bibnamefont {Gontcho}}, \bibinfo {author} {\bibfnamefont {A.~X.}\
  \bibnamefont {Gonzalez-Morales}}, \bibinfo {author} {\bibfnamefont
  {V.}~\bibnamefont {Gonzalez-Perez}}, \bibinfo {author} {\bibfnamefont
  {C.}~\bibnamefont {Gordon}}, \bibinfo {author} {\bibfnamefont
  {D.}~\bibnamefont {Green}}, \bibinfo {author} {\bibfnamefont
  {D.}~\bibnamefont {Gruen}}, \bibinfo {author} {\bibfnamefont
  {R.}~\bibnamefont {Gsponer}}, \bibinfo {author} {\bibfnamefont
  {G.}~\bibnamefont {Gutierrez}}, \bibinfo {author} {\bibfnamefont
  {J.}~\bibnamefont {Guy}}, \bibinfo {author} {\bibfnamefont {B.}~\bibnamefont
  {Hadzhiyska}}, \bibinfo {author} {\bibfnamefont {C.}~\bibnamefont {Hahn}},
  \bibinfo {author} {\bibfnamefont {M.~M.~S.}\ \bibnamefont {Hanif}}, \bibinfo
  {author} {\bibfnamefont {H.~K.}\ \bibnamefont {Herrera-Alcantar}}, \bibinfo
  {author} {\bibfnamefont {K.}~\bibnamefont {Honscheid}}, \bibinfo {author}
  {\bibfnamefont {C.}~\bibnamefont {Howlett}}, \bibinfo {author} {\bibfnamefont
  {D.}~\bibnamefont {Huterer}}, \bibinfo {author} {\bibfnamefont
  {V.}~\bibnamefont {Iršič}}, \bibinfo {author} {\bibfnamefont
  {M.}~\bibnamefont {Ishak}}, \bibinfo {author} {\bibfnamefont
  {S.}~\bibnamefont {Juneau}}, \bibinfo {author} {\bibfnamefont {N.~G.}\
  \bibnamefont {Karaçayli}}, \bibinfo {author} {\bibfnamefont
  {R.}~\bibnamefont {Kehoe}}, \bibinfo {author} {\bibfnamefont
  {S.}~\bibnamefont {Kent}}, \bibinfo {author} {\bibfnamefont {D.}~\bibnamefont
  {Kirkby}}, \bibinfo {author} {\bibfnamefont {A.}~\bibnamefont {Kremin}},
  \bibinfo {author} {\bibfnamefont {A.}~\bibnamefont {Krolewski}}, \bibinfo
  {author} {\bibfnamefont {Y.}~\bibnamefont {Lai}}, \bibinfo {author}
  {\bibfnamefont {T.~W.}\ \bibnamefont {Lan}}, \bibinfo {author} {\bibfnamefont
  {M.}~\bibnamefont {Landriau}}, \bibinfo {author} {\bibfnamefont
  {D.}~\bibnamefont {Lang}}, \bibinfo {author} {\bibfnamefont {J.}~\bibnamefont
  {Lasker}}, \bibinfo {author} {\bibfnamefont {J.~M.~L.}\ \bibnamefont {Goff}},
  \bibinfo {author} {\bibfnamefont {L.~L.}\ \bibnamefont {Guillou}}, \bibinfo
  {author} {\bibfnamefont {A.}~\bibnamefont {Leauthaud}}, \bibinfo {author}
  {\bibfnamefont {M.~E.}\ \bibnamefont {Levi}}, \bibinfo {author}
  {\bibfnamefont {T.~S.}\ \bibnamefont {Li}}, \bibinfo {author} {\bibfnamefont
  {E.}~\bibnamefont {Linder}}, \bibinfo {author} {\bibfnamefont
  {K.}~\bibnamefont {Lodha}}, \bibinfo {author} {\bibfnamefont
  {C.}~\bibnamefont {Magneville}}, \bibinfo {author} {\bibfnamefont
  {M.}~\bibnamefont {Manera}}, \bibinfo {author} {\bibfnamefont
  {D.}~\bibnamefont {Margala}}, \bibinfo {author} {\bibfnamefont
  {P.}~\bibnamefont {Martini}}, \bibinfo {author} {\bibfnamefont
  {M.}~\bibnamefont {Maus}}, \bibinfo {author} {\bibfnamefont {P.}~\bibnamefont
  {McDonald}}, \bibinfo {author} {\bibfnamefont {L.}~\bibnamefont
  {Medina-Varela}}, \bibinfo {author} {\bibfnamefont {A.}~\bibnamefont
  {Meisner}}, \bibinfo {author} {\bibfnamefont {J.}~\bibnamefont
  {Mena-Fernández}}, \bibinfo {author} {\bibfnamefont {R.}~\bibnamefont
  {Miquel}}, \bibinfo {author} {\bibfnamefont {J.}~\bibnamefont {Moon}},
  \bibinfo {author} {\bibfnamefont {S.}~\bibnamefont {Moore}}, \bibinfo
  {author} {\bibfnamefont {J.}~\bibnamefont {Moustakas}}, \bibinfo {author}
  {\bibfnamefont {N.}~\bibnamefont {Mudur}}, \bibinfo {author} {\bibfnamefont
  {E.}~\bibnamefont {Mueller}}, \bibinfo {author} {\bibfnamefont
  {A.}~\bibnamefont {Muñoz-Gutiérrez}}, \bibinfo {author} {\bibfnamefont
  {A.~D.}\ \bibnamefont {Myers}}, \bibinfo {author} {\bibfnamefont
  {S.}~\bibnamefont {Nadathur}}, \bibinfo {author} {\bibfnamefont
  {L.}~\bibnamefont {Napolitano}}, \bibinfo {author} {\bibfnamefont
  {R.}~\bibnamefont {Neveux}}, \bibinfo {author} {\bibfnamefont {J.~A.}\
  \bibnamefont {Newman}}, \bibinfo {author} {\bibfnamefont {N.~M.}\
  \bibnamefont {Nguyen}}, \bibinfo {author} {\bibfnamefont {J.}~\bibnamefont
  {Nie}}, \bibinfo {author} {\bibfnamefont {G.}~\bibnamefont {Niz}}, \bibinfo
  {author} {\bibfnamefont {H.~E.}\ \bibnamefont {Noriega}}, \bibinfo {author}
  {\bibfnamefont {N.}~\bibnamefont {Padmanabhan}}, \bibinfo {author}
  {\bibfnamefont {E.}~\bibnamefont {Paillas}}, \bibinfo {author} {\bibfnamefont
  {N.}~\bibnamefont {Palanque-Delabrouille}}, \bibinfo {author} {\bibfnamefont
  {J.}~\bibnamefont {Pan}}, \bibinfo {author} {\bibfnamefont {S.}~\bibnamefont
  {Penmetsa}}, \bibinfo {author} {\bibfnamefont {W.~J.}\ \bibnamefont
  {Percival}}, \bibinfo {author} {\bibfnamefont {M.}~\bibnamefont {Pieri}},
  \bibinfo {author} {\bibfnamefont {M.}~\bibnamefont {Pinon}}, \bibinfo
  {author} {\bibfnamefont {C.}~\bibnamefont {Poppett}}, \bibinfo {author}
  {\bibfnamefont {A.}~\bibnamefont {Porredon}}, \bibinfo {author}
  {\bibfnamefont {F.}~\bibnamefont {Prada}}, \bibinfo {author} {\bibfnamefont
  {A.}~\bibnamefont {Pérez-Fernández}}, \bibinfo {author} {\bibfnamefont
  {I.}~\bibnamefont {Pérez-Ràfols}}, \bibinfo {author} {\bibfnamefont
  {D.}~\bibnamefont {Rabinowitz}}, \bibinfo {author} {\bibfnamefont
  {A.}~\bibnamefont {Raichoor}}, \bibinfo {author} {\bibfnamefont
  {C.}~\bibnamefont {Ramírez-Pérez}}, \bibinfo {author} {\bibfnamefont
  {S.}~\bibnamefont {Ramirez-Solano}}, \bibinfo {author} {\bibfnamefont
  {M.}~\bibnamefont {Rashkovetskyi}}, \bibinfo {author} {\bibfnamefont
  {C.}~\bibnamefont {Ravoux}}, \bibinfo {author} {\bibfnamefont
  {M.}~\bibnamefont {Rezaie}}, \bibinfo {author} {\bibfnamefont
  {J.}~\bibnamefont {Rich}}, \bibinfo {author} {\bibfnamefont {A.}~\bibnamefont
  {Rocher}}, \bibinfo {author} {\bibfnamefont {C.}~\bibnamefont {Rockosi}},
  \bibinfo {author} {\bibfnamefont {N.~A.}\ \bibnamefont {Roe}}, \bibinfo
  {author} {\bibfnamefont {A.}~\bibnamefont {Rosado-Marin}}, \bibinfo {author}
  {\bibfnamefont {A.~J.}\ \bibnamefont {Ross}}, \bibinfo {author}
  {\bibfnamefont {G.}~\bibnamefont {Rossi}}, \bibinfo {author} {\bibfnamefont
  {R.}~\bibnamefont {Ruggeri}}, \bibinfo {author} {\bibfnamefont
  {V.}~\bibnamefont {Ruhlmann-Kleider}}, \bibinfo {author} {\bibfnamefont
  {L.}~\bibnamefont {Samushia}}, \bibinfo {author} {\bibfnamefont
  {E.}~\bibnamefont {Sanchez}}, \bibinfo {author} {\bibfnamefont
  {C.}~\bibnamefont {Saulder}}, \bibinfo {author} {\bibfnamefont {E.~F.}\
  \bibnamefont {Schlafly}}, \bibinfo {author} {\bibfnamefont {D.}~\bibnamefont
  {Schlegel}}, \bibinfo {author} {\bibfnamefont {M.}~\bibnamefont {Schubnell}},
  \bibinfo {author} {\bibfnamefont {H.}~\bibnamefont {Seo}}, \bibinfo {author}
  {\bibfnamefont {R.}~\bibnamefont {Sharples}}, \bibinfo {author}
  {\bibfnamefont {J.}~\bibnamefont {Silber}}, \bibinfo {author} {\bibfnamefont
  {F.}~\bibnamefont {Sinigaglia}}, \bibinfo {author} {\bibfnamefont
  {A.}~\bibnamefont {Slosar}}, \bibinfo {author} {\bibfnamefont
  {A.}~\bibnamefont {Smith}}, \bibinfo {author} {\bibfnamefont
  {D.}~\bibnamefont {Sprayberry}}, \bibinfo {author} {\bibfnamefont
  {T.}~\bibnamefont {Tan}}, \bibinfo {author} {\bibfnamefont {G.}~\bibnamefont
  {Tarlé}}, \bibinfo {author} {\bibfnamefont {S.}~\bibnamefont {Trusov}},
  \bibinfo {author} {\bibfnamefont {R.}~\bibnamefont {Vaisakh}}, \bibinfo
  {author} {\bibfnamefont {D.}~\bibnamefont {Valcin}}, \bibinfo {author}
  {\bibfnamefont {F.}~\bibnamefont {Valdes}}, \bibinfo {author} {\bibfnamefont
  {M.}~\bibnamefont {Vargas-Magaña}}, \bibinfo {author} {\bibfnamefont
  {L.}~\bibnamefont {Verde}}, \bibinfo {author} {\bibfnamefont
  {M.}~\bibnamefont {Walther}}, \bibinfo {author} {\bibfnamefont
  {B.}~\bibnamefont {Wang}}, \bibinfo {author} {\bibfnamefont {M.~S.}\
  \bibnamefont {Wang}}, \bibinfo {author} {\bibfnamefont {B.~A.}\ \bibnamefont
  {Weaver}}, \bibinfo {author} {\bibfnamefont {N.}~\bibnamefont {Weaverdyck}},
  \bibinfo {author} {\bibfnamefont {R.~H.}\ \bibnamefont {Wechsler}}, \bibinfo
  {author} {\bibfnamefont {D.~H.}\ \bibnamefont {Weinberg}}, \bibinfo {author}
  {\bibfnamefont {M.}~\bibnamefont {White}}, \bibinfo {author} {\bibfnamefont
  {J.}~\bibnamefont {Yu}}, \bibinfo {author} {\bibfnamefont {Y.}~\bibnamefont
  {Yu}}, \bibinfo {author} {\bibfnamefont {S.}~\bibnamefont {Yuan}}, \bibinfo
  {author} {\bibfnamefont {C.}~\bibnamefont {Yèche}}, \bibinfo {author}
  {\bibfnamefont {E.~A.}\ \bibnamefont {Zaborowski}}, \bibinfo {author}
  {\bibfnamefont {P.}~\bibnamefont {Zarrouk}}, \bibinfo {author} {\bibfnamefont
  {H.}~\bibnamefont {Zhang}}, \bibinfo {author} {\bibfnamefont
  {C.}~\bibnamefont {Zhao}}, \bibinfo {author} {\bibfnamefont {R.}~\bibnamefont
  {Zhao}}, \bibinfo {author} {\bibfnamefont {R.}~\bibnamefont {Zhou}}, \ and\
  \bibinfo {author} {\bibfnamefont {H.}~\bibnamefont {Zou}},\ }\href {\doibase
  10.48550/arXiv.2404.03001} {\bibfield  {journal} {\bibinfo  {journal} {arXiv
  e-prints}\ ,\ \bibinfo {pages} {arXiv:2404.03001}} (\bibinfo {year}
  {2024})},\ \Eprint {http://arxiv.org/abs/2404.03001} {arXiv:2404.03001
  [astro-ph.CO]} \BibitemShut {NoStop}%
\bibitem [{\citenamefont {{Abdul-Karim}}\ \emph {et~al.}(2023)\citenamefont
  {{Abdul-Karim}}, \citenamefont {{Armengaud}}, \citenamefont {{Mention}},
  \citenamefont {{Chabanier}}, \citenamefont {{Ravoux}},\ and\ \citenamefont
  {{Luki{\'c}}}}]{abdulkarimMeasumentPxLya2023}%
  \BibitemOpen
  \bibfield  {author} {\bibinfo {author} {\bibfnamefont {M.~L.}\ \bibnamefont
  {{Abdul-Karim}}}, \bibinfo {author} {\bibfnamefont {E.}~\bibnamefont
  {{Armengaud}}}, \bibinfo {author} {\bibfnamefont {G.}~\bibnamefont
  {{Mention}}}, \bibinfo {author} {\bibfnamefont {S.}~\bibnamefont
  {{Chabanier}}}, \bibinfo {author} {\bibfnamefont {C.}~\bibnamefont
  {{Ravoux}}}, \ and\ \bibinfo {author} {\bibfnamefont {Z.}~\bibnamefont
  {{Luki{\'c}}}},\ }\href {\doibase 10.48550/arXiv.2310.09116} {\bibfield
  {journal} {\bibinfo  {journal} {arXiv e-prints}\ ,\ \bibinfo {eid}
  {arXiv:2310.09116}} (\bibinfo {year} {2023})},\ \Eprint
  {http://arxiv.org/abs/2310.09116} {arXiv:2310.09116 [astro-ph.CO]}
  \BibitemShut {NoStop}%
\bibitem [{\citenamefont {{de Belsunce}}\ \emph {et~al.}(2024)\citenamefont
  {{de Belsunce}}, \citenamefont {{Philcox}}, \citenamefont {{Irsic}},
  \citenamefont {{McDonald}}, \citenamefont {{Guy}},\ and\ \citenamefont
  {{Palanque-Delabrouille}}}]{debelsunceP3DLya2024}%
  \BibitemOpen
  \bibfield  {author} {\bibinfo {author} {\bibfnamefont {R.}~\bibnamefont {{de
  Belsunce}}}, \bibinfo {author} {\bibfnamefont {O.~H.~E.}\ \bibnamefont
  {{Philcox}}}, \bibinfo {author} {\bibfnamefont {V.}~\bibnamefont {{Irsic}}},
  \bibinfo {author} {\bibfnamefont {P.}~\bibnamefont {{McDonald}}}, \bibinfo
  {author} {\bibfnamefont {J.}~\bibnamefont {{Guy}}}, \ and\ \bibinfo {author}
  {\bibfnamefont {N.}~\bibnamefont {{Palanque-Delabrouille}}},\ }\href
  {\doibase 10.48550/arXiv.2403.08241} {\bibfield  {journal} {\bibinfo
  {journal} {arXiv e-prints}\ ,\ \bibinfo {eid} {arXiv:2403.08241}} (\bibinfo
  {year} {2024})},\ \Eprint {http://arxiv.org/abs/2403.08241} {arXiv:2403.08241
  [astro-ph.CO]} \BibitemShut {NoStop}%
\bibitem [{\citenamefont {Doux}\ \emph {et~al.}(2016)\citenamefont {Doux},
  \citenamefont {Schaan}, \citenamefont {Aubourg}, \citenamefont {Ganga},
  \citenamefont {Lee}, \citenamefont {Spergel},\ and\ \citenamefont
  {Tr{\'e}guer}}]{douxFirstDetectionCosmic2016}%
  \BibitemOpen
  \bibfield  {author} {\bibinfo {author} {\bibfnamefont {C.}~\bibnamefont
  {Doux}}, \bibinfo {author} {\bibfnamefont {E.}~\bibnamefont {Schaan}},
  \bibinfo {author} {\bibfnamefont {E.}~\bibnamefont {Aubourg}}, \bibinfo
  {author} {\bibfnamefont {K.}~\bibnamefont {Ganga}}, \bibinfo {author}
  {\bibfnamefont {K.-G.}\ \bibnamefont {Lee}}, \bibinfo {author} {\bibfnamefont
  {D.~N.}\ \bibnamefont {Spergel}}, \ and\ \bibinfo {author} {\bibfnamefont
  {J.}~\bibnamefont {Tr{\'e}guer}},\ }\href {\doibase
  10.1103/PhysRevD.94.103506} {\bibfield  {journal} {\bibinfo  {journal}
  {\prd}\ }\textbf {\bibinfo {volume} {94}},\ \bibinfo {pages} {103506}
  (\bibinfo {year} {2016})}\BibitemShut {NoStop}%
\bibitem [{\citenamefont {{Kara{\c{c}}ayl{\i}}}\ \emph
  {et~al.}(2024)\citenamefont {{Kara{\c{c}}ayl{\i}}}, \citenamefont
  {{Martini}}, \citenamefont {{Weinberg}}, \citenamefont {{Ferraro}},
  \citenamefont {{de Belsunce}}, \citenamefont {{Aguilar}}, \citenamefont
  {{Ahlen}}, \citenamefont {{Armengaud}}, \citenamefont {{Brooks}},
  \citenamefont {{Claybaugh}}, \citenamefont {{de la Macorra}}, \citenamefont
  {{Dey}}, \citenamefont {{Doel}}, \citenamefont {{Fanning}}, \citenamefont
  {{Forero-Romero}}, \citenamefont {{Gontcho}}, \citenamefont
  {{Gonzalez-Morales}}, \citenamefont {{Gutierrez}}, \citenamefont {{Guy}},
  \citenamefont {{Honscheid}}, \citenamefont {{Kirkby}}, \citenamefont
  {{Kisner}}, \citenamefont {{Kremin}}, \citenamefont {{Lambert}},
  \citenamefont {{Landriau}}, \citenamefont {{Le Guillou}}, \citenamefont
  {{Levi}}, \citenamefont {{Manera}}, \citenamefont {{Meisner}}, \citenamefont
  {{Miquel}}, \citenamefont {{Mueller}}, \citenamefont
  {{Mu{\~n}oz-Guti{\'e}rrez}}, \citenamefont {{Myers}}, \citenamefont
  {{Newman}}, \citenamefont {{Nie}}, \citenamefont {{Niz}}, \citenamefont
  {{Palanque-Delabrouille}}, \citenamefont {{Percival}}, \citenamefont
  {{Poppett}}, \citenamefont {{Prada}}, \citenamefont {{Ravoux}}, \citenamefont
  {{Rezaie}}, \citenamefont {{Ross}}, \citenamefont {{Rossi}}, \citenamefont
  {{Sanchez}}, \citenamefont {{Schlafly}}, \citenamefont {{Schlegel}},
  \citenamefont {{Seo}}, \citenamefont {{Sprayberry}}, \citenamefont {{Tan}},
  \citenamefont {{Tarl{\'e}}}, \citenamefont {{Weaver}},\ and\ \citenamefont
  {{Zou}}}]{karacayliCmbxLyaP1d2024}%
  \BibitemOpen
  \bibfield  {author} {\bibinfo {author} {\bibfnamefont {N.~G.}\ \bibnamefont
  {{Kara{\c{c}}ayl{\i}}}}, \bibinfo {author} {\bibfnamefont {P.}~\bibnamefont
  {{Martini}}}, \bibinfo {author} {\bibfnamefont {D.~H.}\ \bibnamefont
  {{Weinberg}}}, \bibinfo {author} {\bibfnamefont {S.}~\bibnamefont
  {{Ferraro}}}, \bibinfo {author} {\bibfnamefont {R.}~\bibnamefont {{de
  Belsunce}}}, \bibinfo {author} {\bibfnamefont {J.}~\bibnamefont {{Aguilar}}},
  \bibinfo {author} {\bibfnamefont {S.}~\bibnamefont {{Ahlen}}}, \bibinfo
  {author} {\bibfnamefont {E.}~\bibnamefont {{Armengaud}}}, \bibinfo {author}
  {\bibfnamefont {D.}~\bibnamefont {{Brooks}}}, \bibinfo {author}
  {\bibfnamefont {T.}~\bibnamefont {{Claybaugh}}}, \bibinfo {author}
  {\bibfnamefont {A.}~\bibnamefont {{de la Macorra}}}, \bibinfo {author}
  {\bibfnamefont {B.}~\bibnamefont {{Dey}}}, \bibinfo {author} {\bibfnamefont
  {P.}~\bibnamefont {{Doel}}}, \bibinfo {author} {\bibfnamefont
  {K.}~\bibnamefont {{Fanning}}}, \bibinfo {author} {\bibfnamefont {J.~E.}\
  \bibnamefont {{Forero-Romero}}}, \bibinfo {author} {\bibfnamefont {S.~G.~A.}\
  \bibnamefont {{Gontcho}}}, \bibinfo {author} {\bibfnamefont {A.~X.}\
  \bibnamefont {{Gonzalez-Morales}}}, \bibinfo {author} {\bibfnamefont
  {G.}~\bibnamefont {{Gutierrez}}}, \bibinfo {author} {\bibfnamefont
  {J.}~\bibnamefont {{Guy}}}, \bibinfo {author} {\bibfnamefont
  {K.}~\bibnamefont {{Honscheid}}}, \bibinfo {author} {\bibfnamefont
  {D.}~\bibnamefont {{Kirkby}}}, \bibinfo {author} {\bibfnamefont
  {T.}~\bibnamefont {{Kisner}}}, \bibinfo {author} {\bibfnamefont
  {A.}~\bibnamefont {{Kremin}}}, \bibinfo {author} {\bibfnamefont
  {A.}~\bibnamefont {{Lambert}}}, \bibinfo {author} {\bibfnamefont
  {M.}~\bibnamefont {{Landriau}}}, \bibinfo {author} {\bibfnamefont
  {L.}~\bibnamefont {{Le Guillou}}}, \bibinfo {author} {\bibfnamefont {M.~E.}\
  \bibnamefont {{Levi}}}, \bibinfo {author} {\bibfnamefont {M.}~\bibnamefont
  {{Manera}}}, \bibinfo {author} {\bibfnamefont {A.}~\bibnamefont {{Meisner}}},
  \bibinfo {author} {\bibfnamefont {R.}~\bibnamefont {{Miquel}}}, \bibinfo
  {author} {\bibfnamefont {E.}~\bibnamefont {{Mueller}}}, \bibinfo {author}
  {\bibfnamefont {A.}~\bibnamefont {{Mu{\~n}oz-Guti{\'e}rrez}}}, \bibinfo
  {author} {\bibfnamefont {A.~D.}\ \bibnamefont {{Myers}}}, \bibinfo {author}
  {\bibfnamefont {J.~A.}\ \bibnamefont {{Newman}}}, \bibinfo {author}
  {\bibfnamefont {J.}~\bibnamefont {{Nie}}}, \bibinfo {author} {\bibfnamefont
  {G.}~\bibnamefont {{Niz}}}, \bibinfo {author} {\bibfnamefont
  {N.}~\bibnamefont {{Palanque-Delabrouille}}}, \bibinfo {author}
  {\bibfnamefont {W.~J.}\ \bibnamefont {{Percival}}}, \bibinfo {author}
  {\bibfnamefont {C.}~\bibnamefont {{Poppett}}}, \bibinfo {author}
  {\bibfnamefont {F.}~\bibnamefont {{Prada}}}, \bibinfo {author} {\bibfnamefont
  {C.}~\bibnamefont {{Ravoux}}}, \bibinfo {author} {\bibfnamefont
  {M.}~\bibnamefont {{Rezaie}}}, \bibinfo {author} {\bibfnamefont {A.~J.}\
  \bibnamefont {{Ross}}}, \bibinfo {author} {\bibfnamefont {G.}~\bibnamefont
  {{Rossi}}}, \bibinfo {author} {\bibfnamefont {E.}~\bibnamefont {{Sanchez}}},
  \bibinfo {author} {\bibfnamefont {E.~F.}\ \bibnamefont {{Schlafly}}},
  \bibinfo {author} {\bibfnamefont {D.}~\bibnamefont {{Schlegel}}}, \bibinfo
  {author} {\bibfnamefont {H.}~\bibnamefont {{Seo}}}, \bibinfo {author}
  {\bibfnamefont {D.}~\bibnamefont {{Sprayberry}}}, \bibinfo {author}
  {\bibfnamefont {T.}~\bibnamefont {{Tan}}}, \bibinfo {author} {\bibfnamefont
  {G.}~\bibnamefont {{Tarl{\'e}}}}, \bibinfo {author} {\bibfnamefont {B.~A.}\
  \bibnamefont {{Weaver}}}, \ and\ \bibinfo {author} {\bibfnamefont
  {H.}~\bibnamefont {{Zou}}},\ }\href {\doibase 10.1103/PhysRevD.110.063505}
  {\bibfield  {journal} {\bibinfo  {journal} {\prd}\ }\textbf {\bibinfo
  {volume} {110}},\ \bibinfo {eid} {063505} (\bibinfo {year} {2024})},\ \Eprint
  {http://arxiv.org/abs/2405.14988} {arXiv:2405.14988 [astro-ph.CO]}
  \BibitemShut {NoStop}%
\bibitem [{\citenamefont {Abdalla}\ \emph {et~al.}(2022)\citenamefont {Abdalla}
  \emph {et~al.}}]{Abdalla:2022yfr}%
  \BibitemOpen
  \bibfield  {author} {\bibinfo {author} {\bibfnamefont {E.}~\bibnamefont
  {Abdalla}} \emph {et~al.},\ }\href {\doibase 10.1016/j.jheap.2022.04.002}
  {\bibfield  {journal} {\bibinfo  {journal} {JHEAp}\ }\textbf {\bibinfo
  {volume} {34}},\ \bibinfo {pages} {49} (\bibinfo {year} {2022})},\ \Eprint
  {http://arxiv.org/abs/2203.06142} {arXiv:2203.06142 [astro-ph.CO]}
  \BibitemShut {NoStop}%
\bibitem [{\citenamefont {Audren}\ \emph {et~al.}(2013)\citenamefont {Audren},
  \citenamefont {Lesgourgues}, \citenamefont {Benabed},\ and\ \citenamefont
  {Prunet}}]{Audren:2012wb}%
  \BibitemOpen
  \bibfield  {author} {\bibinfo {author} {\bibfnamefont {B.}~\bibnamefont
  {Audren}}, \bibinfo {author} {\bibfnamefont {J.}~\bibnamefont {Lesgourgues}},
  \bibinfo {author} {\bibfnamefont {K.}~\bibnamefont {Benabed}}, \ and\
  \bibinfo {author} {\bibfnamefont {S.}~\bibnamefont {Prunet}},\ }\href
  {\doibase 10.1088/1475-7516/2013/02/001} {\bibfield  {journal} {\bibinfo
  {journal} {JCAP}\ }\textbf {\bibinfo {volume} {1302}},\ \bibinfo {pages}
  {001} (\bibinfo {year} {2013})},\ \Eprint {http://arxiv.org/abs/1210.7183}
  {arXiv:1210.7183 [astro-ph.CO]} \BibitemShut {NoStop}%
\bibitem [{\citenamefont {Brinckmann}\ and\ \citenamefont
  {Lesgourgues}(2019)}]{Brinckmann:2018cvx}%
  \BibitemOpen
  \bibfield  {author} {\bibinfo {author} {\bibfnamefont {T.}~\bibnamefont
  {Brinckmann}}\ and\ \bibinfo {author} {\bibfnamefont {J.}~\bibnamefont
  {Lesgourgues}},\ }\href {\doibase 10.1016/j.dark.2018.100260} {\bibfield
  {journal} {\bibinfo  {journal} {Phys. Dark Univ.}\ }\textbf {\bibinfo
  {volume} {24}},\ \bibinfo {pages} {100260} (\bibinfo {year} {2019})},\
  \Eprint {http://arxiv.org/abs/1804.07261} {arXiv:1804.07261 [astro-ph.CO]}
  \BibitemShut {NoStop}%
\end{thebibliography}%

\newpage 

\pagebreak
\widetext
\begin{center}
\textbf{\large Supplemental Material}
\end{center}
\setcounter{equation}{0}
\setcounter{figure}{0}
\setcounter{table}{0}
\setcounter{page}{1}
\makeatletter
\renewcommand{\theequation}{S\arabic{equation}}
\renewcommand{\thefigure}{S\arabic{figure}}

\textit{One-loop EFT overview.} 
We describe now the EFT  
for non-linear Ly$\alpha$ correlations
in three dimensions, derived in~\cite{Ivanov:2023yla}. (See also~\cite{Desjacques:2018pfv} for an earlier related work in the context of galaxy bias
with line-of-sight 
selection effects.)
The non-linear flux fluctuation field 
$\delta_F=F/\bar F-1$ in redshift space
at the cubic 
order in the linear matter overdensity $\delta^{(1)}$ is expressed as
\be
\begin{split}
 \delta^{(s)}_F(\k)= &\sum_{n=1}^3 \Big[ \prod_{j=1}^n\int_{}\frac{d^3{\bf k}_j}{(2\pi)^3} \delta^{(1)}(\k_j)\Big]
 K_n(\k_1,...,\k_n)\times (2\pi)^3\delta^{(3)}_D(\k-\k_1-...-\k_j)\,,
 \end{split}
\ee
where $f$ is the 
logarithmic growth factor, $\mu\equiv k_\parallel/k$, $k_\parallel$ is the line-of-sight wavevector component.
The above perturbation theory
kernels for Ly$\alpha$ are given by:
\be
\label{eq:K2full}
\begin{split}
& K_1(\k) = b_1-b_\eta f\mu^2\,,\\
& K_2(\k_1,\k_2)=\frac{b_2}{2}+b_{\mathcal{G}_2}\left(\frac{(\k_1\cdot \k_2)^2}{k_1^2 k_2^2}-1\right)+b_1F_2(\k_1,\k_2)  -b_\eta f\mu^2 G_2(\k_1,\k_2) - fb_{\delta \eta}\frac{\mu_2^2+\mu_1^2}{2} +b_{\eta^2}f^2\mu_1^2\mu_2^2\\
& +b_1f\frac{\mu_1\mu_2}{2}\left(\frac{k_2}{k_1} + \frac{k_1}{k_2}\right)
-b_\eta f^2\frac{\mu_1\mu_2}{2}\left(\frac{k_2}{k_1}\mu_2^2 + \frac{k_1}{k_2}\mu_1^2\right) + b_{(KK)_\parallel}\left(\mu_1\mu_2 \frac{(\k_1\cdot \k_2)}{k_1k_2}
-\frac{\mu_1^2+\mu_2^2}{3}+\frac{1}{9}
\right)\\
& + b_{\Pi^{(2)}_\parallel}\left(\mu_1\mu_2 \frac{(\k_1\cdot \k_2)}{k_1k_2}+\frac{5}{7}\mu^2 \left(1-\frac{(\k_1\cdot \k_2)^2}{k_1^2 k_2^2}\right)\right)\,,
\end{split} 
\ee
with $\mu_i=(\hat{\bf z}\cdot \hat{\k}_i)$, $\hat {\bf z}$ denoting 
the line-of-sight direction unit vector,  
$f=d\ln D_+/d\ln a$ ($D_+$ is the growth factor),
and we have used the
usual density and velocity kernels from standard cosmological perturbation theory:
\be
\begin{split}
F_2(\k_1,\k_2) &= \frac{5}{7} + \frac{2}{7}\frac{(\k_1\cdot \k_2)^2}{k_1^2k_2^2} + \frac{1}{2}\frac{\k_1\cdot \k_2}{k_1k_2}\bigg(\frac{k_1}{k_2} + \frac{k_2}{k_1}\bigg)\,\,, \\
G_2(\k_1,\k_2) &= \frac{3}{7} + \frac{4}{7}\frac{(\k_1\cdot \k_2)^2}{k_1^2k_2^2} + \frac{1}{2}\frac{\k_1\cdot \k_2}{k_1k_2}\bigg(\frac{k_1}{k_2} + \frac{k_2}{k_1}\bigg) \,\,,
\end{split}
\ee
The general expression for the $K_3$ kernel is quite cumbersome. 
Below we present it only for the 
kinematic configurations that appear in the one-loop power spectrum integrals: 
\be 
\label{eq:K3full}
\begin{split}
& \int_{\q} K_3(\k,\q,-\q) 
P_{\rm lin}(q)
\\
=\,&
b_1
\int_{\q} F_3(\k,\q,-\q)P_{\text{lin}}(q)
-
f b_\eta\mu^2
\int_{\q} G_3(\q,-\q,\k)P_{\text{lin}}(q)
\:+ 
\int_{\q}
\left[
1 - \left(\khat\cdot\qhat\right)^2
\right]
P_{\rm lin}(q)
\\
&\:\times
\Bigg\{
\frac{4}{21}
(
5b_{\mathcal{G}_2}
+ 
2b_{\Gamma_3})
\left[\left(\frac{(\k-\q)\cdot\q}{|\k-\q|q}\right)^2 - 1\right]
- 
\frac{2}{21} f b_{\delta\eta}
\left[
   \frac{3(k_\parallel-q_\parallel)^2}{|\k-\q|^2}
   +
   \frac{5q_\parallel^2}{q^2}
\right]
\\
&\quad\:
+\frac47 f^2 b_{\eta^2}
\frac{q_\parallel^2}{q^2}
\frac{(k_\parallel-q_\parallel)^2}{|\k-\q|^2}
+
\frac{20}{21} b_{(KK)_\parallel}
\left[
    \frac{(\k\cdot\q-q^2)(k_\parallel-q_\parallel)q_\parallel}{|\k-\q|^2q^2} 
    - \frac13\frac{(k_\parallel-q_\parallel)^2}{|\k-\q|^2} - \frac13 \frac{q_\parallel^2}{q^2} + \frac19
\right]
\\
&\quad\:+\frac{10}{21}
b_{\Pi_\parallel^{[2]}}
\frac{(\k\cdot\q-q^2)}{|\k-\q|^2}
\frac{(k_\parallel-q_\parallel)^2}{q^2}
+
\frac{10}{21}
\left[
b_{\delta\Pi_\parallel^{[2]}} 
-\frac13 b_{(K\Pi^{[2]})_\parallel}
- 
f b_{\eta\Pi_\parallel^{[2]}} 
\frac{q_\parallel^2}{q^2}
\right]
\frac{(k_\parallel-q_\parallel)^2}{|\k-\q|^2}
\\
&\quad\:+
\frac{10}{21}
b_{(K\Pi^{[2]})_\parallel}
\frac{(\q\cdot\k-q^2)}{q|\k-\q|}
\frac{q_\parallel(k_\parallel-q_\parallel)}{q|\k-\q|}
+
\frac{10}{21}
f b_{\Pi^{[2]}_\parallel}
\frac{q_\parallel(k_\parallel-q_\parallel)^3}{q^2|\k-\q|^2}
\\
&\quad\:
+
(b_{\Pi_\parallel^{[3]}}+2b_{\Pi_\parallel^{[2]}})
\Bigg[
\frac{13}{21}
\frac{\k\cdot\q-q^2}{|\k-\q|^2}
\frac{q_\parallel(k_\parallel-q_\parallel)}{q^2} 
-
\frac{5\mu^2}{9}
\left[\left(\frac{(\k-\q)\cdot\q}{|\k-\q|q}\right)^2 - \frac{1}{3}\right]
\Bigg]
\\
&\quad\:+
\frac{2}{21}f b_1 
\left[
5
\frac{q_\parallel(k_\parallel-q_\parallel)}{q^2}
+
3
\frac{q_\parallel(k_\parallel-q_\parallel)}{|\k-\q|^2}
\right]
-
\frac27 
f^2 
b_\eta
\frac{q_\parallel(k_\parallel-q_\parallel)}{q^2|\k-\q|^2}
\left[
(k_\parallel-q_\parallel)^2
+
q_\parallel^2
\right]
\Bigg\}\,.
\end{split}
\ee
The expressions for the standard fluid kernels $F_3$ and $G_3$ can be found e.g. in~\cite{Bernardeau:2001qr}.

The EFT for Ly$\alpha$ features 
a number of free bias parameters. All 
together, there are  two linear bias parameters, $b_1$ and $b_\eta$,
and 11 non-linear bias coefficients,
\be
\begin{split}
\{b_2\equiv b_{\delta^2},b_{\mathcal{G}_2},b_{(KK)_\parallel},b_{\delta\eta},b_{\eta^2},b_{\Pi^{[2]}_\parallel}\}\,,\quad 
\{b_{\Gamma_3},b_{\delta\Pi^{[2]}_\parallel},b_{\eta\Pi^{[2]}_\parallel},b_{(K\Pi^{[2]})_\parallel},b_{\Pi^{[3]}_\parallel}\}\,,
\end{split}
\ee
where the first and second lines contain coefficients of operators 
quadratic and cubic in linear density field, respectively.
In what follows we will refer to the above bias parameters as $b_{\mathcal{O}}$, $\mathcal{O}=\delta^2,\mathcal{G}_2$ etc.
Assuming that the linear density field 
is Gaussian-distributed with the 
power spectrum $P_{\rm lin}(k)$, 
we obtain the 3D one-loop power spectrum
\be
\label{eq:deltaP1L}
\begin{split}
&P^{\rm 1-loop}_{\rm 3D}(\k=\{k,\mu\})=K_1^2(\k)P_{\rm lin}(k) + 2\int_\q K_2^2(\q,\k-\q)
P_{\text{lin}}(|\k-\q|)P_{\text{lin}}(q)  + 6 K_1(\k)P_{\text{lin}}(k)\int_\q K_3(\k,-\q,\q)P_{\text{lin}}(q)\,.
\end{split}
\ee
We have implemented the calculation of the 
one-loop model above as a module in the 
CLASS-PT code~\cite{Chudaykin:2020aoj}.
The one-loop integrals are decomposed 
into powers of $\mu$ 
and computed using the FFTLog technique
presented in~\cite{Ivanov:2023yla,Simonovic:2017mhp}.
All together, 
135 one loop integrals are 
computed in $\approx 1$ second on one CPU core for a single point 
in cosmological parameter space. This speed is sufficient
for Markov chain Monte Carlo (MCMC) cosmological 
parameter sampling. Note that sampling nuisance parameters
is much faster as it  
does not require a re-calculation of 
the loop corrections.

In principle, higher derivative counterterms should be added to Eq.~\eqref{eq:deltaP1L}
to renormalize the loop integrals. 
These counterterms also capture 
the effects of Jeans smoothing.
However, the UV-sensitivity is mild and the Jeans effects
are weak at the one-loop order so that the counterterm contributions
can be ignored given the current range of scales and 
the precision level of the 
SDSS data~\cite{Ivanov:2023yla}.
In particular, the
the gas smoothing and thermal broadening
scales 
are of the
order of $10~h$Mpc$^{-1}$\cite{Villasenor:2022aiy}
which are significantly 
smaller than the scales utilized in our 
analysis. 
These numerical estimates
are also consistent with the 
direct fits to the Sherwood data 
done in~\cite{Ivanov:2023yla}.
As an explicit check, we include the 
$k^2$-counterterms in the 
additional analysis based
on the re-normalized mean 
flux of Sherwood, and find 
their impact to be quite limited.

The 1D power spectrum of Ly$\alpha$
fluctuations measured along the line-of-sight is given by~\cite{1989MNRAS.238..293L}
\be
\label{eq:p1d}
P_{\rm {1D}}(k_\parallel)= \frac{1}{2\pi}\int_{k_\parallel}^\infty dk~k P_{\rm 3D}(k,k_\parallel)\,.
\ee 
Importantly, the 1D power spectrum 
involves integration over hard momenta that 
are not in the perturbative regime. 
In EFT, one should do the above integral 
up to a certain cutoff $k_{\rm max}$ and then add suitable counterterms 
that remove the cutoff dependence. As was shown in~\cite{Garny:2020rom,Ivanov:2023yla},
the leading order counterterms take the following form: 
\be
P^{\rm ctr}_{\rm 1D}(k_\parallel) = C_0 + C_2 k^2_\parallel\,. 
\ee
This increases the number of free parameters up to 13.
For each point in cosmological parameter space, we 
compute the integral Eq.~\eqref{eq:p1d} numerically, 
assuming $k_{\rm max}=3~\hMpc$ as in~\cite{Ivanov:2023yla}.

\textit{Treatment of Systematics.}
We model 
the SiIII oscillations as a multiplicative
factor~\cite{SDSS:2004kjl,BOSS:2013rpr}
\be 
\kappa_{\rm SiIII}=1+2\left(\frac{f_{\rm SiIII}}{1-\bar F_{\rm fid}}\right)\cos(vk_\parallel)
+\left(\frac{f_{\rm SiIII}}{1-\bar F_{\rm fid}}\right)^2\,,
\ee
with $f_{\rm SiIII}=8.7\cdot 10^{-3}$, $v=2\pi/(0.0028)$ km/s, $\bar F_{\rm fid}=e^{-0.0025(1+z)^{3.7}}$.
Thermal broadening is modeled with the damping kernel $e^{-(k_\parallel/k_s)^2}$
with $k_s=0.11$(km/s)$^{-1}$. 
We note that our main results are insensitive to these particular choices, i.e. we find similar 
constraints when the  
parameters describing the above systematic 
effects are varied.

\begin{figure}[htb!]
\centering
\includegraphics[width=0.4\textwidth]{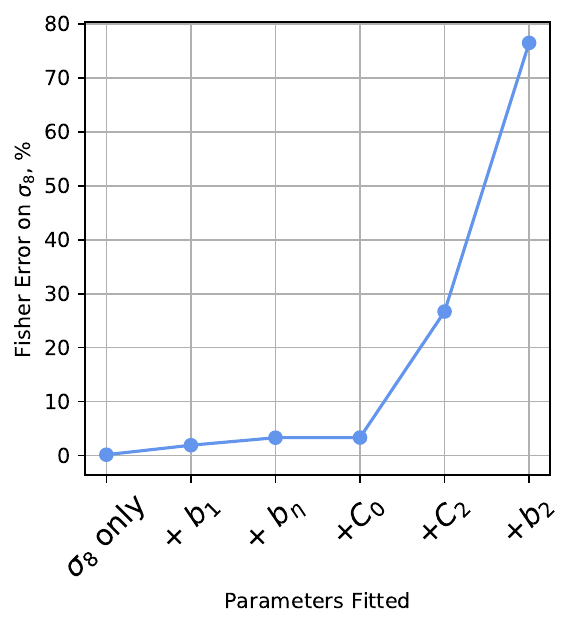}
\includegraphics[width=0.51\textwidth]{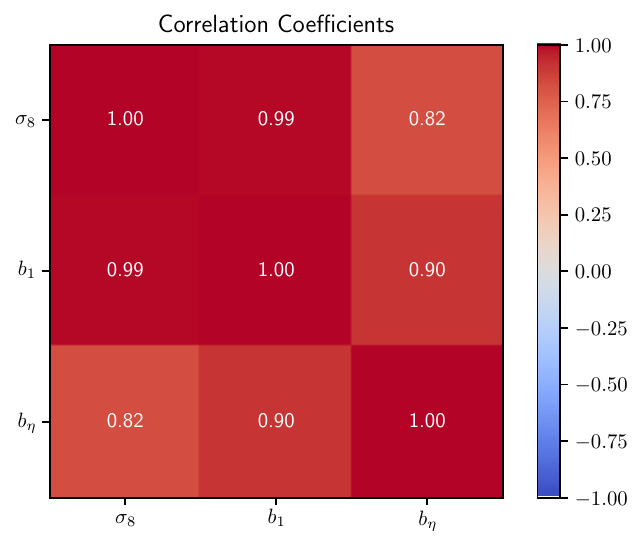}
   \caption{\textit{Left panel:} 
   Marginalized Fisher errors on the mass fluctuation amplitude 
   $\sigma_8$ as a function of a number of free EFT 
   parameters in the fit. Results are reported for a single 
   redshift bin $3.1<z<3.3$. \textit{Right panel:} Fisher 
   correlation coefficients for a fit with only three
   parameters: $\sigma_8,b_1,b_\eta$.
    } \label{fig:paramadded}
\end{figure}

\textit{Fisher matrix analysis.}
We now present details of the Fisher matrix analysis 
that we use to determine the baseline parameters of our likelihood 
analysis. We focus on a single redshift of the data, centered at $z=3.2$.
Our analysis shows that the error on $\sigma_8$ grows 
fast with a number of free parameters in the fit, 
see Fig.~\ref{fig:paramadded}.
In particular, the data loose sensitivity to 
$\sigma_8$ almost entirely if non-linear 
bias parameters are varied. 
We motivates a strategy to keep the linear bias parameters
free, and fix the non-linear ones to values
informed by simulations. Note that even in this case
we find a very significant degeneracy between
$\sigma_8,b_1$ and $b_\eta$, see the right panel of Fig.~\ref{fig:paramadded}. 
This leads to a very significant degradation of 
constraining power due to marginalization:
from $0.1\%$ (un-marginalized limit)
to $3\%$.
Obviously, this degeneracy
would be exact in linear theory.
In our case it gets eventually
broken by non-linear
one-loop
corrections.

As far as the cosmological parameters are concerned, 
our Fisher matrix analysis suggests that with the above baseline settings 
we can obtain constraints competitive with other probes only 
for the mass fluctuation amplitude $\sigma_8$, at the level of few percent.
Specifically, focusing on one redshift, we find that adding $n_s$ to the fit 
yield constraints
 $\{\frac{\sigma_{\sigma_8}}{\sigma_8},\sigma_{n_s}\}=\{0.25,0.06\}$;
while adding to the fit $\omega_{\rm cdm}$ produces uninformative constraints
$\{\frac{\sigma_{\sigma_8}}{\sigma_8},\sigma_{n_s},\frac{\sigma_{\omega_{cdm}}}{\omega_{cdm}}\}=\{0.67,0.28,0.52\}$.
In addition, the cosmological parameter contours are extremely degenerate,  e. g.
for the analysis with free $n_s$ and $\sigma_8$ we find a correlation coefficient $-0.99$,
which indicates that measuring these parameters individually with the 
current EFT settings is challenging. 
We have confirmed this statement 
in a full MCMC analysis as well. 
Better priors on $b_1$ and 
$b_\eta$ from hydrodynamical simulations will be needed in
order to measure $n_s$ and $\sigma_8$ individually from the 1D FPS.

\begin{figure}[htb!]
\centering
\includegraphics[width=0.49\textwidth]{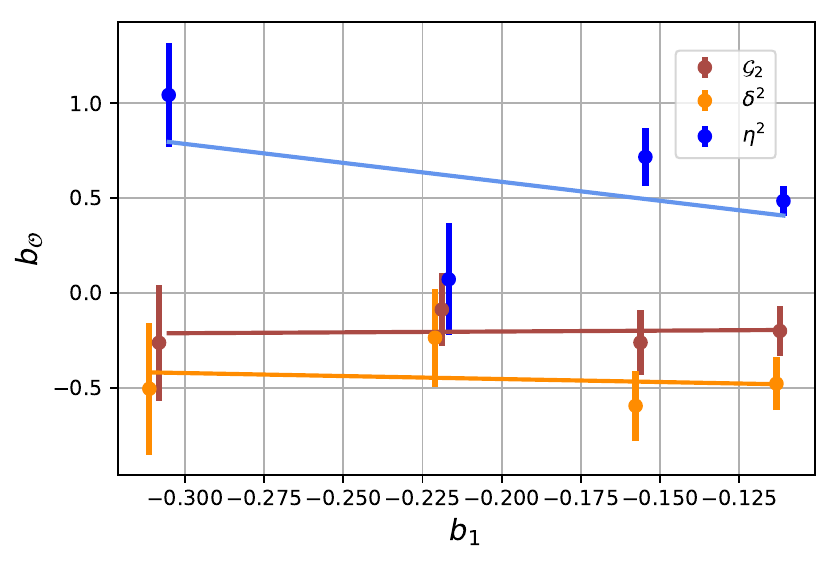}
\includegraphics[width=0.49\textwidth]{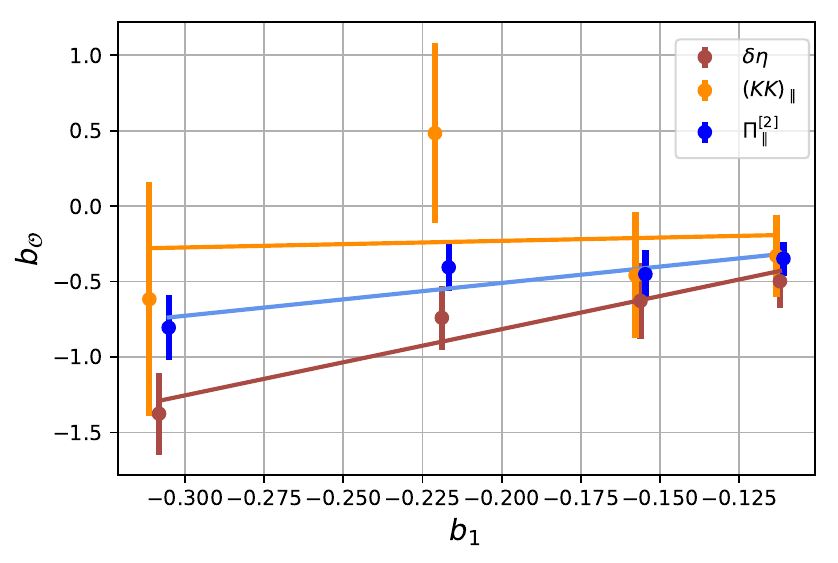}
\includegraphics[width=0.49\textwidth]{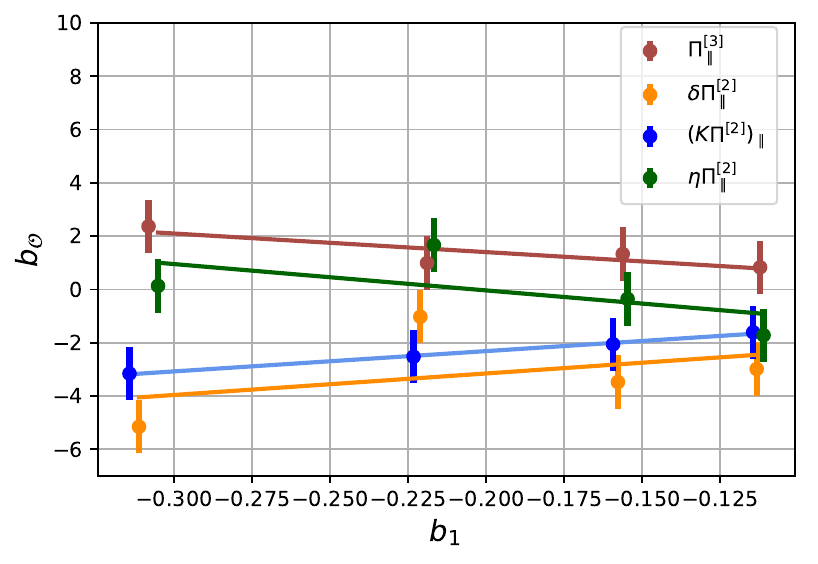}
\includegraphics[width=0.49\textwidth]{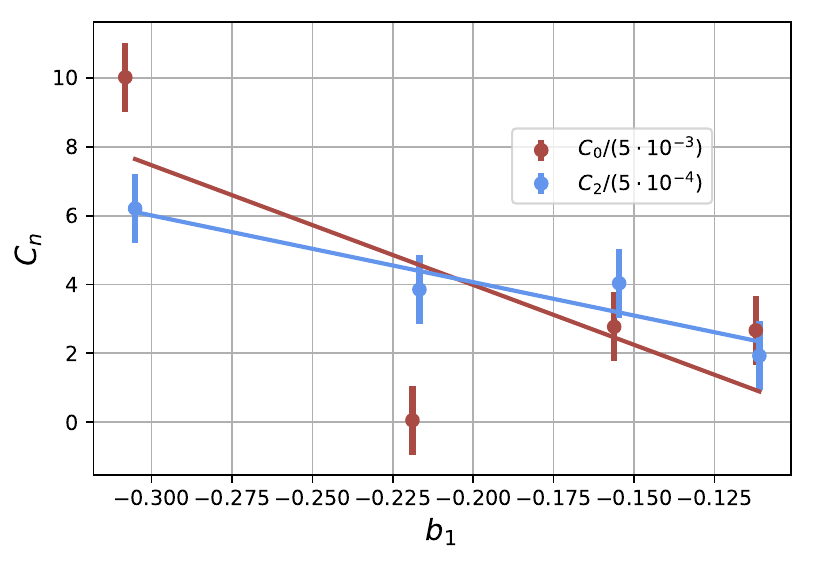}
   \caption{Bias parameters and counterterterms of the Sherwood simulation data.
   Solid lines depict the linear fits to the data.
    } \label{fig:eftparam}
\end{figure}

\textit{EFT bias parameters from the Sherwood simulations.} 
We fit the public 3D power spectra of Ly$\alpha$
fluctuations from the Sherwood simulations using the same 
methodology as in~\cite{Ivanov:2023yla}. The public data cover
four redshift bins with centers at $z=[2,2.4,2.8,3.2]$.
The $z=2.8$ and $z=3.2$ data were fitted before in~\cite{Ivanov:2023yla}.
For the remaining two redshifts, we use the parameter drift plots to 
estimate the data cut $\kmax$ for which the EFT model is applicable.
We also use the same priors on EFT parameters as in~\cite{Ivanov:2023yla},
noting that the resulting constraints 
are much narrower than the priors, expect for $b_{\Gamma_3}$ whose degeneracy
with $b_{\mathcal{G}_2}$
cannot be broken at the power
spectrum level. 
Since this parameter
only affects our results in combination
$2 b_{\Gamma_3}+5b_{\mathcal{G}_2}$,
we set $b_{\Gamma_3}=0$
without loss of generality,
following~\cite{Ivanov:2023yla}.

Our measurements of different nonlinear bias parameters $b_{\mathcal{O}}$
versus corresponding values of $b_1$
are presented in Fig.~\ref{fig:eftparam}, along with the linear fits
to the data points. Error bars on the quadratic parameters follow from the 
Gaussian covariance for the power spectrum. 
The errors on the EFT parameters
depend very strongly on $\kmax$ used in the fits,
and the possible choices of $\kmax$ are limited
by the binning of the public Sherwood power spectra. 
Because of these reasons, for 
some cubic bias parameters the
nominal errorbars appear to be quite small,
which may bias the $b_{\mathcal{O}}(b_1)$
relations extracted from the simulation. 
In order to be conservative and avoid overfitting, 
we assume a flat error $1$ on all 
cubic bias parameters. 
Finally, we also assume flat errors 1 on $C_0/(5\cdot 10^{-3}[\Mpch])$
and $C_2/(5\cdot 10^{-4}[\Mpch]^3)$,
because the covariance for the 1D FPS 
of Sherwood is not available, 
so there is no reliable way to estimate the errors in this case.
We found that the dependencies $b_{\mathcal{O}}(b_1)$
and $C_n(b_1)$ can be well fitted with linear functions, 
\be 
b_{\mathcal{O}}=A_{\mathcal{O}} b_1 + B_\mathcal{O}.
\ee 
The best-fit values of $A_{\mathcal{O}}$ and $B_\mathcal{O}$
are given in Table~\ref{tab:fit}. For $C_n$ we have 
\be 
\frac{C_0}{[\Mpch]} = -0.17 b_1 -0.0148\,,\quad \frac{C_2}{[\Mpch]^3} = -9.7\cdot 10^{-4} b_1 +9.4\cdot 10^{-5}\,.
\ee 

\begin{table}[h]
\centering
\begin{tabular}{|c|c|c|c|c|c|c|c|c|c|c|}
\hline
Operator $\mathcal{O}$ & $\mathcal{G}_2$ & $\delta^2$ & $\eta^2$ & $\delta\eta$ & $(KK)_\parallel$ & $\Pi^{[2]}_\parallel$ & $\Pi^{[3]}_\parallel$ & $\delta \Pi^{[2]}_\parallel$ & $(K\Pi^{[2]})_\parallel$ & $\eta \Pi^{[2]}_\parallel$ \\ \hline
$A_{\mathcal{O}}$ & 0.089 & -0.32 & -2.0 & 4.4 & 0.45 & 2.2 & -7.0 & 8.2 & 7.8 & -9.9 \\ \hline
$B_{\mathcal{O}}$ & -0.18 & -0.52 & 0.19 & 0.063 & -0.14 & -0.074 & 0.0043 & -1.5 & -0.79 & -2.0 \\ \hline
\end{tabular}
\caption{Numerical values for the fit to the Sherwood EFT parameters as functions of the linear bias $b_1$.}
\label{tab:fit}
\end{table}

\textit{Synthetic 1D FPS data from \texttt{LaCE}.} 
We use the {\texttt Cabayol23} neural network emulator available from {\texttt LaCE}\footnote{\url{https://github.com/igmhub/LaCE/tree/main}} to generate 1-dimensional Ly$\alpha$ power spectrum where we vary only $\sigma_T$ and $\gamma$ and scanning over a range from \{0.0,0.5\} and \{0.5,2.0\} respectively. The typical fits to FPS at a single redshift has a 
reduced $\chi^2$ ranging from $\sim$ 1.2 to 1.4 which indicates a reasonable match, but not a good fit. We find that the last point largely arises due to discrepancies at smaller scales as can be appreciated from the r.h.s. of Fig.~\ref{fig:money}. Note that we have not optimized the full set of the \texttt{LaCE}
parameters to precisely reproduce the data. This is done in order to avoid
double counting of information when we use 
the \texttt{LaCE} mocks for re-calibration of 1D EFT counterterms.
The best-fit $C_n$ parameters extracted from our mocks are:
\be 
\frac{C_0}{[\Mpch]} = -0.17 b_1 -0.2053\,,\quad \frac{C_2}{[\Mpch]^3} = -9.7\cdot 10^{-4} b_1 + 2.16\cdot 10^{-2}\,.
\ee 
We see that the typical momentum scales associated with these parameters are around $4~\hMpc$,
which is consistent with the EFT expectation
that they should be of the order of $k_{\rm NL}\sim (5-10)~\hMpc$.

\begin{figure}[htb!]
\centering
\includegraphics[width=0.99\textwidth]{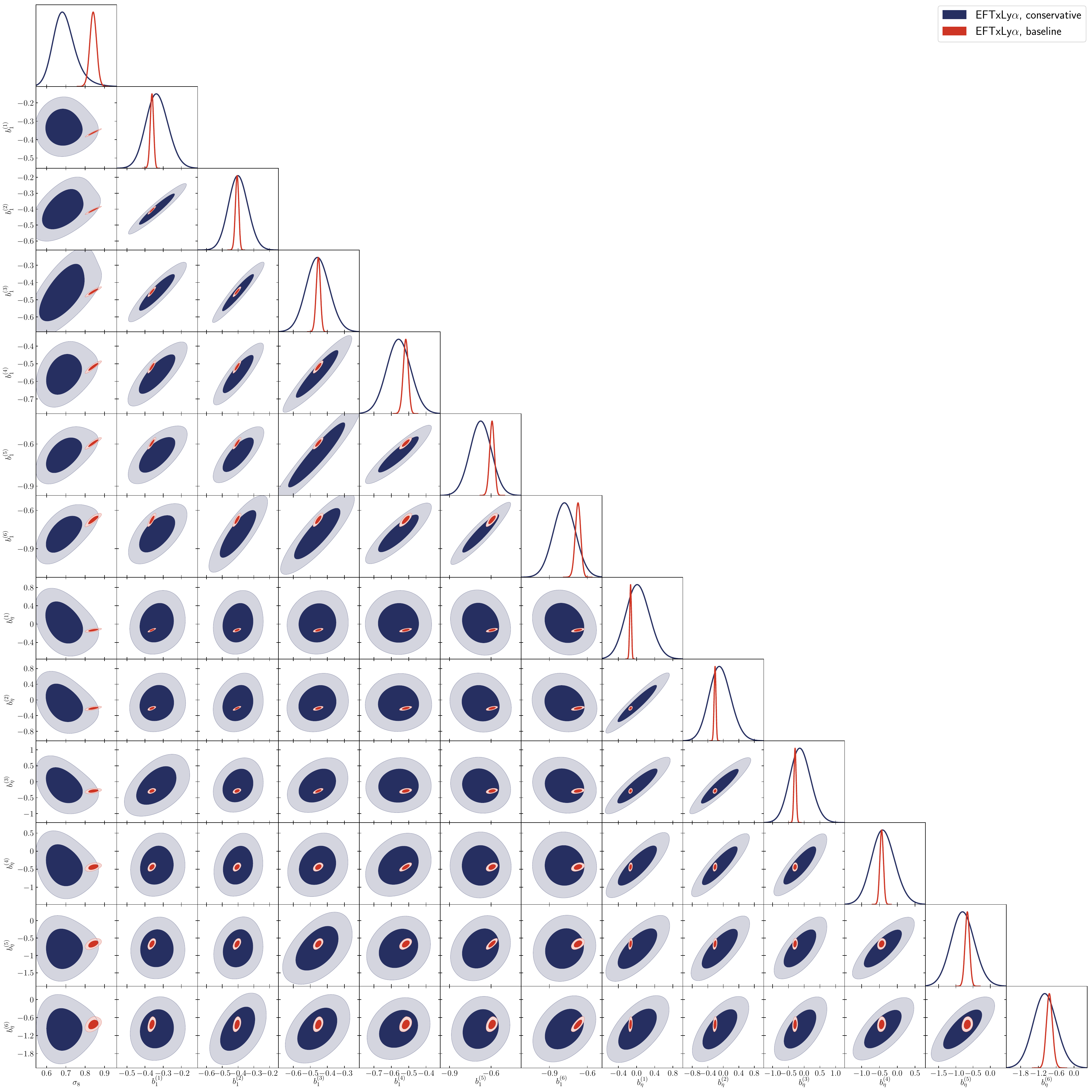}
   \caption{Constraints on $\sigma_8$
   and the linear bias parameters
   from the baseline and conservative analyses. Superscripts $(1),...,(6)$
   correspond to redshift bins 
   with mean redshifts $z=3.2,3.4,3.6,3.8,4,4.2$.
    } \label{fig:param_constraints}
\end{figure}

\textit{Full Parameter Constraints and Conservative Analysis.}
The corner plot 
and 1D marginalized limits 
from our baseline analysis 
are displayed 
fig.~\ref{fig:param_constraints}
and in table~\ref{eq:tab1}. 

It is important to see how much our results change 
if we allow some scatter in the 
EFT priors, i.e. if we marginalize 
EFT parameters around the priors calibrated from Sherwood. 
To that end, 
we have re-run our  
analysis 
with the following
very conservative scatter 
around our EFT priors. 
We allow $\mathcal{O}(1)$
variations of the mean 
of priors of all quadratic EFT 
parameters, 
$\mathcal{O}(3)$
variations of cubic EFT 
parameters, 
plus the 
$\mathcal{O}(5)$
scatter in $C_0$
and the $\mathcal{O}(0.5)$
scatter in $C_2$, in natural 
units of $5\cdot 10^{-3}~[h^{-1}\text{Mpc}]$
and 
$5\cdot 10^{-4}~[h^{-1}\text{Mpc}]^3$ (corresponding to $k_{\rm NL}^{-1}$
and $k_{\rm NL}^{-3}$),
respectively. These results 
are shown in fig.~\ref{fig:param_constraints}
and in table~\ref{eq:tab1}.

In the conservative case we find 
a rather loose 
constraint on $\sigma_8=0.69\pm 0.06$,
with the best-fit value $\sigma_8=0.75$.
This result is also consistent
with Planck, albeit we note that 
this measurement (a) 
is not competitive with 
other cosmological probes
and (b) is affected by prior 
effects as evidenced by
the non-Gaussian shape of the $\sigma_8$
density, which somewhat
obscures its interpretation.
This exercise 
however 
reinforces 
the main message of our work: 
informative priors 
on EFT parameters 
from simulations are necessary 
for competitive 
cosmological constraints. 

We would like to stress
that the main reason for the scatter
observed in fig.~\ref{fig:eftparam}
is the precision of the Sherwood 
power spectrum data.
If we were to use more precise
data, or the field-level technique
for EFT parameter measurements, 
we would expect to see much less 
of a scatter. 
Optimistically, based on  
the examples of 
dark matter halos or galaxies
as in~\cite{Ivanov:2024hgq},
we assume that the fundamental 
bias
relations dictated by the underlying physics are tight,
and our Sherwood measurements
simply represent noisy
realizations of these relations.
(The Lyman alpha 
clouds are not located in 
dark matter halos, and therefore, 
the validity of this
assumption cannot be justified
by the phenomenology of bias
parameters of dark matter halos.)
Our measurements are fully consistent with this
assumption.

\begin{table*}
\begin{tabular}{|l|c|c|c|c|}
 \hline
 \multicolumn{5}{|c|}{Baseline} \\
 \hline
Param & best-fit & mean$\pm\sigma$ & 95\% lower & 95\% upper \\ \hline
$b_1^{(1)}$ &$-0.3602$ & $-0.3625_{-0.0091}^{+0.0099}$ & $-0.3813$ & $-0.3439$ \\
$b_1^{(2)}$ &$-0.4046$ & $-0.4066_{-0.01}^{+0.011}$ & $-0.4282$ & $-0.3852$ \\
$b_1^{(3)}$ &$-0.452$ & $-0.4532_{-0.012}^{+0.013}$ & $-0.4777$ & $-0.429$ \\
$b_1^{(4)}$ &$-0.5141$ & $-0.5173_{-0.014}^{+0.015}$ & $-0.5467$ & $-0.4878$ \\
$b_1^{(5)}$ &$-0.585$ & $-0.5937_{-0.017}^{+0.018}$ & $-0.6287$ & $-0.559$ \\
$b_1^{(6)}$ &$-0.6645$ & $-0.674_{-0.022}^{+0.023}$ & $-0.7175$ & $-0.6305$ \\
$b_\eta^{(1)}$ &$-0.1319$ & $-0.1326_{-0.019}^{+0.02}$ & $-0.1706$ & $-0.09527$ \\
$b_\eta^{(2)}$ &$-0.2168$ & $-0.2131_{-0.024}^{+0.025}$ & $-0.2611$ & $-0.1652$ \\
$b_\eta^{(3)}$ &$-0.3$ & $-0.2889_{-0.036}^{+0.036}$ & $-0.3598$ & $-0.2179$ \\
$b_\eta^{(4)}$ &$-0.4367$ & $-0.4395_{-0.051}^{+0.052}$ & $-0.5398$ & $-0.3382$ \\
$b_\eta^{(5)}$ &$-0.6294$ & $-0.668_{-0.068}^{+0.067}$ & $-0.8012$ & $-0.5347$ \\
$b_\eta^{(6)}$ &$-0.798$ & $-0.8267_{-0.11}^{+0.1}$ & $-1.035$ & $-0.6149$ \\
$\sigma_8$ &$0.841$ & $0.841^{+0.017}_{-0.017}$ & $0.807$ & $0.875$ \\
\hline
 \end{tabular} 
 \begin{tabular}{|l|c|c|c|c|}
 \hline
  \multicolumn{5}{|c|}{Conservative} \\
 \hline
Param & best-fit & mean$\pm\sigma$ & 95\% lower & 95\% upper \\ \hline
$b_1^{(1)}$ &$-0.1951$ & $-0.3027_{-0.054}^{+0.063}$ & $-0.4172$ & $-0.1975$ \\
$b_1^{(2)}$ &$-0.2527$ & $-0.3692_{-0.053}^{+0.062}$ & $-0.486$ & $-0.2622$ \\
$b_1^{(3)}$ &$-0.3132$ & $-0.4255_{-0.056}^{+0.064}$ & $-0.5497$ & $-0.3078$ \\
$b_1^{(4)}$ &$-0.398$ & $-0.5355_{-0.058}^{+0.066}$ & $-0.659$ & $-0.4203$ \\
$b_1^{(5)}$ &$-0.4858$ & $-0.6539_{-0.065}^{+0.07}$ & $-0.7893$ & $-0.5252$ \\
$b_1^{(6)}$ &$-0.5807$ & $-0.7653_{-0.076}^{+0.077}$ & $-0.9165$ & $-0.6244$ \\
$b_\eta^{(1)}$ &$-0.3199$ & $-0.01109_{-0.29}^{+0.2}$ & $-0.4754$ & $0.488$ \\
$b_\eta^{(2)}$ &$-0.4202$ & $-0.1133_{-0.31}^{+0.23}$ & $-0.591$ & $0.4459$ \\
$b_\eta^{(3)}$ &$-0.5205$ & $-0.1298_{-0.35}^{+0.28}$ & $-0.7051$ & $0.4748$ \\
$b_\eta^{(4)}$ &$-0.7458$ & $-0.4443_{-0.32}^{+0.27}$ & $-1.008$ & $0.1495$ \\
$b_\eta^{(5)}$ &$-1.018$ & $-0.8504_{-0.36}^{+0.27}$ & $-1.458$ & $-0.212$ \\
$b_\eta^{(6)}$ &$-1.181$ & $-1.047_{-0.41}^{+0.32}$ & $-1.754$ & $-0.3302$ \\
$\sigma_8$ &$0.747$ & $0.692_{-0.061}^{+0.032}$ & $0.600$ & $0.805$ \\
\hline
 \end{tabular} 
 \caption{Best-fits and 1d marginalized limits for $\sigma_8$ and the linear EFT parameters 
 from the eBOSS data. Superscripts $(1),...,(6)$
   correspond to redshift bins 
   with mean redshifts $z=3.2,3.4,3.6,3.8,4,4.2$.}
 \label{eq:tab1}
 \end{table*}

Let us comment now on the values 
of the EFT parameters. 
We observe a very good agreement
at the level of $b_1$, within
redshifts 
overlapping with SDSS.
The agreement at the level 
of $b_1$ implies an agreement for 
the non-linear bias parameters 
as they are fixed to the $b_{\mathcal{O}}-b_1$
relations dictated by Sherwood.

For $b_\eta$,
our baseline results are quite different
from Sherwood values, 
c.f. fig.~\ref{fig:param_constraints}
and in table~\ref{eq:tab1}.
This is 
expected given that this parameter
can absorb both the errors due to statistical scatter of 
EFT priors
and the mismatch between 
the eBOSS and Sherwood data
discussed in the main text.
As we discuss below, part of this mismatch comes
from using different values 
of the mean flux. Once 
the mean flux is normalized properly, 
we find much better agreement
at the level of $b_\eta$,
confirming the above argument. 
In addition, if we allow 
scatter in the prior
relations dictated by Sherwood,
as in our conservative analysis, 
we also find limits 
on $b_\eta$ consistent with the 
Sherwood values within $95\%$ CL
for the overlapping redshifts. 
All these test 
imply that our $b_\eta$
values are consistent with 
Sherwood when compared in
consistent conditions. 

Let us also point out that 
our conservative analysis is 
fully consistent with the 
emulator-based analysis 
of~\cite{Fernandez:2023grg}
that uses a similar high-redshift data as we do. Our analysis thus 
confirms 
the success of the
simulations paradigm.

\begin{figure}[htb!]
\centering
\includegraphics[width=0.49\textwidth]{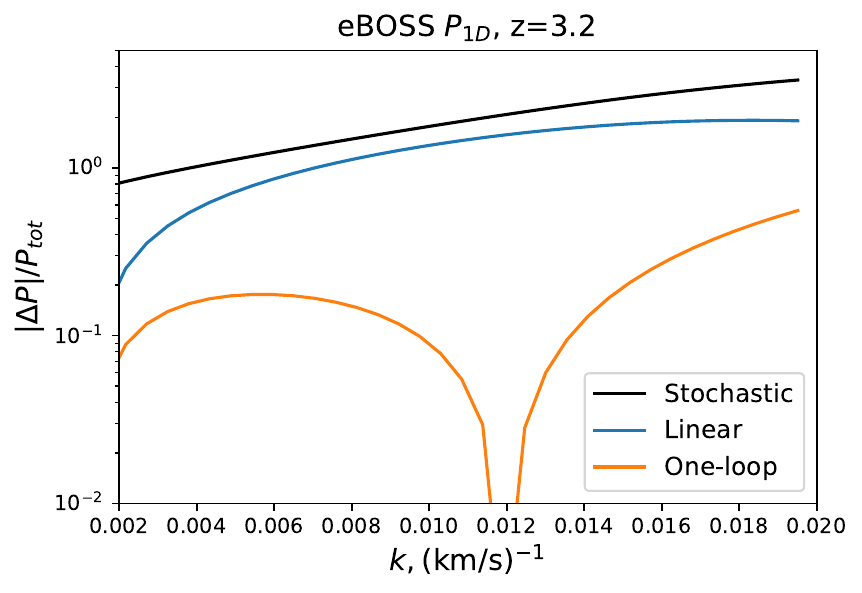}
\includegraphics[width=0.45\linewidth]{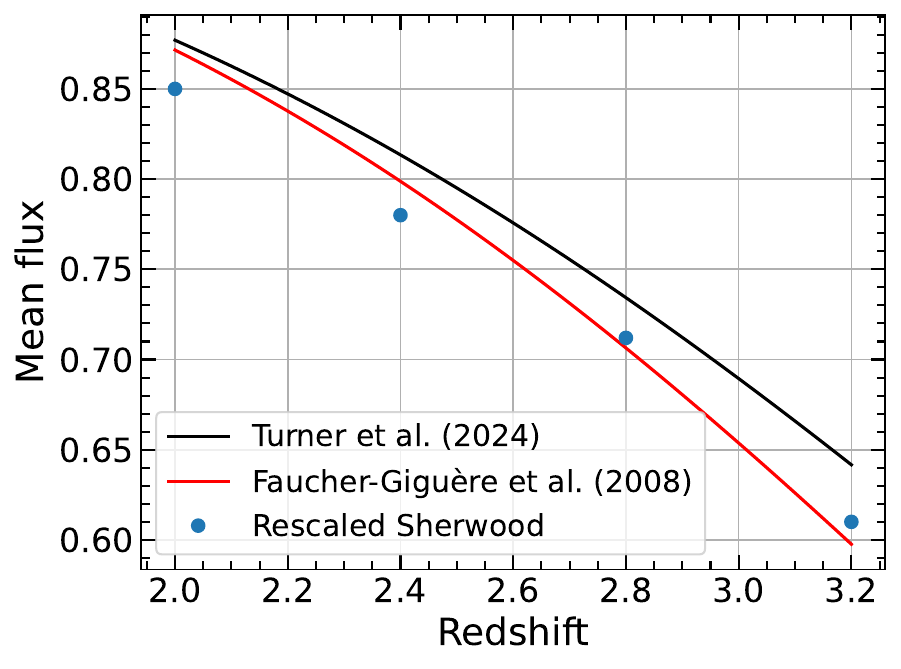}
   \caption{Left panel: Breakdown of different 
   contributions to the 
   1D flux power spectrum
   of the eBOSS data at $z=3.2$. 
   The theory curves are from the corresponding best fitting model. To ease comparison, 
   we omit the 
   Silicon III form factor. \\
    Right panel: 
  The  mean flux after rescaling the optical depths from the Sherwood simulations. We prioritized matching $P_\mathrm{1D}$ over the mean flux, yet these mean flux values are a good match to the literature.
    } \label{fig:breakdown}
\end{figure}

\textit{Breakdown of different contributions 
to the 1D power spectrum.}
It is interesting to study how different 
contributions combine together to form 
the final 1D power spectrum. 
In particular, our results are in a complete agreement
with the scrambling exercise of Ref.~\cite{Irsic:2018hhg},
implying that the 1D signal 
is dominated by the stochastic ``1-absorber'' term, which 
reduces the correlation with the cosmological signal. 
A part of the 
stochastic contributions in EFT 
are captured by $C_n$'s.
However, in EFT, all coefficients
and parameters depend
on the cutoff in order to make
the final result cutoff
independent. 
In particular, 
for 1D integrals 
we use a renormalization scheme
in which all $C_n$'s are
measured for the effective cutoff 
choice 
$k_{\rm max}=3~h$Mpc$^{-1}$. 
The stochastic contributions in 
these prescription contain
UV-dependent pieces that cancel
the cutoff-dependence (``infinite terms'') plus
the actual physical stochastic terms (``finite terms'').
Therefore, the physical cutoff-independent 
stochastic 
contribution is the sum of the 
cutoff-dependent part of the 1D integrals 
over the 3D power spectrum
plus contributions stemming from 
$C_n$'s. 
If there were no constant contributions
to the 3D power spectrum, 
the cutoff dependent part
of the 1D integrals would produce 
a constant piece, which could be 
added to $C_n$'s terms to get
the full stochastic contribution. 
In our case, however, we also have 
one-loop deterministic 
constant contributions
that originate from
the quadratic terms 
in perturbation theory, 
such as the auto-spectrum of
the $\delta^2$ operator. Nevertheless, 
these terms suppressed
in the loop expansion, 
so their total 
contributions 
to the constant power
spectrum is subdominant. 
In addition, it is not easy to 
separate the $k_\parallel^2$ 
UV-dependent contributions 
from the tree-level and one-loop
1D integrals easily, although
given the above arguments, these 
must be treated as stochastic 
terms too. 
With these caveats in mind, 
we have produced 
a plot with contributions 
of linear, one-loop, 
and stochastic EFT terms to the total 
1D power spectrum at $z=3.2$
following from the best-fit parameters
from table~\ref{eq:tab1}
see the left panel of fig.~\ref{fig:breakdown}.
We see that the stochastic term 
(a proxy to the one-absorber term of ~\cite{Irsic:2018hhg})
indeed dominates the signal 
on pretty much all scales probed by eBOSS data, 
in full agreement with the results of 
Ref.~\cite{Irsic:2018hhg}. We also see that the 
liner theory contribution 
dominates over the one-loop
correction, in agreement 
with the EFT power counting. 
Note that having two separate expansions
for the stochastic and 
deterministic contributions
where stochastically is quite large 
is also consistent with EFT, 
see e.g. an example of EFT for 
high-redshift quasars~\cite{Chudaykin:2022nru}.

Let us also note that the scaling of the one-loop correction
with $k$ in fig.~\ref{fig:breakdown} is somewhat mild. This can be 
understood within the scaling Universe approximation~\cite{Ivanov:2023yla},
suggesting that the one-loop corrections to the position space 
density variance should scale as $(k/k_{\rm NL})^{2(3+n)}\sim (k/k_{\rm NL})$ for 
$n\approx -2.5$ relevant at the Lyman-$\alpha$ scales.
Thus, the observed mild behavior is a result of an approximate scale-invariance
of the matter power spectrum at the scales of interest.

\begin{figure}
    \centering
    \includegraphics[width=0.45\linewidth]{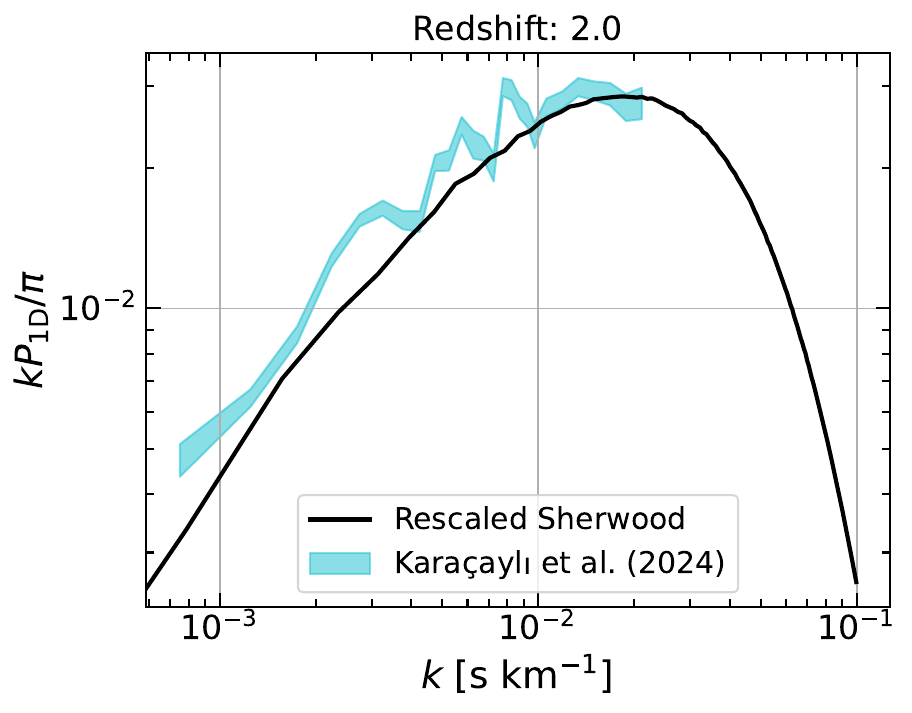}
    \includegraphics[width=0.45\linewidth]{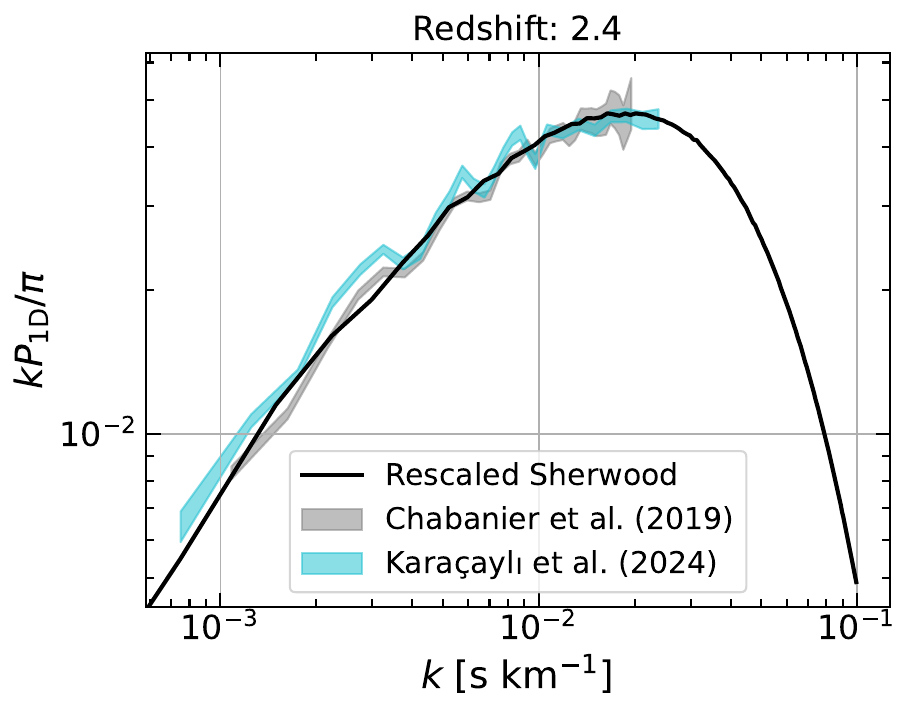} \\
    \includegraphics[width=0.45\linewidth]{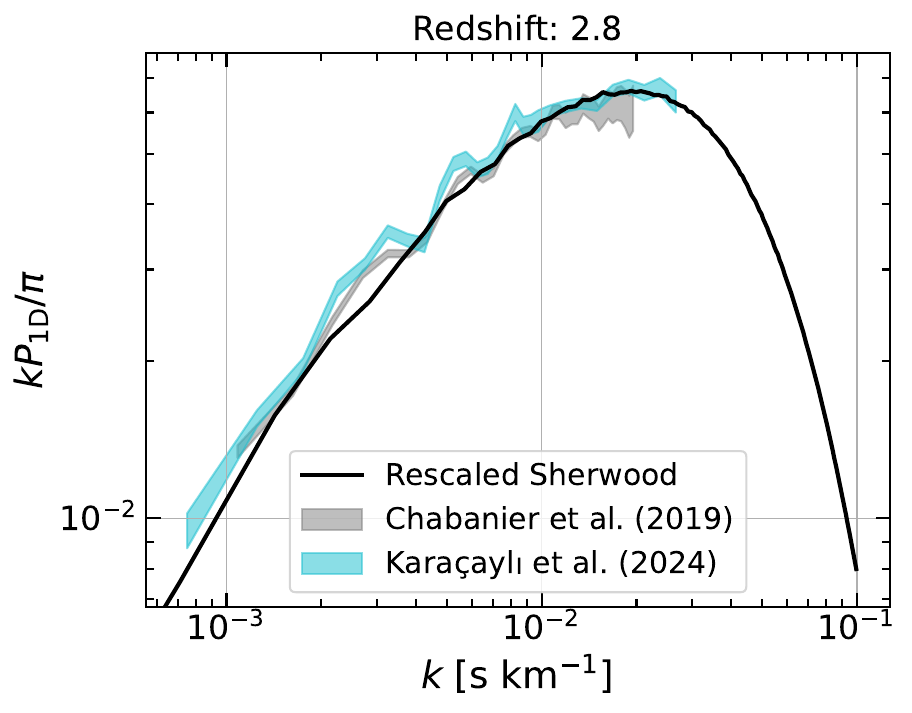}
    \includegraphics[width=0.45\linewidth]{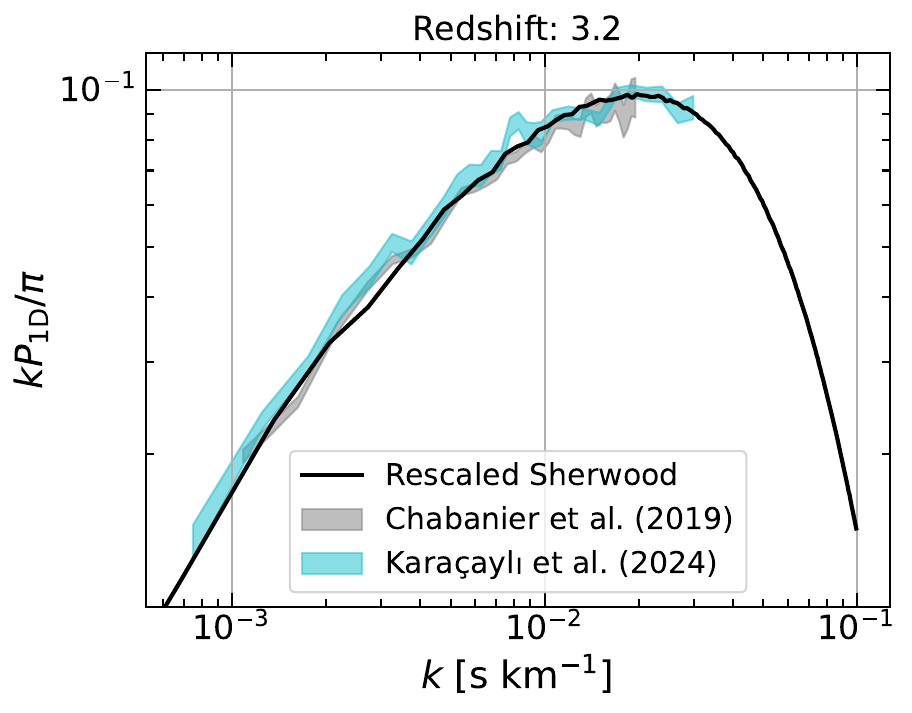}
    \caption{$P_\mathrm{1D}$ after rescaling the optical depth from Sherwood simulations. We obtain a great match between measurement and simulations.}
    \label{fig:rescaled_p1d_sherwood}
\end{figure}

\textit{Analysis based on rescaling the optical depth from Sherwood.} As mentioned in the text, the power spectra and mean flux from \cite{Givans:2022qgb}'s Sherwood simulation extraction do not match the mean flux values in the literature or eBOSS $P_\mathrm{1D}$ measurements. Since the amplitude of UV background fluctuations is unknown a priori, it needs to be tuned in the post-processing by scaling the optical depth to match a target mean flux and $P_\mathrm{1D}$. In what follows, we follow a simple procedure to reach a good match for both eBOSS $P_\mathrm{1D}$ and literature mean flux measurements by trying a few mean flux values by hand. We first calculate the probability distribution function (PDF) $\mathcal{P}(\tau)$ in a given simulation box. The mean flux of rescaled skewers as a function of the scaling factor $\alpha$ is $\bar{F}(\alpha) = \int e^{-\alpha\tau} \mathcal{P}(\tau) \mathrm{d}\tau$. The factor $\alpha$ can be solved using Newton-Raphson iteration for the root of $f(\alpha) = \bar{F}(\alpha) - \bar{F}^* = 0$. To improve the accuracy of the solution for $\alpha$, we perform one last iteration after convergence with the PDF integration by exactly calculating the mean flux with scaled optical depth values. The right panel of Fig.~\ref{fig:breakdown} shows the final mean flux values we settled for a better match in $P_\mathrm{1D}$. These values are close to the literature values by \cite{fauchergiguereMeasurementOpacity2008, turnerLyaForestMeanFluxFromDesiY12024}. Fig.~\ref{fig:rescaled_p1d_sherwood} shows that we obtain a good match in $P_\mathrm{1D}$ between measurement and simulations.
 We note, however, that 
the large-$k_\parallel$ behavior 
of the 1D power spectra at $z=2.8$
and $z=3.2$ is not captured 
perfectly, 
which already suggests 
that the matching the mean 
flux is not sufficient 
to achieve a complete match 
between the Sherwood 
and eBOSS measurements.

\begin{figure}[htb!]
\centering
\includegraphics[width=0.49\textwidth]{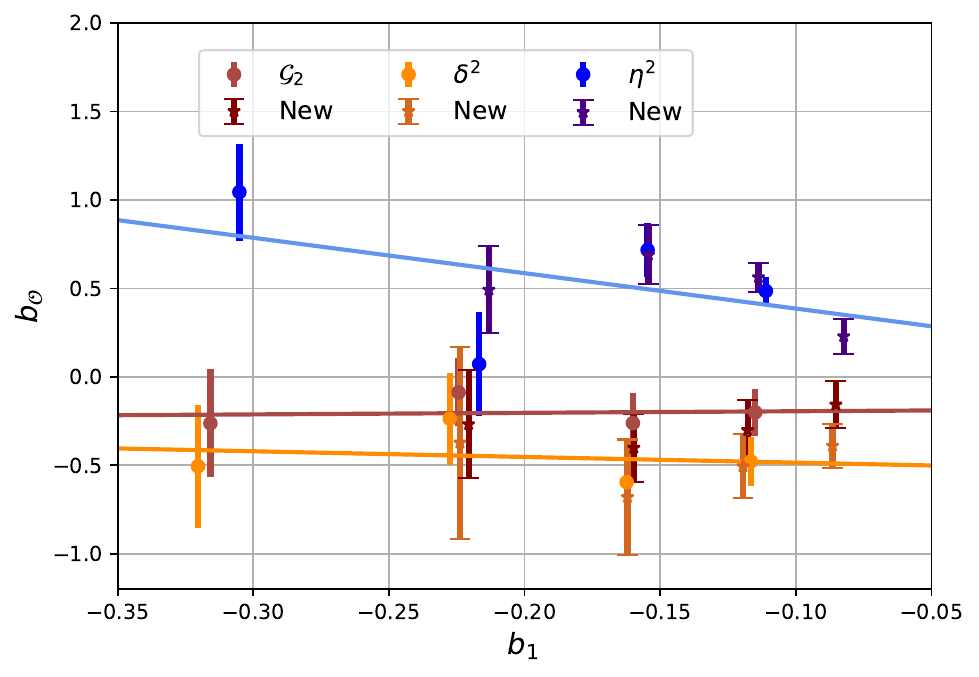}
\includegraphics[width=0.49\textwidth]{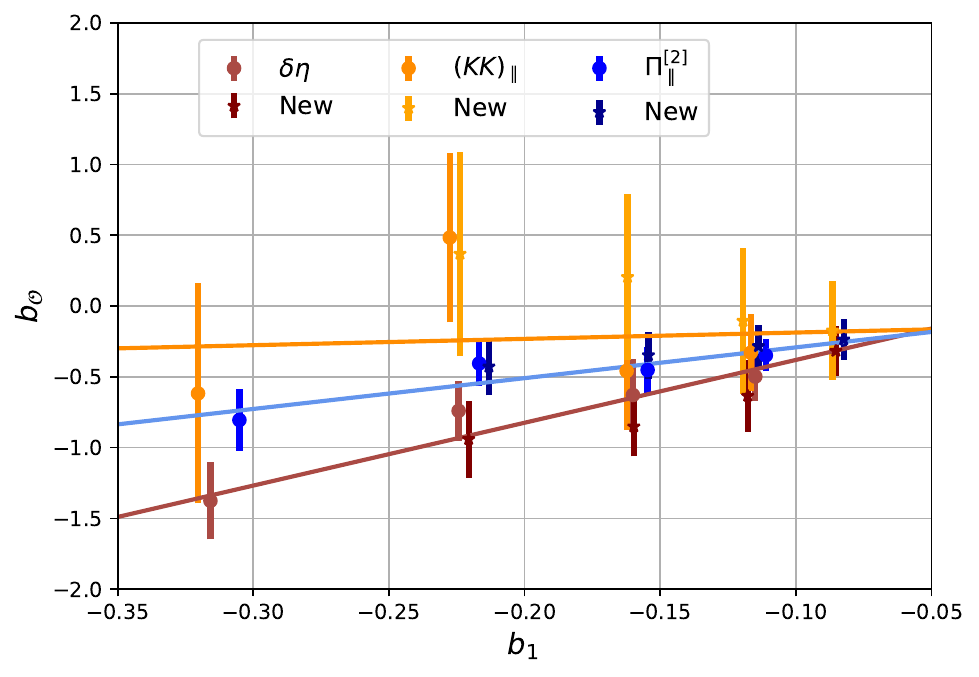}
\includegraphics[width=0.49\textwidth]{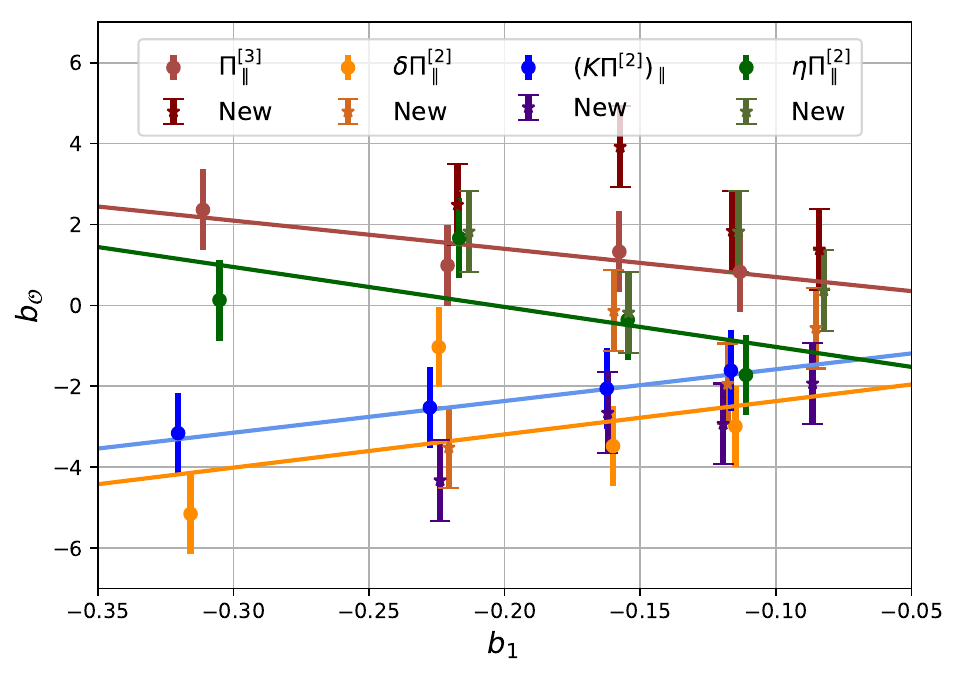}
\includegraphics[width=0.49\textwidth]{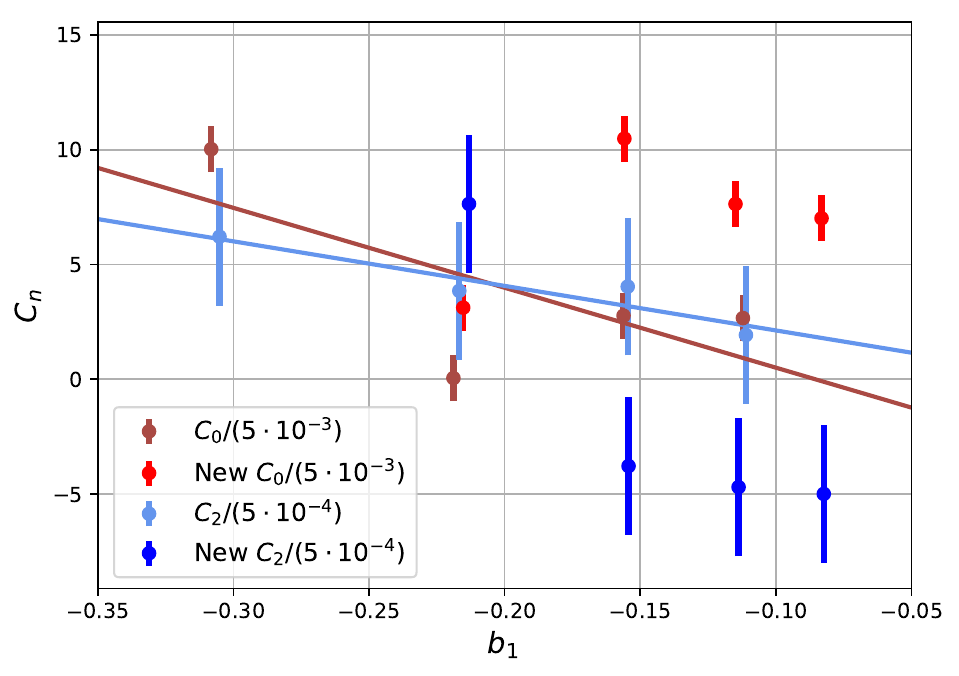}
\includegraphics[width=0.49\textwidth]{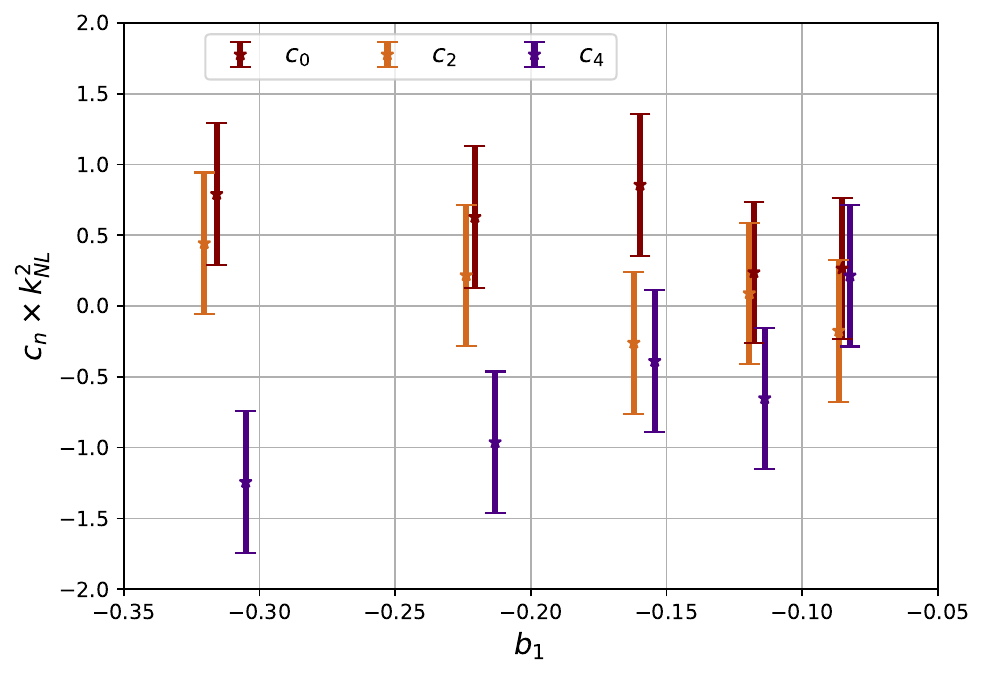}
   \caption{Bias parameters and counterterterms of the Sherwood simulation data.
   Solid lines depict the linear fits to the data. For the $k^2$-counterterms in the bottom panel, we 
   show results from the rescaled mean flux and add additional $z=3.2$
   data points from the original 
   flux normalization 
   in order to increase the range.
    } \label{fig:eftparam2}
\end{figure}

Having matched the flux, we have 
remeasured the EFT parameters
of the 3D 
Lyman alpha power spectrum of the 
Sherwood simulations
for all available redshifts. 
Our results are shown in fig.~\ref{fig:eftparam2}
where we display the new measurements
as well as the old ones extracted from the 
original mean flux. 
In addition, we also display the 
EFT $k^2P_{11}$-type counterterms
defined in ref.~\cite{Ivanov:2023yla}.
While their values 
are in general consistent with zero, 
we have added them in the model in 
the new analysis in order to estimate their 
impact on the final results. 

From fig.~\ref{fig:eftparam2}, we observe
that the quadratic bias parameters
fully agree with the priors $b_{\mathcal{O}}(b_1)$
extracted from the original Sherwood measurements. 
Hence, post-processing does not alter the EFT parameter
priors, which confirms their robustness,
and reinforces our argument
that the priors 
are more reliable than the 
exact values of EFT parameters.
 We see some noticeable scatter for the cubic bias parameters, but their values are still consistent
 with the priors. 
In order to re-do our analysis with the new flux,
we extract the new priors on EFT parameters,
listed in table~\ref{tab:fit2}.

\begin{table}[h]
\centering
\begin{tabular}{|c|c|c|c|c|c|c|c|c|c|c|c|c|c|}
\hline
Operator $\mathcal{O}$ & $\mathcal{G}_2$ & $\delta^2$ & $\eta^2$ & $\delta\eta$ & $(KK)_\parallel$ & $\Pi^{[2]}_\parallel$ & $\Pi^{[3]}_\parallel$ & $\delta \Pi^{[2]}_\parallel$ & $(K\Pi^{[2]})_\parallel$ 
& $\eta \Pi^{[2]}_\parallel$ & $k_{\rm NL}^2 c_0$ & $k_{\rm NL}^2 c_2$ & $k_{\rm NL}^2 c_4$\\ \hline
$A_{\mathcal{O}}$ & 0.154 & 0.061 & -2.84& 4.31& 1.55 & 2.48 & -3.08 & 20.7 & 5.83 & 1.60 & -2.36 & -2.65
& 5.60 \\ \hline
$B_{\mathcal{O}}$ & -0.252 &  -0.480 & 0.11 & -0.0745 & 0.205 & 0.011 & 1.86 & 1.34 & -1.99 & 1.07 & 0.145 & -0.397 & 3.65 \\ \hline
\end{tabular}
\caption{Numerical values for the fit to the Sherwood EFT parameters as functions of the linear bias $b_1$.
$k_{\rm NL}=10~\hMpc$.
}
\label{tab:fit2}
\end{table}

 A significant 
 change, however, takes place for the 
 1D stochastic counter-terms, which require values
 inconsistent 
 with the original Sherwood measurements.
 This is also expected as we have changed
 the overall amplitude of the 1D power spectrum
 which is sensitive to the $C_{0,2}$
 parameters. 
It confirms our main argument
that only these parameters need to be 
re-calibrated in order to match the 
power spectra of the eBOSS data. 

To match the \texttt{LaCE} data, 
we have done such a re-calibration
similar to the one that we have done 
in the baseline analysis. 
The new analysis 
of Sherwood data with the 
rescaled flux suggests that 
the main need for this
re-calibration is the 
uncertainty in the EFT parameter priors. 
The mentioned above mismatch 
in the shape of the
1D power spectrum of eBOSS
is an effect of secondary importance.
The calibration of $C_0$ and $C_2$
priors
from the \texttt{LaCE} emulator 
data gives the new priors: 
\be 
\frac{C_0}{[\Mpch]} = -0.26 b_1 -1.1175\,,\quad \frac{C_2}{[\Mpch]^3} = 1.68\cdot 10^{-3} b_1 + 0.1107\,.
\ee

In addition, in order 
to make our analysis 
more diverse and extended, 
we have added 
priors
on the higher-derivative 
counterterms to it,
extracted from Sherwood
and implemented as priors in
the eBOSS analysis 
in a way identical 
to the other EFT parameters.
We find that they do not have 
a strong impact on the measurement, 
but their presence, in general, may 
increase the flexibility
of the EFT modeling. 
Specifically, we find that the effects of the 
$k^2-$
counterterms are largely absorbed 
into the 1D stochastic 
parameters $C_0$ and $C_2$,
and therefore they are irrelevant for the
results of our baseline analysis.

Using these settings, we have carried out
the analysis of the eBOSS
data using the new priors 
from Sherwood with a re-scaled
mean flux. For the same 
parameter settings as before, 
we find 
\be 
\sigma_8=0.841 \pm  0.009,
\ee
in perfect agreement with the original
analysis results ($\sigma_8 = 0.841 \pm 0.017$), but with the
narrower error bar.
The more narrow error bars can be related to a re-distribution
of power between different
contributions to
the total 1D power spectrum
with the new set of EFT parameters. 
When combined with Planck, 
we also find a somewhat stronger
neutrino mass constraints, 
$\sum m_\nu <0.06$ eV at $95\%$CL. 
While this result is slightly stronger 
than our baseline measurement, 
we prefer to stick to the 
latter for two reasons: 
(a) our original analysis is based
purely on the publicly 
available data and hence can be 
independently reproduced (unlike the new one, for which we had to use 
the raw Sherwood 3D simulation files
that are not in public access), 
(b) our original result
is more conservative. 
All in all, we conclude that 
the neutrino mass bounds
are only weakly sensitive to the 
the mean flux assumptions. 

Finally, a comment
is in order on the values
of the best-fit parameters 
$b_1$ and $b_\eta$. In the new
analysis the optimal values of $b_1$ cover the range $\approx [-0.5,-0.8]$,
while $b_\eta\in [0,0.2]$. While the 
$b_1$ parameter modules are somewhat 
larger than those of Sherwood, 
the $b_\eta$
values
are quite close the Sherwood ones,
implying that a more 
consistent flux calibration
is important to
improve the agreement
between our baseline EFT analysis 
and 
simulations.

\end{document}